\documentclass[a4paper,11pt]{article}
\pdfoutput=1 

\usepackage{jheppub} 

\usepackage[T1]{fontenc} 

\usepackage{tikz}

\newcommand{\AdS}{\text{AdS}}
\newcommand{\CFT}{\text{CFT}}
\newcommand{\AdSST}{\text{AdS}_3\times\text{S}^3\times\text{T}^4}
\newcommand{\Sph}{\text{S}}
\newcommand{\Tor}{\text{T}}

\newcommand{\psu}{\mathfrak{psu}}
\newcommand{\su}{\mathfrak{su}}
\newcommand{\slalg}{\mathfrak{sl}}
\newcommand{\so}{\mathfrak{so}}
\renewcommand{\u}{\mathfrak{u}}
\newcommand{\Ql}{\mathbf{Q}}
\newcommand{\Sl}{\mathbf{S}}
\renewcommand{\sl}{\mathbf{s}}
\newcommand{\Jl}{\mathbf{J}}
\newcommand{\Ll}{\mathbf{L}}
\newcommand{\Qr}{\widetilde{\mathbf{Q}}}
\newcommand{\sr}{\widetilde{\mathbf{s}}}
\newcommand{\Sr}{\widetilde{\mathbf{S}}}
\newcommand{\Jr}{\widetilde{\mathbf{J}}}
\newcommand{\Lr}{\widetilde{\mathbf{L}}}

\newcommand{\Hl}{\mathbf{H}}
\newcommand{\Hr}{\widetilde{\mathbf{H}}}
\newcommand{\C}{\mathbf{C}}
\newcommand{\T}{\mathbf{T}}

\newcommand{\ql}{\mathbf{q}}
\newcommand{\qr}{\widetilde{\mathbf{q}}}
\newcommand{\id}{\mathbf{1}}
\newcommand{\ket}[1]{\left|{#1}\right\rangle}

\newcommand{\sQl}{\mathcal{Q}}
\newcommand{\sQr}{\widetilde{\mathcal{Q}}}

\newcommand{\bs}{\boldsymbol\Sigma}

\newcommand{\h}{\mathbf{h}}

\newcommand{\B}{{\scriptscriptstyle\text{B}}}
\newcommand{\F}{{\scriptscriptstyle\text{F}}}
\renewcommand{\L}{{\scriptscriptstyle\text{L}}}
\newcommand{\R}{{\scriptscriptstyle\text{R}}}
\newcommand{\z}{{\scriptscriptstyle\text{o}}}

\title{Integrable bootstrap for AdS3/CFT2 correlation functions}

\author[a]{Burkhard Eden,}
\author[a]{Dennis le Plat,}
\author[b]{Alessandro Sfondrini}

\affiliation[a]{Institut f\"ur Mathematik und Physik, Humboldt-Universit\"at zu Berlin,\\
Zum gro{\ss}en Windkanal 6, 12489 Berlin, Germany}
\affiliation[b]{Dipartimento di Fisica e Astronomia, Universit\`a degli Studi di Padova,\\
\& Istituto Nazionale di Fisica Nucleare, Sezione di Padova,\\
via Marzolo 8, 35131 Padova, Italy}

\emailAdd{eden@math.hu-berlin.de}
\emailAdd{diplat@physik.hu-berlin.de}
\emailAdd{alessandro.sfondrini@unipd.it}

\abstract{We propose an integrable bootstrap framework for the computation of correlation functions for superstrings in  $\AdS_3\times\Sph^3\times\Tor^4$ backgrounds supported by an arbitrary mixture or Ramond-Ramond and Neveu-Schwarz-Neveu-Schwarz fluxes.
The framework extends the ``hexagon tessellation'' approach which was originally proposed for $\AdS_5\times\Sph^5$ and for the first time it demonstrates its applicability to other (less supersymmetric) setups.
We work out the hexagon form factor for two-particle states, including its dressing factors which follow from those of the spectral problem, and we show that it satisfies non-trivial consistency conditions. We propose a bootstrap principle, slightly different from that of $\AdS_5\times\Sph^5$, which allows to extend the form factor to arbitrarily many particles. Finally, we compare its predictions with some correlation functions of protected operators.
Possible applications of this construction include the study of wrapping corrections, of higher-point correlation functions, and of non-planar corrections.}

\preprint{HU-EP-21/01}

\begin{document} 
\bibliographystyle{JHEP}
\maketitle
\flushbottom

\newpage
\section{Introduction}

Our understanding of theoretical physics has always been shaped by experimental observations on the one side, and by the construction of a theoretical framework which may allow us to compute, compare and study relevant observables on the other side. Some questions having to do with the fundamental behaviour and self-consistency of a physical theory are much more easily answered when we can compute as many observables as possible exactly, without resorting to truncations, approximations or simulations. The study of exactly solvable systems is by now a large branch of physics. In the quantum world, it encompasses integrable spin-chains, lattice models, two-dimensional conformal field theories (CFTs) and integrable quantum field theories (IQFTs). Most recently, this found applications to string theory (which is indeed defined on a two-dimensional worldhseet) as well as their holographic duals in the AdS/CFT correspondence.

One of the best understood examples of AdS/CFT is the correspondence between superstrings on $\AdS_5\times \Sph^5$ and the $\mathcal{N}=4$ supersymmetric Yang-Mills theory in $D=4$ ($\mathcal{N}=4$ SYM) \cite{Maldacena:1997re,Witten:1998qj,Gubser:1998bc}. This is the AdS/CFT setup with the maximal amount of supersymmery. Despite that, both the string theory and the gauge theory look like tough nuts to crack: on the worldsheet, the superstring theory is supported by Ramond-Ramond fluxes, which makes it hard to describe it as a CFT; in the dual, the gauge theory appears to be as involved as any four-dimensional Yang-Mills theory. A remarkable simplification appears in the planar limit~\cite{tHooft:1973alw} (or large-$N$ limit, where $N$ is the number of colours), as it was first noticed by Minahan and Zarembo~\cite{Minahan:2002ve}: the one-loop spectrum of anomalous dimensions of certain operators is exactly solvable in terms of a Bethe ansatz~\cite{Bethe:1931hc,Faddeev:1996iy}. In fact, this can be extended to the all-loop planar spectrum of all local operators, though a complete understanding of the problem requires to overcome the issue of wrapping corrections. To this end it is crucial to reformulate the problem on the string worldsheet~\cite{Ambjorn:2005wa} and to introduce a mirror model~\cite{Arutyunov:2007tc}. Eventually, the whole spectral problem could be encoded in a powerful system of equations known as the quantum spectral curve~\cite{Gromov:2013pga}.  We refer the reader to refs.~\cite{Arutyunov:2009ga,Beisert:2010jr} for reviews of the AdS${}_5$/CFT${}_4$ spectral problem.
Quite remarkably, integrability survives even beyond the strict $N\to\infty$ limit. In particular, the hexagon tessellation program was introduced in reference~\cite{Basso:2015zoa} as a way to turn the computation of three-point functions of generic (non-protected) operators into the computation of an integrable form factor. In broad strokes, the idea is that we may entirely constrain, by symmetry arguments, the relevant form factor for the case of one and two particles, and bootstrap it for arbitrarily many particles. The resulting form factor takes a relatively simple form~\cite{Basso:2015zoa} in terms of the S~matrix that was found by Beisert~\cite{Beisert:2005tm} for the spectral problem.
Accounting for wrapping corrections is more problematic and, up to date, an open issue. It is possible to account for L\"uscher-type~\cite{Luscher:1985dn,Luscher:1986pf} order-by-order corrections to wrapping, see for instance refs.~\cite{Basso:2015zoa, Eden:2015ija,Basso:2015eqa,Basso:2017muf}, but it seems to be hard to push these computations much further, at least for generic functions. On the other hand, it is possible to extend such computations to more complicated correlation functions: to higher-point planar correlation functions~\cite{Eden:2016xvg,Fleury:2016ykk}, as well as to non-planar correlators~\cite{Eden:2017ozn,Bargheer:2017nne,Bargheer:2019kxb}. The main aim of this paper is to extend this very promising program beyond the case of AdS$_5$/CFT$_4$.

Another extremely interesting holographic setup is the AdS$_3$/CFT$_2$ correspondence~\cite{Maldacena:1997re}. In terms of superstring backgrounds, there are three families of backgrounds that one may consider: $\AdSST$, $\AdS_3\times\Sph^3\times \text{K3}$ and $\AdS_3\times\Sph^3\times \Sph^3\times\Sph^1$. Each of these preserves sixteen real supersymmetries, half of the amount of $\AdS_5\times\Sph^5$. The focus of this paper is on what is arguably the simplest of these setups, $\AdSST$. The case of K3 follows to a large extent from that, at least for orbifold K3s, while the case of $\AdS_3\times\Sph^3\times \Sph^3\times\Sph^1$ is interesting in its own right, but beyond the scope of our discussion. For these backgrounds there are effectively \textit{two} parameters that determine the spectrum in the planar limit: loosely speaking, this is because the background can be supported by a combination of Ramond-Ramond (RR) and Neveu-Schwarz-Neveu-Schwarz fluxes (NSNS), which affect the spectrum very differently --- see ref.~\cite{OhlssonSax:2018hgc} for a discussion of the moduli of this background and their effect on the spectrum.
When there is no NSNS flux, the background is most similar to the case of $\AdS_{5}\times\Sph^5$. When there is no RR flux, the spectrum becomes very degenerate, and a continuum appears corresponding to the so-called long strings.
For such a background (and only in this case), a simple worldsheet CFT description exists in the Ramond-Neveu-Schwarz formalism. For the $\AdS_3$ part it can be given in terms of  a supersymmetric $\mathfrak{sl}(2,\mathbb{R})$ level-$k$ Wess-Zumino-Witten model~\cite{Maldacena:2000hw,Maldacena:2000kv,Maldacena:2001km}, which can then be coupled to an $\su(2)$ WZW model and free Bosons to account for the remaining compact spaces. The special case $k=1$ requires slightly different worldsheet-CFT techniques~\cite{Berkovits:1999im}, but it is very interesting because it seems to be the only point of the whole moduli space where one has a firm handle on the CFT dual~\cite{Giribet:2018ada,Gaberdiel:2018rqv, Eberhardt:2018ouy, Eberhardt:2020bgq,Gaberdiel:2020ycd}, which should be the symmetric-product orbifold CFT of four free Bosons and as many Fermions, $\text{Sym}^N(\Tor^4)$.

Remarkably, both the pure-RR and pure-NSNS background, as well as anything in between, are classically integrable~\cite{Babichenko:2009dk, Cagnazzo:2012se}. Integrability, expressed in terms of factorised scattering, seems also to hold at the quantum level. Largely by analogy with the case of $\AdS_5\times\Sph^5$, it was first understood for pure-RR backgrounds, see ref.~\cite{Sfondrini:2014via} for a review, and then extended to mixed-flux backgrounds~\cite{Hoare:2013pma,Hoare:2013lja,Lloyd:2014bsa}. Also in this case, wrapping corrections are not entirely under control~\cite{Abbott:2015pps}. A notable exception is the case of pure-NSNS backgrounds, where the spectrum was computed by means of integrability and showed to match with the WZW prediction~\cite{Baggio:2018gct,Dei:2018mfl}, including at the special value of the level $k=1$~\cite{Sfondrini:2020ovj}.

It is natural to ask whether we may use integrability to compute three- and higher-point correlation functions of generic operators on $\AdSST$. For anything but pure-NSNS background, this would be a major advance as there is currently no technique to do so. Conversely, for pure-NSNS backgrounds, this could yield a nice comparison with the worldsheet-CFT (or RNS) approach, and possibly shine new light on how the hexagon approach relates to the CFT machinery. This is particularly interesting because the hexagon approach is formulated in terms of the target-space Fermions, and hence should be closer to the dual theory. (Worldsheet functions have recently been studied in ref.~\cite{Eberhardt:2019ywk} at level $k=1$ precisely to map them to the their holographic counterparts.)
We will see below that indeed, the hexagon approach can be used also for $\AdSST$, with arbitrary background fluxes. We will also work out the hexagon form factor for one- and two-particle states, based on symmetry, and use a factorisation principle to bootstrap the form factor for arbitrarily-many particles. This lays the basis for a systematic investigation of correlation function in the whole moduli space of $\AdSST$.

The article is structured as it follows. We decided to dedicate section~\ref{sec:review} to a rather detailed review of $\AdSST$ integrability, given that we will need many results for the spectral problem which are somewhat scattered over the literature; we take this occasion to try to fix some conventions and correct some minor misprints that are floating in the older literature. The main part of the paper is section~\ref{sec:hexagon} where we set up the hexagon program for $\AdSST$. In section~\ref{sec:check} we check our construction against a result easily available in the literature, \textit{i.e.}\ the three-point functions of certain protected operators; it should be stressed that here, unlike in $\AdS_{5}\times\Sph^5$, the spectrum of protected operators is quite rich, and their correlation functions quite non-trivial. Finally, we discuss our result and outlook in section~\ref{sec:conclusions}. We also have spelled out the full $\AdSST$ S~matrix in appendix~\ref{app:Smatrix} for the readers' convenience (again, the whole result was not explicitly written in any given article to the best of our knowledge).

\section{Review of integrability for \texorpdfstring{$\AdSST$}{AdS3xS3xT4}}
\label{sec:review}
The $\AdSST$ metric superisometries are given by
\begin{equation}
\label{eq:globalsusy}
\psu(1,1|2)\oplus \psu(1,1|2)\,,
\end{equation}
where each copy contains eight real supercharges and a Bosonic subalgebra $\su(1,1)\oplus\su(2)$. In total, the Bosonic subalgebra is then $[\su(1,1)\oplus\su(1,1)]\oplus[\su(2)\oplus\su(2)]\cong\so(2,2)\oplus \so(4)$. The factorisation of the isometries like in eq.~\eqref{eq:globalsusy} is a key feature of the $\AdS_3/\CFT_2$ duality. In particular, each of the two copies of the non-compact subalgebra $\su(1,1)\cong\slalg(2,\mathbb{R})$ correpsonds to the chiral and antichiral part of the global conformal algebra in the dual $\CFT_2$. In addition to these isometries, we have a four shift isometries $\u(1)^{\oplus4}$ corresponding to the~$\Tor^4$ directions. Finally, the flat manifold $\Tor^4$ enjoys a \emph{local} $\so(4)_{\T^4}$ rotational symmetry, which for later convenience we also decompose as $\so(4)_{\Tor^4}\cong\su(2)_{\bullet}\oplus\su(2)_{\circ}$. This \textit{is not} a symmetry of the whole theory (it is broken by the boundary conditions of the $\Tor^4$ fields) but it will play an important role nonetheless, for two reasons. First of all, locally the Killing spinors will be charged under $\so(4)_{\Tor^4}$, which will make this algebra useful to group the symmetry generators; secondly but importantly, states with no momentum or winding along $\Tor^4$ are blind to its global features, and as long as we restrict to those (as we will do), this rotational isometry will be important.

The complete type IIB superstring background features additional fields beyond the $\AdSST$ metric. Generically, the background will involve  Neveu-Schwarz-Neveu-Schwarz (NSNS) and Ramond-Ramond (RR) fluxes. To be concrete, we will consider a background with a NSNS three-form flux $H=dB$ proportional to the volume form of $\AdS_3\times\Sph^3$ and a RR three form flux $F$ also proportional to the same volume form. Such a background can be thought of as arising from the F1-NS5-D1-D5 system. Its dynamics is dominated by two parameters: the amount of NSNS fluxes and the amount of RR fluxes (in units where the $\AdS_3$ radius is set to one). Within this two-parameter space, there are two interesting limits: the case where RR fluxes are absent, which can be described by a supersymmetric $\slalg(2,\mathbb{R})$ WZW model and corresponds to the F1-NS5 system, and the case where $H=B=0$ which corresponds to the D1-D5 system. The latter is most similar to $\AdS_5\times\Sph^5$. The whole two-parameter case is classically integrable~\cite{Babichenko:2009dk,Cagnazzo:2012se} and is believed to be integrable at the quantum level, see \cite{Sfondrini:2014via} for a review. It is worth emphasising that the F1-NS5-D1-D5 system has more moduli than the two we just introduced~\cite{Larsen:1999uk}. However, in the near horizon limit and when restricting to states with no momentum or winding along $\Tor^4$, only these two modules end up being important, see ref.~\cite{OhlssonSax:2018hgc} for a detailed discussion.

The classical integrability of strings on $\AdSST$ was discussed in~\cite{Babichenko:2009dk,Cagnazzo:2012se}. The study of the integrable S matrix was initiated in~\cite{Borsato:2013qpa,Hoare:2013lja} and completed, for the matrix part, in refs.~\cite{Borsato:2014exa,Borsato:2014hja,Lloyd:2014bsa}. The dressing factors were studied in \cite{Borsato:2013hoa,Borsato:2016xns,Fontanella:2019baq} for backgrounds without NS-NS fluxes. The S matrix (including dressing factors) for the case of NS-NS fluxes \textit{only}, is also known and was worked out in \cite{Baggio:2018gct,Dei:2018mfl}, where it was also shown that the resulting mirror TBA reproduces the WZW spectrum. It is worth emphasising that taking the pure-NSNS limit in the S~matrix is subtle because all excitations become massless then. 
Besides, the dressing factors for the generic mixed-flux backgrounds are not known (see \cite{Babichenko:2014yaa} for work in this direction).
Even if most of the integrability construction is reviewed in \cite{Sfondrini:2014via}, some details and especially the latest developments are scattered over the literature. Hence we find it useful to review its main features below and collect some formulae in the appendices.

\subsection{Supersymmetry algebra}
\label{sec:algebra}
The supersymmetry algebra is given by two copies of $\psu(1,1|2)$ as in eq.~\eqref{eq:globalsusy}. The first copy, which we label ``left'', is given by eight supercharges $\Ql_{m \alpha a}$ ($\alpha=\pm, a=\pm, m=\pm$), three $\su(2)$ (R-symmetry) generators $\Jl_{\alpha}$ ($\alpha=\pm,3$) and three $\su(1,1)$ generators $\Ll_{m}$ ($m=\pm,0$). Notice that the supercharges carry an index $a=\pm$, due to the fact that they transform in the fundamental representation of an $\su(2)$ automorphism. In fact, it can be seen geometrically~\cite{Borsato:2014hja} that such an automorphism is a subalgebra of $\so(4)_{\Tor^4}$, and in what follows we will label it $\su(2)_\bullet$. The second copy, which we call ``right'' and denote with tildas, is given by $\Qr_{\dot{m}\dot{\alpha}a}$, $\Jr_{\dot{\alpha}}$ and $\Lr_{\dot{m}}$. Note however that these charges are charged under the \textit{same} $\su(2)_{\bullet}$ automorphism as the ``left'' ones. The names ``left'' and ``right'' correspond to the interpretation of these charges as symmetries of the two-dimensional dual superconformal symmetry with $\mathcal{N}=(4,4)$ symmetry. As remarked above, the two (R-symmetry) $\su(2)$ algebras give the isometries of the three sphere, $\su(2)\oplus\su(2)\cong\so(4)$, and the two $\su(1,1)$ give the $\AdS_3$ isometries, $\su(1,1)\oplus\su(1,1)\cong\so(2,2)$.

\subsubsection{Matrix realisation and Weyl-Cartan basis for the ``left algebra''}
A convenient matrix realisation of $\su(1,1|2)$ is given by $(2|2)$ supermatrices, which can be written as blocks
\begin{equation}
    M=\left(\begin{array}{c|c}
        m &\theta  \\\hline
         \eta& n
    \end{array}\right),
\end{equation}
where Latin letters are bosonic blocks and Greek ones are fermionic. The matrix $M$ must satisfy
\begin{equation}
    M^\dagger+ \Sigma\,M\,\Sigma = 0\,,\qquad
    \Sigma=\text{diag}=(1,-1,1,1)\,.
\end{equation}
We can introduce the following explicit parametrisations for the complexified algebra
\begin{equation}
    \Ll_0 = \frac{1}{2}\left(\begin{array}{cc|cc}
-1 & 0 & 0 & 0  \\
 0 & 1 & 0 & 0  \\\hline
 0 & 0 & 0 & 0  \\
 0 & 0 & 0 & 0  
    \end{array}\right),\qquad    \Ll_+ = \left(\begin{array}{cc|cc}
 0 & 0 & 0 & 0  \\
 1 & 0 & 0 & 0  \\\hline
 0 & 0 & 0 & 0  \\
 0 & 0 & 0 & 0  
    \end{array}\right),\qquad    \Ll_- = \left(\begin{array}{cc|cc}
 0 & 1 & 0 & 0  \\
 0 & 0 & 0 & 0  \\\hline
 0 & 0 & 0 & 0  \\
 0 & 0 & 0 & 0  
    \end{array}\right),
\end{equation}
\begin{equation}
    \Jl_3 = \frac{1}{2}\left(\begin{array}{cc|cc}
 0 & 0 & 0 & 0  \\
 0 & 0 & 0 & 0  \\\hline
 0 & 0 & 1 & 0  \\
 0 & 0 & 0 &-1  
    \end{array}\right),\qquad    \Jl_+ = \left(\begin{array}{cc|cc}
 0 & 0 & 0 & 0  \\
 0 & 0 & 0 & 0  \\\hline
 0 & 0 & 0 & 1  \\
 0 & 0 & 0 & 0  
    \end{array}\right),\qquad    \Jl_- = \left(\begin{array}{cc|cc}
 0 & 0 & 0 & 0  \\
 0 & 0 & 0 & 0  \\\hline
 0 & 0 & 0 & 0  \\
 0 & 0 & 1 & 0  
    \end{array}\right),
\end{equation}
\begin{equation}
    \Ql_{--1} = \left(\begin{array}{cc|cc}
 0 & 0 & 0 & 0  \\
 0 & 0 & 0 & 0  \\\hline
 0 & 0 & 0 & 0  \\
 0 &-1 & 0 & 0  
    \end{array}\right),\qquad    \Ql_{++2} = \left(\begin{array}{cc|cc}
 0 & 0 & 0 & 0  \\
 0 & 0 & 0 & 1  \\\hline
 0 & 0 & 0 & 0  \\
 0 & 0 & 0 & 0  
    \end{array}\right),
\end{equation}
\begin{equation}
    \Ql_{++1} = \left(\begin{array}{cc|cc}
 0 & 0 & 0 & 0  \\
 0 & 0 & 0 & 0  \\\hline
 1 & 0 & 0 & 0  \\
 0 & 0 & 0 & 0  
    \end{array}\right),\qquad    \Ql_{--2} = \left(\begin{array}{cc|cc}
 0 & 0 &-1 & 0  \\
 0 & 0 & 0 & 0  \\\hline
 0 & 0 & 0 & 0  \\
 0 & 0 & 0 & 0  
    \end{array}\right),
\end{equation}
\begin{equation}
    \Ql_{+-1} = \left(\begin{array}{cc|cc}
 0 & 0 & 0 & 0  \\
 0 & 0 & 0 & 0  \\\hline
 0 & 0 & 0 & 0  \\
 1 & 0 & 0 & 0  
    \end{array}\right),\qquad    \Ql_{-+2} = \left(\begin{array}{cc|cc}
 0 & 0 & 0 & 1  \\
 0 & 0 & 0 & 0  \\\hline
 0 & 0 & 0 & 0  \\
 0 & 0 & 0 & 0  
    \end{array}\right),
\end{equation}
\begin{equation}
    \Ql_{-+1} = \left(\begin{array}{cc|cc}
 0 & 0 & 0 & 0  \\
 0 & 0 & 0 & 0  \\\hline
 0 &-1 & 0 & 0  \\
 0 & 0 & 0 & 0  
    \end{array}\right),\qquad    \Ql_{+-2} = \left(\begin{array}{cc|cc}
 0 & 0 & 0 & 0  \\
 0 & 0 &-1 & 0  \\\hline
 0 & 0 & 0 & 0  \\
 0 & 0 & 0 & 0  
    \end{array}\right),
\end{equation}
These satisfy the commutation relations
\begin{equation}
\begin{aligned}
    &[\Ll_0,\Ll_\pm]=\pm\Ll_\pm\,,\quad &&[\Ll_+,\Ll_-]=2\Ll_0\,,\\
    &[\Jl_3,\Jl_\pm]=\pm\Jl_\pm\,,\quad &&[\Jl_+,\Jl_-]=2\Jl_3\,,\\
    &[\Ll_0,\Ql_{\pm\alpha A}]=\pm\frac{1}{2}\Ql_{\pm\alpha A}\,,\qquad
    &&[\Ll_\pm,\Ql_{\mp\alpha A}]=\Ql_{\pm\alpha A}\,,\\
    &[\Jl_3,\Ql_{a\pm A}]=\pm\frac{1}{2}\Ql_{a\pm A}\,,\qquad
    &&[\Jl_\pm,\Ql_{a\mp A}]=\Ql_{a\pm A}\,,\\
    &\{\Ql_{\pm + A},\Ql_{\pm + B}\}=\pm\epsilon_{AB}\Ll_{\pm}\,,\qquad
    &&\{\Ql_{+\pm A},\Ql_{-\pm B}\}=\mp\epsilon_{AB}\Jl_{\pm}\,,
\end{aligned}
\end{equation}
and finally
\begin{equation}
    \{\Ql_{+\pm A},\Ql_{-\mp B}\}=\epsilon_{AB}\big(-\Ll_0\pm \Jl_3\big)\,.
\end{equation}

The Weyl-Cartan basis of the algebra is as it follows:
\begin{equation}
    [\mathfrak{h}_i,\mathfrak{h}_j]=0\,,\qquad
    [\mathfrak{e}_i,\mathfrak{f}_j]=\delta_{ij}\mathfrak{h}_j,\qquad
    [\mathfrak{h}_i,\mathfrak{e}_j]=A_{ij}\mathfrak{e}_j,\quad
    [\mathfrak{h}_i,\mathfrak{f}_j]=-A_{ij}\mathfrak{f}_j,
\end{equation}
with
\begin{equation}
    \begin{aligned}
    &\mathfrak{h}_1=-\Ll_0-\Jl_3,&\quad&\mathfrak{e}_1=+\Ql_{+-1}&\quad&\mathfrak{f}_1=+\Ql_{-+2},\\
    &\mathfrak{h}_2=2\Jl_3,&&\mathfrak{e}_2=+\Jl_+&&\mathfrak{f}_2=+\Jl_-,\\
    &\mathfrak{h}_3=-\Ll_0-\Jl_3,&&\mathfrak{e}_3=+\Ql_{+-2}&&\mathfrak{f}_3=-\Ql_{-+1}\,,
    \end{aligned}
\end{equation}
with Cartan matrix
\begin{equation}
    A=\left(\begin{array}{ccc}
     0 &  -1  &  0 \\
    -1 &  +2  & -1 \\
     0 &  -1  &  0
    \end{array}\right)\,.
\end{equation}

\subsubsection{BPS condition and ``left Hamiltonian''}
In this notation, the BPS condition for the algebra is
\begin{equation}
    -\Ll_0-\Jl_3 \geq 0\,.
\end{equation}
This positive-semidefinite operator can therefore be used to define a ``Hamiltonian'' for the left algebra. Indeed it can be shown that this is precisely the contribution of left charges to the  light-cone Hamiltonian~\cite{Borsato:2013qpa,Borsato:2014hja}
\begin{equation}
    \Hl \equiv -\Ll_0-\Jl_3\,.
\end{equation}
The supercharges that commute with $\Hl$ are
\begin{equation}
\label{eq:simplerootsleft}
\begin{aligned}
    &\Ql^{1}\equiv \mathfrak{f}_1=+\Ql_{-+2},\quad
    &&\Ql^{2}\equiv\mathfrak{f}_3=-\Ql_{-+1},\\
    &\Sl_{1}\equiv\mathfrak{e}_1=+\Ql_{+-1},\quad
    &&\Sl_{2}\equiv\mathfrak{e}_3=+\Ql_{+-2}.
\end{aligned}
\end{equation}
They form the algebra $\psu(1|1)^{\oplus 2}\oplus \mathfrak{u}(1)$
\begin{equation}
    \big\{ \Ql^{A},\,\Sl_{B}\big\}= \delta^{A}{}_{B}\,\Hl\geq 0\,.
\end{equation}

The charges satisfy the Hermiticity conditions
\begin{equation}
\begin{aligned}
    &(\Ql_{+-1})^\dagger = +\Ql_{-+2}\,,\qquad
    &&(\Ql_{-+2})^\dagger = +\Ql_{+-1}\,,\\
    &(\Ql_{-+1})^\dagger = - \Ql_{+-2}\,,\qquad
    &&(\Ql_{+-2})^\dagger = - \Ql_{-+1}\,,
\end{aligned}
\end{equation}
or equivalently
\begin{equation}
    (\Ql^{A})^\dagger = \Sl_{A},\qquad
    (\Sl_{A})^\dagger = \Ql^{A},
\end{equation}

\subsubsection{Weyl-Cartan basis for the ``right algebra''}
Let us now come to the second copy of $\psu(1,1|2)$, the ``right'' algebra. The construction of the matrix representation is identical. We will however pick a slightly different Weyl-Cartan basis. Denoting the generators by tildes, we have
\begin{equation}
    \begin{aligned}
    &\tilde{\mathfrak{h}}_1=+\Lr_0+\Jr_3,& \quad&\tilde{\mathfrak{e}}_1=+\Qr_{-+1}& \quad&\tilde{\mathfrak{f}}_1=+\Qr_{+-2},\\
    &\tilde{\mathfrak{h}}_2=-2\Lr_0,& &\tilde{\mathfrak{e}}_2=+\Lr_+& &\tilde{\mathfrak{f}}_2=-\Lr_-,\\
    &\tilde{\mathfrak{h}}_3=+\Lr_0+\Jr_3,& &\tilde{\mathfrak{e}}_3=+\Qr_{-+2}& &\tilde{\mathfrak{f}}_3=-\Qr_{+-1}\,,
    \end{aligned}
\end{equation}
with Cartan matrix
\begin{equation}
    \widetilde{A}=\left(\begin{array}{ccc}
     0 &  +1  &  0 \\
    +1 &  -2  & +1 \\
     0 &  +1  &  0
    \end{array}\right)\,.
\end{equation}
Running a little ahead of ourselves, let us motivate this choice. Given that the algebra~\eqref{eq:globalsusy} is factorised, we can choose the positive roots in either copy of the algebra independently. It will turn out however that, when considering a certain class of ``off-shell'' observables, the symmetries will be extended by two central charges that couple the left and right algebras~\cite{Borsato:2013hoa}. In that case, our choice of positive roots will prove convenient.

\subsubsection{BPS condition and ``right Hamiltonian''}
Once again we define
\begin{equation}
    \Hr\equiv -\Lr_0-\Jr_3\geq0\,,
\end{equation}
and
\begin{equation}
\label{eq:simplerootsright}
\begin{aligned}
    &\Qr_{1}\equiv +\mathfrak{e}_1=+\Qr_{-+1},\quad
    &&\Qr_{2}\equiv+\mathfrak{e}_3=+\Qr_{-+2},\\
    &\Sr^{1}\equiv-\mathfrak{f}_1=-\Qr_{+-2},\quad
    &&\Sr^{2}\equiv-\mathfrak{f}_3=+\Qr_{+-1},
\end{aligned}
\end{equation}
so that
\begin{equation}
    \{\Qr_A,\Sr^B\}=\delta_A{}^B\,\Hr\geq0\,.
\end{equation}
Like before we have that
\begin{equation}
    (\Qr^{A})^\dagger = \Sr_{A},\qquad
    (\Sr_{A})^\dagger = \Qr^{A}.
\end{equation}

\subsection{Centrally extended off-shell symmetry algebra}
\label{sec:centrallyextendedalgebra}
Much like in the case of $\AdS_5/\CFT_4$~\cite{Beisert:2005tm, Arutyunov:2006ak}, the algebra relevant for integrability features a \emph{central extension} with respect to the superisometry algebra. This central extension annihilates all physical states. However, it acts nontrivially on the individual worldsheet excitations that make up a physical state (or, in spin-chain language, on the magnons that make up the Bethe state). We refer the reader to~\cite{Arutyunov:2009ga,Beisert:2010jr} for reviews of the construction in the $\AdS_5/\CFT_4$ setup.

For our purposes, it will be sufficient at this stage to recall~\cite{Borsato:2013qpa} how the algebra of symmetries commuting with $\Hl$ and $\Hr$ may be extended. In the notation just introduced, the algebra is
\begin{equation}
        \Big\{\Ql^A,\, \Sl_B\Big\} = \Hl\,\delta^A{}_B\,,\qquad
        \Big\{\Qr_A,\, \Sr^{B}\Big\} = \Hr\,\delta_A{}^B\,,
\end{equation}
This allows for a central extension. It is possible to check semiclassically that the central extension appears for $\AdSST$ backgrounds with Ramond-Ramond flux~\cite{Borsato:2014hja}. Introducing two central charges $\mathbf{P}$ and $\mathbf{K}$ we have
\begin{equation}
    \Big\{\Ql^A,\, \Qr_B\Big\} = \mathbf{P}\,\delta^A{}_B\,,\qquad
    \Big\{\Sl_A\,, \Sr^{B}\Big\} = \mathbf{K}\,\delta_A{}^B\,,
\end{equation}
where the reality conditions discussed above imply that for a unitary representation $\mathbf{K}^\dagger=\mathbf{P}$ and $\mathbf{P}^\dagger=\mathbf{K}$.
In presence of this central extension, our choice of simple roots \eqref{eq:simplerootsleft} and \eqref{eq:simplerootsright} appears natural. Since $\mathbf{K}$ is central, if $\Ql^A$ is a \textit{negative} root, then $\Qr_A$ needs to be a \textit{positive} root, and similarly for $\Sr^A$ and $\Sl_A$.

\subsubsection{Factorisation of the centrally extended algebra}
\label{sec:algebrafactorisation}
The algebra above is $\psu(1|1)^{\oplus4}$ centrally extended, which plays a role similar to $\su(2|2)^{\oplus2}$ in $\AdS_5/\CFT_4$. In the latter case, the factorisation in $\su(2|2)^{\oplus2}$ was quite useful in simplifying many computations: in particular, it was sufficient to work out a $\su(2|2)$-invariant S~matrix~\cite{Beisert:2005tm} which served as a building block of the full $\su(2|2)^{\oplus2}$-invariant S~matrix. To emphasise the similarity in the factorised structure, we introduce the $\psu(1|1)^{\oplus2}$ centrally extended algebra, given by
\begin{equation}
\label{eq:smallalgebra}
    \Big\{\ql,\, \sl\Big\} = \Hl,\quad
    \Big\{\qr, \sr\Big\} = \Hr\,,
    \qquad
    \Big\{\ql,\, \qr\Big\} = \mathbf{P},\quad
    \Big\{\sl, \sr\Big\} = \mathbf{K}\,,
\end{equation}
The algebra in eq.~\eqref{eq:smallalgebra} plays the role that $\su(2|2)$ plays for $\AdS_5/\CFT_4$. 
We can then obtain the larger algebra by setting
\begin{equation}
\label{eq:productform}
    \Ql^1\equiv \ql\otimes \id\,,\quad \Ql^2\equiv \bs\otimes \ql,\qquad
    \Sl_1\equiv \sl\otimes \id\,,\quad\Sl_2\equiv \bs\otimes \sl\,,
\end{equation}
where $\bs$ is the graded identity, $\bs=\delta_{ij}(-1)^{F_j}$ and (note the lowered indices)
\begin{equation}
    \Qr_1\equiv \qr\otimes \id\,,\quad \Qr_2\equiv \bs\otimes \qr,\qquad
    \Sr^{1}\equiv \sr\otimes \id\,,\quad\Sr^{2}\equiv \bs\otimes \sr\,.
\end{equation}
To show that this gives the same $\psu(1|1)^{\oplus4}$ centrally extended as above, note that on any supercharge we have e.g.
\begin{equation}
    \bs\, \ql\, \bs = - \ql\,.
\end{equation}
Indices are raised and lowered with the Levi-Civita symbol with $\varepsilon^{12}=-\varepsilon_{12}=1$.

\subsection{Short representations of the light-cone symmetry algebra}
\label{sec:representations}

Having identified the algebra that commutes with the left and right Hamiltonians $\Hl$ and $\Hr$, as well as its central extension, it is time to construct its short representation. Worldsheet excitations will transform in these representations~\cite{Borsato:2013hoa,Borsato:2014hja}.
It is convenient to start from the smaller algebra~\eqref{eq:smallalgebra}.

\subsubsection{Short representations of \texorpdfstring{$\psu(1|1)^{\oplus2}$}{psu(1|1)**2} centrally extended}
We are interested in the short representations of the smaller algebra~\eqref{eq:smallalgebra}. Let $\ket{\phi}$ be a highest weight state, and let us say that $\ql$ is a lowering operator. Then it must be $\sl\ket{\phi}=0$, because $\sl$ is a raising operator. From the commutation relation involving $\mathbf{P}$ we see that $\qr$ must also act as a raising operator on $\ql\ket{\phi}$. The representation is short if we can assume that  $\qr\ket{\phi}=0$ and $\sr(\ql\ket{\phi})=0$, so that no new states are generated and the representation is two-dimensional. In that case we can write
\begin{equation}
\begin{aligned}
    0&=\big[\sr\qr\ql-\sr\qr\ql\big]\ket{\phi}=
    \big[(\sr\qr+\qr\sr)\ql-\qr\sr\ql-\sr(\qr\ql+\ql\qr)\big]\ket{\phi}\\
    &= \big[\Hr\ql-\qr\sr\ql-\sr\mathbf{P}\big]\ket{\phi} =  \big[\Hr\ql-\sr\mathbf{P}\big]\ket{\phi} \,.
\end{aligned}
\end{equation}
By taking the anti-commutator of this expression with $\sl$ we can find a condition which depends only on the central charges and therefore applies to the whole representation (not only to $\ket{\phi}$).
\begin{equation}
\label{eq:shortening}
    \Hl\,\Hr=\mathbf{P}\,\mathbf{K}\,,\qquad \text{on the representation.}
\end{equation}
Interestingly, if $\mathbf{P}=0$ it must be either $\Hl=0$ or $\Hr=0$, \textit{i.e.}, the representation is chiral.
(When $\Hl=\Hr=0$ the representation decomposes in two unidimensional singlet representations.)
We will see that such chiral representations appear for all physical states, as well as for any state in a theory with no RR fluxes.
In conclusion, the only short representations (besides singlets) are two dimensional, they consist of a Boson and one Fermion, and we indicate them as $\mathbf{(1|1)}$.

A short representation with highest weigth state $|\phi\rangle$  is parametrised by the eigenvalues of the central charges, $(P,K, H,\widetilde{H})$. The shortening condition~\eqref{eq:shortening} implies that, if the representation is unitary,
\begin{equation}
    P\,K\geq 0\,.
\end{equation}
For this reason for unitary representations we will henceforth indicate
\begin{equation}
    \C \equiv \mathbf{P}\,,\qquad\C^\dagger\equiv \mathbf{K}\,.
\end{equation}
The representation has the form
\begin{equation}
    \ql\, |\phi\rangle = a\,|\varphi\rangle,\quad
    \sl\, |\varphi\rangle = a^*\,|\phi\rangle,\quad
    \sr\, |\phi\rangle = b^*\,|\varphi\rangle,\quad
    \qr\, |\varphi\rangle = b\,|\phi\rangle\,,
\end{equation}
where $a,b\in\mathbb{C}$. Note that on this representation
\begin{equation}
    |\phi\rangle = \text{highest-weight state}\,,\qquad
    |\varphi\rangle = \text{lowest-weight state}\,,
\end{equation}
Note that $|\phi\rangle$ and $|\varphi\rangle$ must have opposite statistics. We get two distinct type of representations by setting $|\phi\rangle$ to be a Boson or a Fermion,
\begin{equation}
    \phi\to\phi^\B \equiv \text{Boson}\,,\qquad
    \varphi\to \varphi^\F\equiv\text{Fermion}\,,
\end{equation}
or viceversa
\begin{equation}
    \phi\to\phi^\F \equiv \text{Fermion}\,,\qquad
    \varphi\to \varphi^\B\equiv\text{Boson}.
\end{equation}

Finally the central charges take the form:
\begin{equation}
\begin{aligned}
\C=C\,\id= ab\,\id,&\qquad
\C^\dagger=C^*\,\id= (ab)^*\,\id,\\
\Hl=H\,\id= |a|^2\,\id,&\qquad
\Hr=\widetilde{H}\,\id= |b|^2\,\id.
\end{aligned}
\end{equation}
One can solve for $a,b$ as a functions of $(C, H,\widetilde{H})$.

\subsubsection{Physical values of the central charges}
We can parametrise the central charges themselves in terms of the coupling constants and the momentum~$p$ of the magnon~\cite{Borsato:2013qpa}
\begin{equation}
    C=+i\frac{h}{2}\,(e^{ip}-1)e^{2i\xi}\,,\qquad
   C^*=-i\frac{h}{2}\,(e^{-ip}-1)e^{-2i\xi}\,,
\end{equation}
where $\xi$ is a representation coefficient related to an automorphism of the algebra. As explained in ref.~\cite{Arutyunov:2009ga}, $\xi$  arises from the boundary conditions of the fields, and is important to establish the coproduct of the algebra.
Notice that $C=C^*=0$ when $p=0$ $\text{mod}2\pi$, which is the case for physical states. Here  $h\geq0$ is a property of the background: the amount of RR background flux. In what follows, we will be interested in the ``most-symmetric coproduct''~\cite{Borsato:2013qpa}, and we will set 
\begin{equation}
\label{eq:Cisreal}
C=C^*=-h\,\sin(p/2)\,.
\end{equation}

Coming to the remaining central charges, let us consider the combinations
\begin{equation}
\label{eq:centralcharges}
\begin{aligned}
&\mathbf{E}\equiv \Hl+\Hr=-\Ll_0-\Lr_0-\Jl_{3}-\Jr_{3}\geq0\,,\\
&\mathbf{M}\equiv \Hl-\Hr=-\Ll_0+\Lr_0-\Jl_{3}+\Jr_{3}\,,
\end{aligned}
\end{equation}
For physical states ($p=0$ mod$2\pi$), the eigenvalues of $\mathbf{M}$ should be quantised in integers. For Bosonsic states this is obvious as the $\AdS_3$ and $\Sph^3$ spins are integer. For Fermionic states, both spins are half-integer, so that the total spin in $\mathbf{M}$ is integer.
It turns out that it is~\cite{Hoare:2013lja,Lloyd:2014bsa}
\begin{equation}
\label{eq:mcentralcharge}
M=|a|^2+|b|^2=\frac{k}{2\pi}p + m\,,\qquad m\in\mathbb{Z}\,.
\end{equation} 
Here $k=1,2,3,\dots$ is a property of the string background and measures the amount of NSNS flux, which is quantised. In the special case where $h=0$ and $k>0$, then $k$ is precisely the level of the supersymmetric WZW model describing the worldsheet theory. Before commenting more on $m$, let us use the shortening condition~\eqref{eq:shortening} to express the last central charge $\mathbf{E}$ as
\begin{equation}
\label{eq:dispersion}
    E^2 = M^2+|C|^2\,,\qquad
    E =\sqrt{\left(m+\frac{k}{2\pi}p\right)^2+h^2\sin^2\frac{p}{2}}\,.
\end{equation}
It is clear that $m$ plays the role of a mass in the dispersion relation. Therefore we introduce the following nomenclature:
\begin{itemize}
\item $m=0$: we call the representation massless. Here we have that $E=0$ at $p=0$ for any value of $h$ and~$k$.
\item $m=+1,+2,\dots$: we call these representations ``left'' because at $p=0$ we have that $E=m$ and $M=m$, which implies $H=m>0$ and $\widetilde{H}=0$.
\item $m=-1,-2,\dots$: we call these representations ``right'' because at $p=0$ we have that $E=-m$ and $M=m$, which implies $H=0$ and $\widetilde{H}=-m>0$.
\end{itemize}
We can further distinguish the case $|m|=1$ which corresponds to fundamental particles, from the case of $|m|=2,3,\dots$, which corresponds to bound states thereof~\cite{Borsato:2013hoa,Hoare:2013lja}.
It is worth emphasising that, unlike what happens in $\AdS_{5}\times\Sph^5$, bound-state modules have the same dimension as fundamental particle modules --- they are two-dimensional.

\subsubsection{Four irreducible representations of \texorpdfstring{$\psu(1|1)^{\oplus2}$}{psu(1|1)**2} c.e.}
\label{sec:irrepssmallalgebra}

Four irreducible representations will be important in what follows. We denote them by
\begin{equation}
    \rho_{\L}=(\phi_{\L}^\B|\varphi_{\L}^\F)\,,\qquad
    \rho_{\R}=(\phi_{\R}^\F|\varphi_{\R}^\B)\,,\qquad
    \rho_\z=(\phi_{\z}^\B|\varphi_{\z}^\F)\,,\qquad
    \rho_\z'=(\phi_{\z}^\F|\varphi_{\z}^\B)\,,
\end{equation}
where the first state is always the highest-weight state,
\begin{equation}
    |\phi_*^*\rangle = \text{highest-weight state}\,,\qquad
    |\varphi_*^*\rangle = \text{lowest-weight state}\,.
\end{equation}
The representations take the same form up to relabeling the representation coefficients:
\begin{equation}
\begin{aligned}
    &\ql\, |\phi_\L^\B\rangle = a_\L\,|\varphi_\L^\F\rangle,\ 
    &&\sl\, |\varphi_\L^\F\rangle = a^*_\L\,|\phi_\L^\B\rangle,\ 
    &&\sr\, |\phi_\L^\B\rangle = b^*_\L\,|\varphi_\L^\F\rangle,\ 
    &&\qr\, |\varphi_\L^\F\rangle = b_\L\,|\phi_\L^\B\rangle\,,\\[0.2cm]
    &\ql\, |\phi_\R^\F\rangle = a_\R\,|\varphi_\R^\B\rangle,\ 
    &&\sl\, |\varphi_\R^\B\rangle = a^*_\R\,|\phi_\R^\F\rangle,\ 
    &&\sr\, |\phi_\R^\F\rangle = b^*_\R\,|\varphi_\R^\B\rangle,\ 
    &&\qr\, |\varphi_\R^\B\rangle = b_\R\,|\phi_\R^\F\rangle\,,\\[0.2cm]
    &\ql\, |\phi_\z^\B\rangle = a_\z\,|\varphi_\z^\F\rangle,\ 
    &&\sl\, |\varphi_\z^\F\rangle = a^*_\z\,|\phi_\z^\B\rangle,\ 
    &&\sr\, |\phi_\z^\B\rangle = b^*_\z\,|\varphi_\z^\F\rangle,\ 
    &&\qr\, |\varphi_\z^\F\rangle = b_\z\,|\phi_\z^\B\rangle\,,\\[0.2cm]
    &\ql\, |\phi_\z^\F\rangle = a_\z\,|\varphi_\z^\B\rangle,\ 
    &&\sl\, |\varphi_\z^\B\rangle = a^*_\z\,|\phi_\z^\F\rangle,\ 
    &&\sr\, |\phi_\z^\F\rangle = b^*_\z\,|\varphi_\z^\B\rangle,\ 
    &&\qr\, |\varphi_\z^\B\rangle = b_\z\,|\phi_\z^\F\rangle\,.
    \label{eq:irreducibleRepresentations}
\end{aligned}    
\end{equation}

The explicit form of the representation coefficients can be given in terms of Zhukovski variables, much like in AdS5. 
We will be able to describe all representation  parameters by introducing different sets of Zhukovsky variables:
\begin{equation}
\label{eq:abparam}
\begin{aligned}
&a_\L=e^{i\xi}\eta_{\L,p} \,,\ 
&&b_\L=-e^{i\xi}\frac{e^{-i p/2}}{x^-_{\L,p}}\eta_{\L,p}\,,
\quad
&&a^*_\L=e^{-i\xi}e^{-ip/2}\eta_{\L,p} \,,\ 
&&b^*_\L=-e^{-i\xi}\frac{1}{x^{+}_{\L,p}}\eta_{\L,p},\\
&b_\R=e^{i\xi}\eta_{\R,p} \,,\ 
&&a_\R=-e^{i\xi}\frac{e^{-i p/2}}{x^-_{\R,p}}\eta_{\R,p}\,,
\quad
&&b^*_\R=e^{-i\xi}e^{-ip/2}\eta_{\R,p} \,,\ 
&&a^*_\R=-e^{-i\xi}\frac{1}{x^{+}_{\R,p}}\eta_{\R,p},\\
&a_\z=e^{i\xi}\eta_{\z,p} \,,\ 
&&b_\z=-e^{i\xi}\frac{e^{-i p/2}}{x^-_{\z,p}}\eta_{\z,p}\,,
\quad
&&a^*_\z=e^{-i\xi}e^{-ip/2}\eta_{\z,p} \,,\ 
&&b^*_\z=-e^{-i\xi}\frac{1}{x^{+}_{\z,p}}\eta_{\z,p}.
\end{aligned}
\end{equation}
The $\eta$ parameter is always
\begin{equation}
\label{eq:etaparameter}
    \eta_{*,p}=e^{ip/4}\sqrt{\frac{ih}{2}(x^-_{*,p}-x^+_{*,p})}\,,
\end{equation}
where we indicated with ``$*$'' the symbols L, R, o. The Zhukovsky variables, instead, satisfy
\begin{equation}
\label{eq:xpm-shortening}
\begin{aligned}
x^+_{\L,p}+\frac{1}{x^+_{\L,p}}-x^-_{\L,p}-\frac{1}{x^-_{\L,p}}&=&\frac{2i\,\big(1+\tfrac{k}{2\pi}p\big)}{h}\,,\\
x^+_{\R,p}+\frac{1}{x^+_{\R,p}}-x^-_{\R,p}-\frac{1}{x^-_{\R,p}}&=&\frac{2i\,\big(1-\tfrac{k}{2\pi}p\big)}{h}\,,\\
x^+_{\z,p}+\frac{1}{x^+_{\z,p}}-x^-_{\z,p}-\frac{1}{x^-_{\z,p}}&=&\frac{2i\,\big(0+\tfrac{k}{2\pi}p\big)}{h}\,,
\end{aligned}
\end{equation}
and can be parametrised as it follows
\begin{equation}
\label{eq:xpm}
\begin{aligned}
    x^{\pm}_{\L,p} &=& \frac{e^{\pm i p /2}}{2h\,\sin \big(\tfrac{p}{2}\big)} \Bigg(\big(1+\tfrac{k}{2\pi}p\big)+\sqrt{\big(1+\tfrac{k}{2\pi}p\big)^2+4h^2\sin^2\big(\tfrac{p}{2}\big)}\Bigg)\,,\\
    x^{\pm}_{\R,p} &=& \frac{e^{\pm i p /2}}{2h\,\sin \big(\tfrac{p}{2}\big)} \Bigg(\big(1-\tfrac{k}{2\pi}p\big)+\sqrt{\big(1-\tfrac{k}{2\pi}p\big)^2+4h^2\sin^2\big(\tfrac{p}{2}\big)}\Bigg)\,,\\
    x^{\pm}_{\z,p} &=& \frac{e^{\pm i p /2}}{2h\,\sin \big(\tfrac{p}{2}\big)} \Bigg(\big(0+\tfrac{k}{2\pi}p\big)+\sqrt{\big(0+\tfrac{k}{2\pi}p\big)^2+4h^2\sin^2\big(\tfrac{p}{2}\big)}\Bigg)\,.
\end{aligned}
\end{equation}
It satisfies~\eqref{eq:xpm-shortening} as well as
\begin{equation}
    x^+_{*,p}-\frac{1}{x^+_{*,p}}-x^-_{*,p}+\frac{1}{x^-_{*,p}}=\frac{2i\,E}{h}\,.
\end{equation}

It is worth noting that the left and right representation are not simply related by sending $m\to-m$ as one may have na\"ively expected. Instead, the paramterisation of the representation coefficients and of the Zhukovski variables is genuinely different. This is done so that $|x^\pm_{*,p}|\geq1 $ for physical particles~\cite{Borsato:2012ud,Borsato:2012ss}.

Notice further that for the massless representation we have defined
\begin{equation}
    a_{\z} = \lim_{m\to 0} a_{\L}\,,\qquad
    b_{\z} = \lim_{m\to 0} b_{\L}\,,\qquad
    x^\pm_{\z,p} = \lim_{m\to 0} x^\pm_{\L,p}\,,
\end{equation}
We could have used the right representation instead. Practically, this amounts to flipping the sign of $p$ in~\eqref{eq:xpm} and switching $a_{\z}\leftrightarrow b_{\z}$ in~\eqref{eq:abparam}. This is actually allowed and does not introduce any new physics because the central charges, and in particular $M$ in eq.~\eqref{eq:mcentralcharge}, are unchanged. Hence the two representations obtained in the two limits must be isomorphic. This may be seen through a change of basis, \textit{e.g.}\ by rescaling the lowest-weight state (but not the highest-weight one) as it follows:
\begin{equation}
    |\varphi\rangle\ \to\ \sigma_p\, |\varphi\rangle\,,
\qquad
\sigma_p\equiv
    \Big[\frac{a_{\L}}{a_{\R}}\Big]_{m\to0} = -\text{sgn}\Big[\sin p/2\Big]\,.
\end{equation}
It is also useful to note the following identity
\begin{equation}
\label{eq:masslessidentity}
    \lim_{m\to 0} \big(x^{\pm}_{\L}(p) \, x^{\mp}_{\R}(p)\big)=1\,,
\end{equation}
which is valid for any~$k$ and generalises the fact, valid at $k=0$, that $x^+_{\z}(p)=1/x^-_{\z}(p)$.

Let us finally comment on the $h\to0$ limit, which corresponds to the WZW model. The Zhukovsky variables are divergent in this limit
\begin{equation}
    x^\pm_{*,p} = \frac{e^{\pm i p/2}}{h\,\sin(p/2)}\frac{M+|M|}{2}+O(h^0)\,,\qquad
    \eta_p = e^{i p/4} \sqrt{\frac{M+|M|}{2}}+O(h^1)\,,\qquad
\end{equation}
The leading order in $h$ of the Zhukovsky variables depends on the sign of $(2\pi m+k p)$, \textit{i.e.}\ on the branch of the dispersion relation 
\begin{equation}
    E(p)= \Big|m+\frac{k}{2\pi}p\Big|\,.
\end{equation}
Therefore, particles moving in the same or in opposite directions have starkly different limits~\cite{Baggio:2018gct}.

\subsubsection{Parametrisation after crossing}
\label{sec:crossingparam}
It will be useful in what follows to consider particles whose momentum analytically is continued to the crossed region, \textit{i.e.}
\begin{equation}
    p\to -p\,,\qquad E(p) \to - E(p)\,.
\end{equation}
This is the analogue of going from the $s$- to the $t$- channel in a relativistic theory. Following the notation of ref.~\cite{Basso:2015zoa} we indicate the crossed momentum as $p^{2\gamma}$. This is justified by the fact that $p^\gamma$ represents the continuation of momentum to the \textit{mirror} region~\cite{Arutyunov:2007tc} which loosely speaking corresponds to ``half crossing''. For a comprehensive discussion of the mirror and crossed regions we refer the readers to~\cite{Arutyunov:2009ga} and, in the context of $\AdSST$, to \cite{Borsato:2013hoa, Lloyd:2014bsa, Borsato:2016xns}.
Under the crossing transformation we have~\cite{Lloyd:2014bsa}
\begin{equation}
    x^{\pm}_{\L}(p^{2\gamma})=\frac{1}{x^{\pm}_{\R}(p)}\,,\qquad
    x^{\pm}_{\R}(p^{2\gamma})=\frac{1}{x^{\pm}_{\L}(p)}\,.
\end{equation}
Hence, the Zhukovsky variables and any rational function thereof map to themselves under a $4\gamma$-shift. Instead, the functions $\eta_{\L}(p)$ and $\eta_{\R}(p)$ behave as it follows,
\begin{equation}
    \eta_{\L}(p^{\pm2\gamma})=\frac{\pm i}{x^{+}_{\R}(p)}\eta_{\R}(p)\,,\qquad
    \eta_{\R}(p^{\pm2\gamma})=\frac{\pm i}{x^{+}_{\L}(p)}\eta_{\L}(p)\,.
\end{equation}
Crossing for massless modes is essentially given by the $m\to0$ limit of the massless case and by recalling the identity~\eqref{eq:masslessidentity}. We have
\begin{equation}
    x^{\pm}_{\z}(p^{2\gamma})=x^{\mp}_{\z}(p)\,,\qquad
    \eta_{\z}(p^{\pm 2\gamma})=\mp i \sigma_p \,e^{-ip/2}\eta_{\z}(p)\,.
\end{equation}

\subsection{Particle content of the theory}
\label{sec:particlecontent}

\renewcommand{\arraystretch}{1.5}
\begin{table}
\begin{center}
\begin{tabular}{l|cccc}
						& $\AdS_{3}$ Bosons	& $\Sph^{3}$ Bosons	& $\Tor^{4}$ Bosons	& Fermions\\
\hline
Left, $m=+1$ 	&	$Z(p)$	&	$Y(p)$	&	& $\Psi^A(p)$ \\
Right $m=-1$ 	&	$\tilde{Z}(p)$	&	$\tilde{Y}(p)$	&	& $\tilde{\Psi}^A(p)$ \\
Massless $m=0$	&	&	& $T^{A\dot{A}}(p)$	& $\chi^{\dot{A}}(p),\ \tilde{\chi}^{\dot{A}}(p)$
\end{tabular}
\end{center}
\caption{\label{tab:particles}
The fundamental particles of $\AdSST$ are eight Bosons and eight Fermions. In this table we arrange them according to which representation they belong (this depends on the sign of the central charge $\mathbf{M}$ at momentum $p=0$, $m=\mathbf{M}|_{p=0}$) and to their geometrical interpretation.}
\end{table}
\renewcommand{\arraystretch}{1}

The fundamental particle content of the theory is summarised in Table~\ref{tab:particles}. They can be arranged in representations constructed out of the $\rho_\pm$, $\rho_0$, $\rho_0'$ representations discussed above.
Let $\rho$ be any short representation of $\psu(1|1)^{\oplus2}$ c.e., which as we saw is two dimensional and takes the form $\mathbf{(1|1)}$. We want to use it to construct representations of $\psu(1|1)^{\oplus4}$ c.e., which we call $\varrho$. Clearly we can set $\varrho = \rho_\pm\otimes\rho_\pm$, or $\varrho = \rho_0\otimes\rho_0$ (or indeed, in this last formula, swap $\rho_0$ for $\rho_0'$, as it will turn out to be the case). However, a representation of the form \textit{e.g.}~$\varrho = \rho_\pm\otimes\rho_\mp$ or $\varrho = \rho_\pm\otimes\rho_0$ would \textit{not} be a representation of the algebra introduced in Section~\ref{sec:centrallyextendedalgebra}. Indeed to obtain a valid representation it is necessary that the two $\psu(1|1)^{\oplus2}$ representations appearing in the tensor product have the same central charge.

\subsubsection{The left representation}
To construct the left representation we consider
\begin{equation}
    \varrho_+ = \rho_+ \otimes\rho_+\,.
\end{equation}
We define the following states
\begin{equation}
\begin{gathered}
    |Y(p)\rangle=|\phi_\L^\B(p) \otimes \phi_\L^\B(p)\rangle\,,\\
    |\Psi^1(p)\rangle=|\varphi_\L^\F(p) \otimes \phi_\L^\B(p)\rangle\,,\qquad\qquad
    |\Psi^2\rangle=|\phi_\L^\B(p) \otimes \varphi_\L^\F(p)\rangle\,,\\
    |Z(p)\rangle=|\varphi_\L^\F(p) \otimes \varphi_\L^\F(p)\rangle\,,
\end{gathered}
\end{equation}
By using this definition we see that the supercharges act is as it follows
\begin{equation}
\begin{aligned}
\begin{tikzpicture}
\node (Y) at (0,1.5) {$|Y(p)\rangle$};
\node (psi1) at (-2,0) {$|\Psi^1(p)\rangle$};
\node (psi2) at (2,0) {$|\Psi^2(p)\rangle$};
\node (Z) at (0,-1.5) {$|Z(p)\rangle$};
\draw[->, thick] (Y) -- (psi1) node[pos=0.3,left] {$\Ql^1,\Sr^1$};
\draw[->, thick] (Y) -- (psi2) node[pos=0.3,right] {$\,\Ql^2,\Sr^2$};
\draw[->, thick] (psi1) -- (Z) node[pos=0.7,left] {$\Ql^2,\Sr^2\ $};
\draw[->, thick] (psi2) -- (Z) node[pos=0.7,right] {$\Ql^1,\Sr^1$};
\end{tikzpicture}
\end{aligned}
\end{equation}
To avoid cluttering the figure we only indicated the lowering operators, and not the raising ones. By using the definitions of section~\ref{sec:algebrafactorisation} we get the following action of the supercharges:
\begin{equation}
    \begin{aligned}
        &\Ql^{A}|Y(p)\rangle&=&\,a_{\L}(p)\,|\Psi^A(p)\rangle\,,\qquad
        &&\Ql^{A}|\Psi^B(p)\rangle&=&\,\varepsilon^{AB}a_{\L}(p)\,|Z(p)\rangle\,,\\
        &\Sl_{A}|\Psi^B(p)\rangle&=&\,\delta_A{}^B\,a_{\L}^*(p)\,|Y(p)\rangle\,,\qquad
        &&\Sl_{A}|Z(p)\rangle&=&-\varepsilon_{AB}\,a_{\L}^*(p)\,|\Psi^{B}(p)\rangle\,,\\
        &\Sr^{A}|Y(p)\rangle&=&\,b_{\L}^*(p)\,|\Psi^A(p)\rangle\,,\qquad
        &&\Sr^{A}|\Psi^B(p)\rangle&=&\,\varepsilon^{AB}b^*_{\L}(p)\,|Z(p)\rangle\,,\\
        &\Qr_{A}|\Psi^B(p)\rangle&=&\,\delta_A{}^B\,b_{\L}(p)\,|Y(p)\rangle\,,\qquad
        &&\Qr_{A}|Z(p)\rangle&=&-\varepsilon_{AB}\,b_{\L}(p)\,|\Psi^{B}(p)\rangle\,,
    \end{aligned}
\end{equation}
where we omitted the vanishing actions and we recall our convention $\varepsilon^{12}=-\varepsilon_{12}=+1$.

\subsubsection{The right representation}
For the right representation
\begin{equation}
    \varrho_{-}= \rho_-\otimes\rho_-\,,
\end{equation}
and we define
\begin{equation}
\begin{gathered}
    |\tilde{Z}(p)\rangle= |\phi_\R^\F(p) \otimes \phi_\R^\F(p)\rangle\,,\\
    |\tilde{\Psi}^1(p)\rangle=|\varphi_\R^\B(p) \otimes \phi_\R^\F(p)\,,\qquad\qquad
    |\tilde{\Psi}^2(p)\rangle=-|\phi_\R^\F(p) \otimes \varphi_\R^\B(p)\rangle\,,\\
    |\tilde{Y}(p)\rangle=|\varphi_\R^\B(p) \otimes \varphi_\R^\B(p)\rangle\,,
\end{gathered}
\end{equation}
where the reason for the minus sign is that ``right'' supercharges are canonically defined with lower $\su(2)_\bullet$ indices, see section~\ref{sec:algebrafactorisation}.
Arranging the representation in this way we see that the lowering operator act in the same fashion as above
\begin{equation}
\begin{aligned}
\begin{tikzpicture}
\node (Y) at (0,1.5) {$|\tilde{Z}(p)\rangle$};
\node (psi1) at (-2,0) {$|\tilde{\Psi}^1(p)\rangle$};
\node (psi2) at (2,0) {$|\tilde{\Psi}^2(p)\rangle$};
\node (Z) at (0,-1.5) {$|\tilde{Y}(p)\rangle$};
\draw[->, thick] (Y) -- (psi1) node[pos=0.3,left] {$\Ql^1,\Sr^1$};
\draw[->, thick] (Y) -- (psi2) node[pos=0.3,right] {$\,\Ql^2,\Sr^2$};
\draw[->, thick] (psi1) -- (Z) node[pos=0.7,left] {$\Ql^2,\Sr^2\ $};
\draw[->, thick] (psi2) -- (Z) node[pos=0.7,right] {$\Ql^1,\Sr^1$};
\end{tikzpicture}
\end{aligned}
\end{equation}
where the representation takes the form
\begin{equation}
    \begin{aligned}
        &\Ql^{A}|\tilde{Z}(p)\rangle&=&\,b_{\R}(p)\,|\tilde{\Psi}^A(p)\rangle\,,\qquad
        &&\Ql^{A}|\tilde{\Psi}^B(p)\rangle&=&-\varepsilon^{AB}b_{\R}(p)\,|\tilde{Y}(p)\rangle\,,\\
        &\Sl_{A}|\tilde{\Psi}^B(p)\rangle&=&\,\delta_A{}^B\,b_{\R}^*(p)\,|\tilde{Z}(p)\rangle\,,\qquad
        &&\Sl_{A}|\tilde{Y}(p)\rangle&=&\,\varepsilon_{AB}\,b_{\R}^*(p)\,|\tilde{\Psi}^{B}(p)\rangle\,,\\
        &\Sr^{A}|\tilde{Y}(p)\rangle&=&\,a_{\R}^*(p)\,|\tilde{\Psi}^A(p)\rangle\,,\qquad
        &&\Sr^{A}|\tilde{\Psi}^B(p)\rangle&=&-\varepsilon^{AB}a^*_{\R}(p)\,|\tilde{Z}(p)\rangle\,,\\
        &\Qr_{A}|\tilde{\Psi}^B(p)\rangle&=&\,\delta_A{}^B\,a_{\R}(p)\,|\tilde{Y}(p)\rangle\,,\qquad
        &&\Qr_{A}|\tilde{Z}(p)\rangle&=&\,\varepsilon_{AB}\,a_{\R}(p)\,|\tilde{\Psi}^{B}(p)\rangle\,.
    \end{aligned}
\end{equation}
Notice that there is a discrete left-right symmetry~\cite{Borsato:2012ud,Lloyd:2014bsa} when swapping the particles
\begin{equation}
    |Y\rangle\leftrightarrow |\tilde{Y}\rangle\,,
    \qquad
    |\Psi^A\rangle \leftrightarrow |\tilde{\Psi}_A\rangle\,,
    \qquad
    |Z\rangle\leftrightarrow |\tilde{Z}\rangle\,.
\end{equation}

\subsubsection{The massless representations}
%
There are actually two massless representations, which carry a charge under another $\su(2)$ algebra, which commutes with all symmetries thus far introduced. This is the $\su(2)_{\circ}$ that emerged from the decomposition of~$\so(4)_{\Tor^4}$. We write
\begin{equation}
    \varrho_0^{\dot{A}}= \big(\rho_0\otimes\rho_0'\big)\boldsymbol\oplus\big(\rho_0'\otimes\rho_0\big)\,,\qquad\dot{A}=1,2\,,
\end{equation}
with the understanding that the two modules $\rho_0\otimes\rho_0'$ and $\rho_0'\otimes\rho_0$ must also fit into a doublet of $\su(2)_{\circ}$. This is not in contradiction with the fact that $\su(2)_{\circ}$ commutes with $\psu(1|1)^{\oplus4}$ centrally extended because, as $\psu(1|1)^{\oplus4}$ c.e.~representations, $\rho_0\otimes\rho_0'\cong\rho_0'\otimes\rho_0$.
In fact, in reference~\cite{Borsato:2014exa} the same representation $\rho_0\otimes\rho_0'$ was used for both $\dot{A}=1$ and $\dot{A}=2$. This amounts to a change of basis.
We now have eight states
\begin{equation}
\begin{gathered}
    |\chi^{\dot{1}}(p)\rangle=|\phi^{\B}_0(p)\otimes\phi^{\F}_0(p)\rangle\,,\\
    |T^{\dot{1}1}(p)\rangle=|\varphi^{\F}_0(p)\otimes\phi^{\F}_0(p)\rangle\,,\qquad\qquad
    |T^{\dot{1}2}(p)\rangle=|\phi^{\B}_0(p)\otimes\varphi^{\B}_0(p)\rangle\,,\\
    |\tilde{\chi}^{\dot{1}}(p)\rangle=|\varphi^{\F}_0(p)\otimes\varphi^{\B}_0(p)\rangle\,,
\end{gathered}
\end{equation}
and
\begin{equation}
\begin{gathered}
    |\chi^{\dot{2}}(p)\rangle=i|\phi^{\F}_0(p)\otimes\phi^{\B}_0(p)\rangle\,,\\\
   |T^{\dot{2}1}(p)\rangle=i|\varphi^{\B}_0(p)\otimes\phi^{\B}_0(p)\rangle\,,\qquad\qquad
    |T^{\dot{2}2}(p)\rangle=-i|\phi^{\F}_0(p)\otimes\varphi^{\F}_0(p)\rangle\,,\\
    |\tilde{\chi}^{\dot{2}}(p)\rangle=-i|\varphi^{\B}_0(p)\otimes\varphi^{\F}_0(p)\rangle\,,
\end{gathered}
\end{equation}
Note that we have introduced an overall~$i$ in the latter representation. This is a matter of convenience that can be also addressed by introducing suitable normalisations later on.
Arranging the representation in this way we see that the lowering operator act in the same fashion as above for either module
\begin{equation}
\begin{aligned}
\begin{tikzpicture}
\node (Y) at (0,1.5) {$|\chi^{\dot{A}}(p)\rangle$};
\node (psi1) at (-2,0) {$|T^{\dot{A}1}(p)\rangle$};
\node (psi2) at (2,0) {$|T^{\dot{A}2}(p)\rangle$};
\node (Z) at (0,-1.5) {$|\tilde{\chi}^{\dot{A}}(p)\rangle$};
\draw[->, thick] (Y) -- (psi1) node[pos=0.3,left] {$\Ql^1,\Sr^1$};
\draw[->, thick] (Y) -- (psi2) node[pos=0.3,right] {$\,\Ql^2,\Sr^2$};
\draw[->, thick] (psi1) -- (Z) node[pos=0.7,left] {$\Ql^2,\Sr^2\ $};
\draw[->, thick] (psi2) -- (Z) node[pos=0.7,right] {$\Ql^1,\Sr^1$};
\end{tikzpicture}
\end{aligned}
\end{equation}
and regardless of the value of the index~$\dot{A}=1,2$ the representation takes the form
\begin{equation}
    \begin{aligned}
        &\Ql^{A}|\chi^{\dot{A}}(p)\rangle &=&\,a_{\z}(p)\,|T^{\dot{A}A}(p)\rangle\,,\qquad
        &&\Ql^{A}|T^{\dot{A}B}(p)\rangle &=&\,\varepsilon^{AB}a_{\z}(p)\,|\tilde{\chi}^{\dot{A}}(p)\rangle\,,\\
        &\Sl_{A}|T^{\dot{A}B}(p)\rangle &=&\,\delta_A{}^B\,a_{\z}^*(p)\,|\chi^{\dot{A}}(p)\rangle\,,\qquad
        &&\Sl_{A}|\tilde{\chi}^{\dot{A}}(p)\rangle &=&-\varepsilon_{AB}\,a_{\z}^*(p)\,|T^{\dot{A}B}(p)\rangle\,,\\
        &\Sr^{A}|\chi^{\dot{A}}(p)\rangle&=&\,b_{\z}^*(p)\,|T^{\dot{A}A}(p)\rangle\,,\qquad
        &&\Sr^{A}|T^{\dot{A}B}(p)\rangle&=&\,\varepsilon^{AB}b^*_{\z}(p)\,|\tilde{\chi}^{\dot{A}}(p)\rangle\,,\\
        &\Qr_{A}|T^{\dot{A}B}(p)\rangle&=&\,\delta_A{}^B\,b_{\z}(p)\,|\chi^{\dot{A}}(p)\rangle\,,\qquad
        &&\Qr_{A}|\tilde{\chi}(p)\rangle&=&-\varepsilon_{AB}\,b_{\z}(p)\,|T^{\dot{A}B}(p)\rangle\,.
    \end{aligned}
\end{equation}

\subsection{Scattering matrix}
\label{sec:Smatrix}
Up to the dressing factors, the S matrix of $\AdSST$ can be constructed by tensoring two S~matrices of $\psu(1|1)^{\oplus2}$ c.e., which in turn can be determined by imposing commutation with the symmetries discussed above~\cite{Borsato:2013qpa}. All in all, this closely resembles what happens with $\AdS_5\times\Sph^5$, with two main differences: firstly, rather than dealing with the algebra $\su(2|2)$ c.e.\ we have here $\psu(1|1)^{\oplus2}$ c.e.; secondly, instead of dealing with a single irreducble representation here we have \textit{four} irreducible representations. (Recall that in $\AdS_5\times\Sph^5$ we have four-dimensional representations of $\su(2|2)$ leading to $4^2=16$ dimensional representations of $\su(2|2)^{\oplus2}$; here we start from two-dimensional representations instead.) As a result of having four irreducible representations, the S~matrix will consist of sixteen blocks with as many dressing factors. Fortunately unitarity and other symmetries reduce the number of independent dressing factors to four.

For the reader's convenience, we introduce below the scattering matrices between $\psu(1|1)^{\oplus2}$ c.e.\ representations that play a role in what follows.

\subsubsection{Left-left scattering}
Here we report the scattering matrix for particles in the $\rho_{\L}$ representation of $\psu(1|1)^{\oplus2}$ centrally extended,
\begin{equation}
\begin{aligned}
    S|\phi_{\L,p}^\B\phi_{\L,q}^\B\rangle &= A^{\L\L}_{pq}\,|\phi_{\L,q}^\B\phi_{\L,p}^\B\rangle,&\quad
    S|\phi_{\L,p}^\B\varphi_{\L,q}^\F\rangle&= B^{\L\L}_{pq}|\varphi_{\L,q}^\F\phi_{\L,p}^\B\rangle + C^{\L\L}_{pq}|\phi_{\L,q}^\B\varphi_{\L,p}^\F\rangle,\\
    S|\varphi_{\L,p}^\F\varphi_{\L,q}^\F\rangle &= F^{\L\L}_{pq}\,|\varphi_{\L,q}^\F\varphi_{\L,p}^\F\rangle,&\quad
    S|\varphi_{\L,p}^\F\phi_{\L,q}^\B\rangle&= D^{\L\L}_{pq}|\phi_{\L,q}^\B\varphi_{\L,p}^\F\rangle + E^{\L\L}_{pq}|\varphi_{\L,q}^\F\phi_{\L,p}^\B\rangle,
\end{aligned}
\end{equation}
where the matrix elements are determined up to an overall prefactor $\Sigma_{pq}^{\L\L}$,
\begin{equation}
\label{eq:SLLexplicit}
\begin{aligned}
    &A_{pq}^{\L\L}=\Sigma_{pq}^{\L\L}\,,\qquad
    &&B_{pq}^{\L\L}=\Sigma_{pq}^{\L\L}e^{-\frac{i}{2}p}\frac{x^+_{\L,p} - x^+_{\L,q}}{x^-_{\L,p} - x^+_{\L,q}}\,,\\
    &C_{pq}^{\L\L}=\Sigma_{pq}^{\L\L}e^{-\frac{i}{2}p}e^{+\frac{i}{2}q}\frac{x^-_{\L,q} - x^+_{\L,q}}{x^-_{\L,p} - x^+_{\L,q}}\frac{\eta_{\L,p}}{\eta_{\L,q}}\,,\qquad
    &&D_{pq}^{\L\L}=\Sigma_{pq}^{\L\L}e^{+\frac{i}{2}q}\frac{x^-_{\L,p} - x^-_{\L,q}}{x^-_{\L,p} - x^+_{\L,q}}\,,\\
    &E_{pq}^{\L\L}=C_{pq}\,,\qquad
    &&F_{pq}^{\L\L}= -\Sigma_{pq}^{\L\L}e^{-\frac{i}{2}p}e^{+\frac{i}{2}q} \frac{x^+_{\L,p} - x^-_{\L,q}}{x^-_{\L,p} - x^+_{\L,q}}\,.
\end{aligned}
\end{equation}
Notice that we include a minus sign in $F^{\L\L}_{pq}$ to account for the Fermion permutation (in other words, in the free-theory limit our S~matrix reduces to the graded permutation operator).

\subsubsection{Right-right scattering}
Here we have
\begin{equation}
\begin{aligned}
    S|\varphi_{\R,p}^\B\varphi_{\R,q}^\B\rangle &= A^{\R\R}_{pq}\,|\varphi_{\R,q}^\B\varphi_{\R,p}^\B\rangle,&\quad
    S|\varphi_{\R,p}^\B\phi_{\R,q}^\F\rangle&= B^{\R\R}_{pq}|\phi_{\R,q}^\F\varphi_{\R,p}^\B\rangle + C^{\R\R}_{pq}|\varphi_{\R,q}^\B\phi_{\R,p}^\F\rangle,\\
    S|\phi_{\R,p}^\F\phi_{\R,q}^\F\rangle &= F^{\R\R}_{pq}\,|\phi_{\R,q}^\F\phi_{\R,p}^\F\rangle,&\quad
    S|\phi_{\R,p}^\F\varphi_{\R,q}^\B\rangle&= D^{\R\R}_{pq}|\varphi_{\R,q}^\B\phi_{\R,p}^\F\rangle + E^{\R\R}_{pq}|\phi_{\R,q}^\F\varphi_{\R,p}^\B\rangle,
\end{aligned}
\end{equation}
with
\begin{equation}
\begin{aligned}
    &A_{pq}^{\R\R}=\Sigma_{pq}^{\R\R}\,,\qquad
    &&B_{pq}^{\R\R}=\Sigma_{pq}^{\R\R}e^{-\frac{i}{2}p}\frac{x^+_{\R,p} - x^+_{\R,q}}{x^-_{\R,p} - x^+_{\R,q}}\,,\\
    &C_{pq}^{\R\R}=\Sigma_{pq}^{\R\R}e^{-\frac{i}{2}p}e^{+\frac{i}{2}q}\frac{x^-_{\R,q} - x^+_{\R,q}}{x^-_{\R,p} - x^+_{\R,q}}\frac{\eta_{\R,p}}{\eta_{\R,q}}\,,\qquad
    &&D_{pq}^{\R\R}=\Sigma_{pq}^{\R\R}e^{+\frac{i}{2}q}\frac{x^-_{\R,p} - x^-_{\R,q}}{x^-_{\R,p} - x^+_{\R,q}}\,,\\
    &E_{pq}^{\R\R}=C_{pq}\,,\qquad
    &&F_{pq}^{\R\R}= -\Sigma_{pq}^{\R\R}e^{-\frac{i}{2}p}e^{+\frac{i}{2}q} \frac{x^+_{\R,p} - x^-_{\R,q}}{x^-_{\R,p} - x^+_{\R,q}}\,.
\end{aligned}
\end{equation}

\subsubsection{Left-right scattering}
Here we have
\begin{equation}
\begin{aligned}
    S|\phi_{\L,p}^\B\varphi_{\R,q}^\B\rangle&= A^{\L\R}_{pq}|\varphi_{\R,q}^\B\phi_{\L,p}^\B\rangle + B^{\L\R}_{pq}|\phi_{\R,q}^\F\varphi_{\L,p}^\F\rangle,
    &\quad
    S|\phi_{\L,p}^\B\phi_{\R,q}^\F\rangle &= C^{\L\R}_{pq}\,|\phi_{\R,q}^\F\phi_{\L,p}^\B\rangle,\\
    S|\varphi_{\L,p}^\F\phi_{\R,q}^\F\rangle&= E^{\L\R}_{pq}|\phi_{\R,q}^\F\varphi_{\L,p}^\F\rangle + F^{\L\R}_{pq}|\varphi_{\R,q}^\B\phi_{\L,p}^\B\rangle,&\quad
    S|\varphi_{\L,p}^\F\varphi_{\R,q}^\B\rangle &= D^{\L\R}_{pq}\,|\varphi_{\R,q}^\B\varphi_{\L,p}^\F\rangle,
\end{aligned}
\end{equation}
with
\begin{equation}
   \begin{aligned}
    &A_{pq}^{\L\R}=\Sigma_{pq}^{\L\R} e^{-\frac{i}{2}p}\frac{1-x^+_{\L,p} x^-_{\R,q}}{1-x^-_{\L,p} x^-_{\R,q}}\,,\qquad
    &&B_{pq}^{\L\R}= \Sigma_{pq}^{\L\R}e^{-\frac{i}{2}p}e^{-\frac{i}{2}q} \frac{2 i}{h} \frac{\eta_{\L,p}\eta_{\R,q}}{1-x^-_{\L,p} x^-_{\R,q}}\,,\\
    &C_{pq}^{\L\R}=\Sigma_{pq}^{\L\R}\,,\qquad
    &&D_{pq}^{\L\R}=\Sigma_{pq}^{\L\R}e^{-\frac{i}{2}p}e^{-\frac{i}{2}q}\frac{1- x^+_{\L,p} x^+_{\R,q}}{1-x^-_{\L,p} x^-_{\R,q}}\,,\\
    &E_{pq}^{\L\R}= -\Sigma_{pq}^{\L\R} e^{-\frac{i}{2}q}\frac{1-x^-_{\L,p} x^+_{\R,q}}{1-x^-_{\L,p} x^-_{\R,q}}\,,\qquad
    &&F_{pq}^{\L\R}=-B_{pq}^{\L\R}\,.
\end{aligned}
\end{equation}

\subsubsection{Right-left scattering}
The right-left S~matrix is related to the left-right one by unitarity. It reads
\begin{equation}
\begin{aligned}
    S|\varphi_{\R,p}^\B\phi_{\L,q}^\B\rangle&= A^{\R\L}_{pq}|\phi_{\L,q}^\B\varphi_{\R,p}^\B\rangle + B^{\R\L}_{pq}|\varphi_{\L,q}^\F\phi_{\R,p}^\F\rangle,
    &\quad
    S|\varphi_{\R,p}^\B\varphi_{\L,q}^\F\rangle &= C^{\R\L}_{pq}\,|\varphi_{\L,q}^\F\varphi_{\R,p}^\B\rangle,\\
    S|\phi_{\R,p}^\F\varphi_{\L,q}^\F\rangle&= E^{\R\L}_{pq}|\varphi_{\L,q}^\F\phi_{\R,p}^\F\rangle + F^{\R\L}_{pq}|\phi_{\L,q}^\B\varphi_{\R,p}^\B\rangle,&\quad
    S|\phi_{\R,p}^\F\phi_{\L,q}^\B\rangle &= D^{\R\L}_{pq}\,|\phi_{\L,q}^\B\phi_{\R,p}^\F\rangle,
\end{aligned}
\end{equation}
with
\begin{equation}
   \begin{aligned}
     &A_{pq}^{\R\L}=\Sigma_{pq}^{\R\L} e^{+\frac{i}{2}q}\frac{1-x^+_{\R,p} x^-_{\L,q}}{1-x^+_{\R,p} x^+_{\L,q}}\,,\qquad
    &&B_{pq}^{\R\L}=\Sigma_{pq}^{\R\L} \frac{2 i}{h} \frac{\eta_{\R,p}\eta_{\L,q}}{1-x^+_{\R,p} x^+_{\L,q}}\,,\\
    &C_{pq}^{\R\L}=\Sigma_{pq}^{\R\L} e^{+\frac{i}{2}p}e^{+\frac{i}{2}q}\frac{1-x^-_{\R,p} x^-_{\L,q}}{1-x^+_{\R,p} x^+_{\L,q}}\,,\qquad
    &&D_{pq}^{\R\L}=\Sigma_{pq}^{\R\L}\,,\\
    &E_{pq}^{\R\L}=-\Sigma_{pq}^{\R\L} e^{+\frac{i}{2}p}\frac{1-x^-_{\R,p} x^+_{\L,q}}{1-x^+_{\R,p} x^+_{\L,q}}\,,\qquad
    &&F_{pq}^{\R\L}=-B_{pq}^{\R\L}\,.
\end{aligned}
\end{equation}

\subsubsection{Massless scattering.}
Because the massless representation coefficients may be obtained either from $\rho_{\L}$ or $\rho_{\R}$ by taking the $m\to0$ limit, so can the relevant S-matrix elements (up to the dressing factor: those do not follow immediately from symmetry, so that a more cautious analysis is required~\cite{Borsato:2013hoa,Borsato:2016xns}). Here we will choose to obtain the massless S-matrix elements from the left-left scattering. The only additional caution in this case is relative to the statistics of the exictations, since in massless representations we may encounter Fermionic highest-weight states. This leads to different signs, which we spell out here, starting by recalling the standard scattering matrix.
\begin{equation}
\begin{aligned}
    S|\phi_{\z,p}^\B\phi_{\z,q}^\B\rangle &= A^{\L\L}_{pq}\,|\phi_{\z,q}^\B\phi_{\z,p}^\B\rangle,&\quad
    S|\phi_{\z,p}^\B\varphi_{\z,q}^\F\rangle&= B^{\L\L}_{pq}|\varphi_{\z,q}^\F\phi_{\z,p}^\B\rangle + C^{\L\L}_{pq}|\phi_{\z,q}^\B\varphi_{\z,p}^\F\rangle,\\
    S|\varphi_{\z,p}^\F\varphi_{\z,q}^\F\rangle &= F^{\L\L}_{pq}\,|\varphi_{\z,q}^\F\varphi_{\z,p}^\F\rangle,&\quad
    S|\varphi_{\z,p}^\F\phi_{\z,q}^\B\rangle&= D^{\L\L}_{pq}|\phi_{\z,q}^\B\varphi_{\z,p}^\F\rangle + E^{\L\L}_{pq}|\varphi_{\z,q}^\F\phi_{\z,p}^\B\rangle,
\end{aligned}
\end{equation}
When both particles are in the $\tilde{\rho}_{\L}$ representation we have, instead
\begin{equation}
\begin{aligned}
    S|\varphi_{\z,p}^\B\varphi_{\z,q}^\B\rangle &= -F^{\L\L}_{pq}\,|\varphi_{\z,q}^\B\varphi_{\z,p}^\B\rangle,&\quad
    S|\varphi_{\z,p}^\B\phi_{\z,q}^\F\rangle&= D^{\L\L}_{pq}|\phi_{\z,q}^\F\varphi_{\z,p}^\B\rangle - E^{\L\L}_{pq}|\varphi_{\z,q}^\B\phi_{\z,p}^\F\rangle,\\
    S|\phi_{\z,p}^\F\phi_{\z,q}^\F\rangle &= -A^{\L\L}_{pq}\,|\phi_{\z,q}^\F\phi_{\z,p}^\F\rangle,&\quad
    S|\phi_{\z,p}^\F\varphi_{\z,q}^\B\rangle&= B^{\L\L}_{pq}|\varphi_{\z,q}^\B\phi_{\z,p}^\F\rangle - C^{\L\L}_{pq}|\phi_{\z,q}^\F\varphi_{\z,p}^\B\rangle.
\end{aligned}
\end{equation}
Note that we could have also defined, in analogy with the above, $A^{\tilde{\L}\tilde{\L}}\equiv-F^{\L\L}$, $B^{\tilde{\L}\tilde{\L}}\equiv D^{\L\L}$, $C^{\tilde{\L}\tilde{\L}}\equiv -E^{\L\L}$, $D^{\tilde{\L}\tilde{\L}}\equiv B^{\L\L}$, $E^{\tilde{\L}\tilde{\L}}\equiv -C^{\L\L}$, and  $F^{\tilde{\L}\tilde{\L}}\equiv -A^{\L\L}$. Similarly, in the mixed case we have
\begin{equation}
\label{eq:Slefttildeleft}
\begin{aligned}
    S|\phi_{\z,p}^\B\varphi_{\z,q}^\B\rangle&= \;\ B^{\L\L}_{pq}|\varphi_{\z,q}^\B\phi_{\z,p}^\B\rangle - C^{\L\L}_{pq}|\phi_{\z,q}^\F\varphi_{\z,p}^\F\rangle,
    &\quad
    S|\phi_{\z,p}^\B\phi_{\z,q}^\F\rangle &= \;\ A^{\L\L}_{pq}\,|\phi_{\z,q}^\F\phi_{\z,p}^\B\rangle,\\
    S|\varphi_{\z,p}^\F\phi_{\z,q}^\F\rangle&= -D^{\L\L}_{pq}|\phi_{\z,q}^\F\varphi_{\z,p}^\F\rangle + E^{\L\L}_{pq}|\varphi_{\z,q}^\B\phi_{\z,p}^\B\rangle,&\quad
    S|\varphi_{\z,p}^\F\varphi_{\z,q}^\B\rangle &= -F^{\L\L}_{pq}\,|\varphi_{\z,q}^\B\varphi_{\z,p}^\F\rangle,
\end{aligned}
\end{equation}
and finally
\begin{equation}
\label{eq:Sleftlefttilde}
\begin{aligned}
    S|\varphi_{\z,p}^\B\phi_{\z,q}^\B\rangle&= \;\ D^{\L\L}_{pq}|\phi_{\z,q}^\B\varphi_{\z,p}^\B\rangle + E^{\L\L}_{pq}|\varphi_{\z,q}^\F\phi_{\z,p}^\F\rangle,
    &\quad
    S|\varphi_{\z,p}^\B\varphi_{\z,q}^\F\rangle &= -F^{\L\L}_{pq}\,|\varphi_{\z,q}^\F\varphi_{\z,p}^\B\rangle,\\
    S|\phi_{\z,p}^\F\varphi_{\z,q}^\F\rangle&= -B^{\L\L}_{pq}|\varphi_{\z,q}^\F\phi_{\z,p}^\F\rangle - C^{\L\L}_{pq}|\phi_{\z,q}^\B\varphi_{\z,p}^\B\rangle,&\quad
    S|\phi_{\z,p}^\F\phi_{\z,q}^\B\rangle &= \;\ A^{\L\L}_{pq}\,|\phi_{\z,q}^\B\phi_{\z,p}^\F\rangle.
\end{aligned}
\end{equation}

\subsubsection{Mixed-mass scattering}
In a similar way as the above, we may obtain the mixed-mass S~matrix by considering the massless limit of the representation parameters only for one of the variables. Additionally, we have to account for the various signs that may arise due to the grading of the highest weight state. Below we list those related to the $\rho_0'\otimes\rho_-$ and $\rho_-\otimes\rho_0'$ representations, since the ones related to $\rho_0'\otimes\rho_+$ and $\rho_+\otimes\rho_0'$ are the same as eqs.~\eqref{eq:Sleftlefttilde} and~\eqref{eq:Slefttildeleft}, respectively.
We have
\begin{equation}
\begin{aligned}
    S|\varphi_{\R,p}^\B\varphi_{\z,q}^\B\rangle&=
    +C^{\R\z}_{pq}\,|\varphi_{\z,q}^\B\varphi_{\R,p}^\B\rangle,
    &\quad
    S|\varphi_{\R,p}^\B\phi_{\z,q}^\F\rangle &=
    +A^{\R\z}_{pq}|\phi_{\z,q}^\F\varphi_{\R,p}^\B\rangle - B^{\R\z}_{pq}|\varphi_{\z,q}^\B\phi_{\R,p}^\F\rangle,\\
    S|\phi_{\R,p}^\F\phi_{\z,q}^\F\rangle&=
    -D^{\R\z}_{pq}|\phi_{\z,q}^\F\phi_{\R,p}^\F\rangle,&\quad
    S|\phi_{\R,p}^\F\varphi_{\z,q}^\B\rangle &=
    -E^{\R\z}_{pq}|\varphi_{\z,q}^\B\phi_{\R,p}^\F\rangle + F^{\R\z}_{pq}|\phi_{\z,q}^\F\varphi_{\R,p}^\B\rangle,
\end{aligned}
\end{equation}
and (correcting a misprint in~\cite{Borsato:2014hja})
\begin{equation}
\begin{aligned}
    S|\varphi_{\z,p}^\B\varphi_{\R,q}^\B\rangle &= +D^{\z\R}_{pq}\,|\varphi_{\R,q}^\B\varphi_{\z,p}^\B\rangle,&\quad
    S|\varphi_{\z,p}^\B\phi_{\R,q}^\F\rangle&= -E^{\z\R}_{pq}|\phi_{\R,q}^\F\varphi_{\z,p}^\B\rangle - F^{\z\R}_{pq}|\varphi_{\R,q}^\B\phi_{\z,p}^\F\rangle,\\
    S|\phi_{\z,p}^\F\phi_{\R,q}^\F\rangle &= -C^{\z\R}_{pq}\,|\phi_{\R,q}^\F\phi_{\z,p}^\F\rangle,&\quad
    S|\phi_{\z,p}^\F\varphi_{\R,q}^\B\rangle&= +A^{\z\R}_{pq}|\varphi_{\R,q}^\B\phi_{\z,p}^\F\rangle + B^{\z\R}_{pq}|\phi_{\R,q}^\F\varphi_{\z,p}^\B\rangle.
\end{aligned}
\end{equation}

\subsubsection{Dressing factors}
The pre-factors introduced above must obey crossing and unitarity constraints, besides having the correct analytic structure to give a sensible S-matrix for the full $\psu(1|1)^{\oplus}$ c.e.\ S~matrix. It is possible to write the solutions in the form~\cite{Borsato:2013hoa,Borsato:2016xns}
\begin{equation}
    \begin{aligned}
    &\left( \Sigma_{pq}^{\L\L} \right)^2 = \left( \Sigma_{pq}^{\R\R} \right)^2 = \frac{e^{i(p-q)}}{\sigma^{* *}(p,q)^2} \frac{x^-_{*,p}- x^+_{*,q}}{x^+_{*,p}-x^-_{*,q}}
    \frac{1-\frac{1}{x^-_{*,p} x^+_{*,q}}}{1-\frac{1}{x^+_{*,p}x^-_{*,q}}}\,,\\
   &\left( \Sigma_{pq}^{\L\R} \right)^2 = \frac{e^{i p}}{\sigma^{\L\R}(p,q)^2} \frac{1- x^-_{\L,p} x^-_{\R,q} }{1- x^+_{\L,p} x^+_{\R,q}} \frac{1-\frac{1}{x^-_{\L,p} x^+_{\R,q}}}{1-\frac{1}{x^+_{\L,p} x^-_{\R,q}}}\,, \\
    &\left( \Sigma_{pq}^{\R\L} \right)^2 = \frac{e^{-i q}}{\sigma^{\R\L}(p,q)^2} \frac{1- x^+_{\R,p} x^+_{\L,q}}{1- x^-_{\R,p} x^-_{\L,q} } \frac{1-\frac{1}{x^-_{\R,p} x^+_{\L,q}}}{1-\frac{1}{x^+_{\R,p}x^-_{\L,q}}}\,.\label{eq:NormalizationSigma}
    \end{aligned}
\end{equation}
and for the massless case
\begin{equation}
    \left( \Sigma^{\z\z}_{pq} \right)^2 = - \frac{e^{\frac{i}{2}(p-q)}}{\sigma^{\z\z}(p,q)^2} \frac{x^-_{\z,p}-x^+_{\z,q}}{x^+_{\z,p}-x^-_{\z,q}}\,.
\end{equation}
Notice that our normalisation differs by an overall minus sign from that of~\cite{Lloyd:2014bsa}. This does not affect the crossing equations and leads to a consistent limit in the near-BMN expansion.
For the mixed-mass cases we have \begin{equation}
\begin{aligned}
    \Sigma^{\bullet \circ}_{\L \z}(p,q)^2 &= e^{+i \frac{p}{2}} \frac{x^-_{\L,p}-x^+_{\z,q}}{x^+_{\L,p}-x^+_{\z,q}} \zeta(p,q) \frac{1}{\sigma^{\bullet \circ}_{\L \z}(p,q)^2} \,, \\
    \Sigma^{\circ \bullet}_{\z \L}(p,q)^2 &= e^{-i \frac{q}{2}} \frac{x^-_{\z,p}-x^+_{\L,q}}{x^-_{\z,p}-x^-_{\L,q}} \zeta(p,q) \frac{1}{\sigma^{\circ \bullet}_{\z \L}(p,q)^2} \,, \\
    \Sigma^{\bullet \circ}_{\R \z}(p,q)^2 &= e^{-i(\frac{p}{2}+q)} \frac{(1-x^-_{\R,p} x^+_{\z,q})(1-x^+_{\R,p} x^+_{\z,q})}{(1-x^-_{\R,p} x^-_{\z,q})^2} \tilde{\zeta}(p,q) \frac{1}{\sigma^{\bullet \circ}_{\R \z}(p,q)^2} \,,\\
    \Sigma^{\circ \bullet}_{\z \R}(p,q)^2 &= e^{+i(p+\frac{q}{2})} \frac{(1-x^-_{\z,p} x^+_{\R,q})(1-x^-_{\z,p} x^-_{\R,q})}{(1-x^+_{\z,p} x^+_{\R,q})^2} \tilde{\zeta}(p,q) \frac{1}{\sigma^{\circ \bullet}_{\z \R}(p,q)^2} \,,
\end{aligned}
\end{equation}
where we introduced the functions
\begin{equation}
\begin{aligned}
    \zeta(p,q) &= \sqrt{\frac{x^-_{*,p}-x^-_{*,q}}{x^+_{*,p}-x^-_{*,q}} \frac{x^+_{*,p}-x^+_{*,q}}{x^-_{*,p}-x^+_{*,q}}} \,,\\ \tilde{\zeta}(p,q)&= \sqrt{\frac{1-x^+_{*,p} x^+_{*,q}}{1-x^+_{*,p} x^-_{*,q}} \frac{1-x^-_{*,p} x^-_{*,q}}{1-x^-_{*,p} x^+_{*,q}}}.
    \end{aligned}
\end{equation}
All of the above formulae are written in terms of some functions~$\sigma^{**}$ which have branch cuts on the Zhukovski plane. The transformations of the Zhukovski variables are given in section~\ref{sec:crossingparam}, while for the detailed description of the cuts of the dressing factors we refer the reader to the review~\cite{Sfondrini:2014via} and the original literature~\cite{Borsato:2013hoa,Borsato:2016xns}.
In what follows we will not use the explicit form of the dressing factors, but we will use their properties under crossing and unitarity.
We have
\begin{equation}
    \begin{aligned}
    &\sigma^{\L\L}(p^{+2\gamma},q)^2 \sigma^{\R\L}(p,q)^2 = g^{\R\L}(p,q) \,,\quad 
    &&\sigma^{\L\L}(p,q)^2 \sigma^{\R\L}(p^{+2\gamma},q)^2 = \tilde{g}^{\L\L}(p,q) \,, \\
    &\sigma^{\R\R}(p^{+2\gamma},q)^2 \sigma^{\L\R}(p,q)^2 = g^{\L\R}(p,q) \,,\quad 
    &&\sigma^{\R\R}(p,q)^2 \sigma^{\L\R}(p^{2\gamma},q)^2 = \tilde{g}^{\R\R}(p,q) \,, \\
    &\sigma^{\L\L}(p,q^{-2\gamma})^2 \sigma^{\L\R}(p,q)^2 = \frac{1}{\tilde{g}^{\L\L}(q^{2\gamma},p)} \,,\quad 
    &&\sigma^{\L\L}(p,q)^2 \sigma^{\L\R}(p,q^{-2\gamma})^2 = \frac{1}{g^{\R\L}(q^{2\gamma},p)} \,, \\
    &\sigma^{\R\R}(p,q^{-2\gamma})^2 \sigma^{\R\L}(p,q)^2 = \frac{1}{\tilde{g}^{\R\R}(q^{2\gamma},p)} \,,\quad
    &&\sigma^{\R\R}(p,q)^2 \sigma^{\R\L}(p,q^{-2\gamma})^2 =  \frac{1}{g^{\L\R}(q^{2\gamma},p)} \,,
    \end{aligned}
\end{equation}
where the rational functions $g(p,q)$ and $\tilde{g}(p,q)$ are given by
\begin{equation}
    \begin{aligned}
    &g^{**}(p,q) = e^{-2iq} \frac{ \left( 1-\frac{1}{x^+_{*,p}x^+_{*,q}} \right) \left(1-\frac{1}{x^-_{*,p}x^-_{*,q}}\right) }{ \left(1-\frac{1}{x^+_{*,p}x^-_{*,q}} \right)^2} \frac{x^-_{*,p}-x^+_{*,q}}{x^+_{*,p}-x^-_{*,q}} \,, \\
    &\tilde{g}^{**}(p,q) = e^{-2iq} \frac{(x^-_{*,p}-x^+_{*,q})^2}{(x^+_{*,p}-x^+_{*,q})(x^-_{*,p}-x^-_{*,q})} \frac{1-\frac{1}{x^-_{*,p}x^+_{*,q}}}{1-\frac{1}{x^+_{*,p}x^-_{*,q}}}\,.
    \end{aligned}
\end{equation}
Similarly, for the massless phase we have that
\begin{equation}
    \sigma^{\z\z}(p^{2\gamma},q)^2 \sigma^{\z\z}(p,q)^2 = \frac{x^+_{\z,p}-x^-_{\z,q}}{x^+_{\z,p}-x^+_{\z,q}} \frac{x^-_{\z,p}-x^+_{\z,q}}{x^-_{\z,p}-x^-_{\z,q}} \,,
\end{equation}
and for the mixed-mass phases, 
\begin{equation}
\begin{aligned}
    \sigma^{\bullet \circ}_{\R \z}(p^{2\gamma},q)^2 \sigma^{\bullet \circ}_{\L \z}(p,q)^2 &= \frac{x^-_{\L,p}-x^+_{\z,q}}{x^+_{\L,p}-x^+_{\z,q}}\frac{x^+_{\L,p}-x^-_{\z,q}}{x^-_{\L,p}-x^-_{\z,q}} &&= \sigma^{\circ \bullet}_{\z \L}(q^{2\gamma},p)^2 \sigma^{\circ \bullet}_{\z \L}(q,p)^2 \,,\\
    \sigma^{\bullet \circ}_{\L \z}(p^{2\gamma},q)^2 \sigma^{\bullet \circ}_{\R \z}(p,q)^2 &= \frac{1-\frac{1}{x^+_{\R,p} x^+_{\z,q}}}{1-\frac{1}{x^+_{\R,p} x^-_{\z,q}}} \frac{1-\frac{1}{x^-_{\R,p} x^-_{\z,q}}}{1-\frac{1}{x^-_{\R,p} x^+_{\z,q}}} &&= \sigma^{\circ \bullet}_{\z \R}(q^{2\gamma},p)^2 \sigma^{\circ \bullet}_{\z \R}(q,p)^2 \,.
\end{aligned}
\end{equation}

\section{Integrability for three-point functions and the hexagon operator}
\label{sec:hexagon}

\begin{figure}[t]
 \centering
 \includegraphics[width=0.8\textwidth]{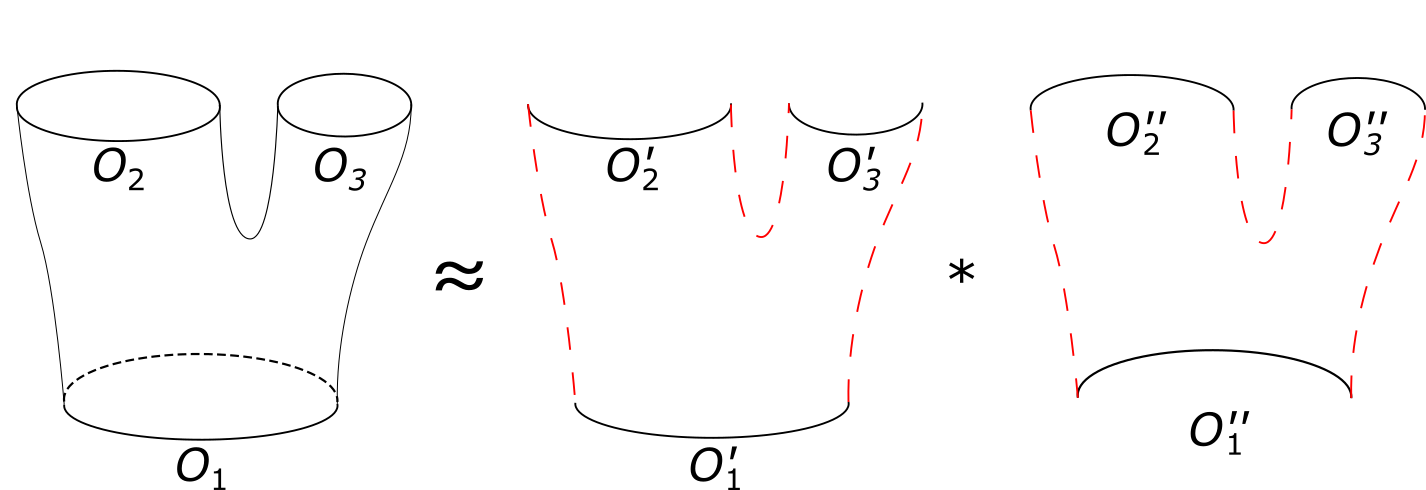}
 \caption{The main idea of ref.~\cite{Basso:2015zoa} is to ``cut open'' the three-point function in string theory to get two patches of worldsheet with six distinct edges (hexagon tesellation). This is the analogue of considering the infinite-volume worldsheet theory for the spectral problem (\textit{i.e.}, cutting open a cylinder into a plane).}
 \label{fig:Hexagon}
\end{figure}

\begin{figure}[t]
 \centering
 \hspace*{1cm}
 \includegraphics[width=0.45\textwidth]{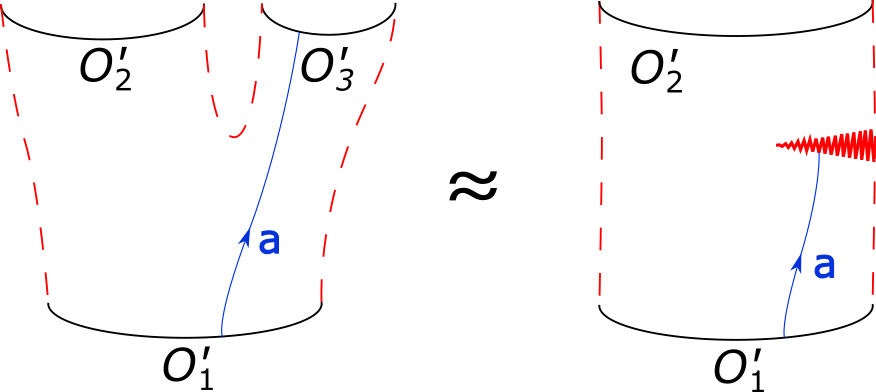}
 \caption{We can represent each hexagonal worldsheet patch as an ordinary two-dimensional theory with the insertion of  a non-local ``hexagon operator'' which creates an excess angle (the red zig-zag line). This operators may absorb excitations (like particle $a$ in the figure) yielding a non-zero result --- its form factor. This form factor is what we are interested in determining starting from the symmetries preserved by the configuration of three operators $O_1, O_2$ and~$O_3$.}
 \label{fig:conicalexcess}
\end{figure}

It was proposed that, for $\AdS_{5}\times \Sph^5$ superstrings, three-~\cite{Basso:2015zoa} and higher-point functions~\cite{Eden:2016xvg, Fleury:2016ykk} of generic operators may be constructed using integrability techniques. The setup is easiest to understand for three-point functions~\cite{Basso:2015zoa} by bearing in mind the approach used for the spectral problem. In the spectral problem, one goes from a closed string (a finite-volume worldsheet) to a decompactified worldsheet where the S~matrix may be defined~\cite{Arutyunov:2009ga}. For three-point functions, too, one wants to consider a decompactification of the ``pair of pants'' topology by cutting it open in two hexagonal patches. Without reviewing this construction in full detail (we refer the reader to~\cite{Basso:2015zoa}) it suffices to say that either patch contains a piece of each of the three closed-string states whose correlator we are interested in computing, see figure~\ref{fig:Hexagon}. We are interested in representing each hexagonal patch as an ordinary worldsheet where a non-local ``hexagon'' operator has been inserted, see figure~\ref{fig:conicalexcess}. What is remarkable is that, in the case of $\AdS_{5}\times \Sph^5$, it is possible~\cite{Basso:2015zoa} to bootstrap the form factors of these operators starting from the light-cone gauge symmetries that helped determine the S~matrix.
It is therefore natural to ask whether a similar construction may be applied to more general setups, and in particular to $\AdSST$. This is what we will discuss in this section.

\subsection{Symmetries of the three-point function}
\label{sec:hexagonalgebra}
In the case of the S~matrix, the original $\psu(1,1|2)^{\oplus2}$ supersymmetry was broken by gauge fixing. Such a gauge fixing relies on the choice of a 1/2-BPS geodesic and, in the dual CFT, amounts to picking a reference two-point function involving one 1/2-BPS operator $O(0)$ and its conjugate $O^\dagger(\infty)$. In the case of three-point functions we need three-operator, sitting at \textit{three} distinct points. It is useful to construct such an operator following ref.~\cite{Drukker:2008pi}, starting from a reference BPS operator and considering its image under translation.

\subsubsection{The supertranslation operator}
Given a 1/2-BPS operator $O(0)$ at $z=0$, we are interested in constructing translated operators $O(z)$. To be concrete, let us say that $O(0)$ is a 1/2-BPS scalar operator which is the highest-weight state in the representation with
\begin{equation}
-\mathbf{L}_0=\mathbf{J}^3= j\,,\qquad
-\widetilde{\mathbf{L}}_0=\widetilde{\mathbf{J}}^3= j\,,
\end{equation}
see section~\ref{sec:algebra}.
In terms of the $\psu(1,1|2)^{\oplus2}$ generators, translations are given by
\begin{equation}
    \T = i\Ll_{-} + i\Lr_{-}\,.
\end{equation}
We are interested in constructing \textit{three} such operators in such a way as to preserve as much (super)symmetry as possible. We expect this to break some of the $\psu(1|1)^{\oplus4}$ centrally extended symmetry described in section~\ref{sec:centrallyextendedalgebra}.
It is easy to see that this requires combining the translation with an R-symmetry rotation~\cite{Drukker:2008pi,Basso:2015zoa}.
Hence we introduce the supertranslation generator
\begin{equation}
\label{eq:supertranslation}
    \T_{\kappa} = i\Ll_{-} + i\Lr_{-}+\kappa\, \Jl_{-}+\kappa\, \Jr_{-}\,,
\end{equation}
where $\kappa\in\mathbb{C}$ is some constant to be determined. It is worth noting that, while we may introduce a distinct $\kappa$ and $\tilde{\kappa}$ for the left and right part of the algebra, we will be able to carry out the bootstrap procedure with a single $\kappa=\tilde{\kappa}$. 
Using $\T_{\kappa}$ we may construct a one parameter family of operators starting from $O(0)$, namely
\begin{equation}
O_{t,\kappa}=e^{t\,\T_{\kappa}}\, O(0)\,e^{-t\,\T_{\kappa}}\,,
\end{equation}
which by construction sits at position~$t$. At the same time, we have that the operator is $t$-rotated in R-symmetry space. For instance, taking $t=\infty$ yields $O^\dagger(\infty)$. A generic configuration of images of $O(0)$ sitting at $t_1, t_2, t_3, \dots$ will be jointly annihilated by the stabilizer of $T_{\kappa}$ in $\psu(1|1)^{\oplus 4}$ centrally extended.
By direct inspection, this supertranslation operator preserves four supercharges in $\psu(1|1)^{\oplus4}$, namely
\begin{equation}
\begin{aligned}
&\Ql_{+-A}-\frac{i}{\kappa}\,\Ql_{-+A}= \Sl_A-\frac{i}{\kappa}\,\epsilon_{AB}\,\Ql^B,\\
&\Qr_{+-A}-\frac{i}{\kappa}\,\Qr_{-+A}
=-\epsilon_{AB}\Sr^B-\frac{i}{\kappa}\Qr_A\,.
\end{aligned}
\end{equation}

\subsubsection{The hexagon subalgebra}
It is convenient to introduce the notation
\begin{equation}
\label{eq:Hexagonalgebra}
\begin{aligned}
&\sQl_A = \Sl_A-\frac{i}{\kappa}\ \epsilon_{AB}\,\Ql^B,\\
&\sQr_A = \Qr_A-i\kappa\,\epsilon_{AB}\,\Sr^B\,,
\end{aligned}
\end{equation}
to indicate the four supercharges that commute with the supertranslation generator $\T_{\kappa}$~\eqref{eq:supertranslation}. By direct inspection, using the relations of section~\ref{sec:centrallyextendedalgebra}, we find
\begin{equation}
    \{\sQl_A,\sQl_A\}=0,\qquad
    \{\sQl_1,\sQl_2\}=-\frac{i}{\kappa}\Big(\{\Sl_1,\epsilon_{21}\Ql^1\}+
    \{\epsilon_{12}\Ql^2,\Sl_2\}\Big)=0\,.
\end{equation}
Moreover, we have
\begin{equation}
\label{eq:centralchareresidual}
   \big\{\sQl_A,\sQr_B\big\}=
   -i\kappa
   \{\Sl_A,\epsilon_{BC}\Sr^C\}- \frac{i}{\kappa}\{\epsilon_{AC}\Ql^C,\Qr_B\}=-\frac{i}{\kappa}\epsilon_{AB}\Big(\mathbf{P}-\kappa^2\mathbf{K}\Big)
\end{equation}
Here $\mathbf{P}$ and $\mathbf{K}$ are the central extensions of the $\psu(1|1)^{\oplus4}$ algebra which are not in $\psu(1,1|2)^{\oplus2}$. In fact, for a unitary representation of the $\psu(1|1)^{\oplus4}$  algebra we should take $\mathbf{P}$ and  $\mathbf{K}$ to be Hermitian conjugate to each other; in fact, as reviewed in section~\ref{sec:centrallyextendedalgebra} it is possible and convenient to take them to be real, \textit{cf.}~\eqref{eq:Cisreal}. Introducing the central charge
\begin{equation}
\mathcal{C}\equiv -\frac{i}{\kappa}\Big(\mathbf{P}-\kappa^2\mathbf{K}\Big)\,,
\end{equation}
we have that on a multi-excitation state involving momenta $p_1,\dots p_N$,
\begin{equation}
\label{eq:curlyCeigenvalue}
\mathcal{C}\,|p_1,\dots p_N\rangle= \frac{(\kappa^2-1)h}{i\kappa}\,\sin\left(\frac{p_1+\cdots +p_N}{2}\right)\,|p_1,\dots p_N\rangle\,.
\end{equation}

\subsubsection{Bootstrap principle}
Let us now specialise to the case of three-point functions. We therefore want to consider three images of the 1/2-BPS operator $O(0)$. For this purpose --- without loss of generality owing to conformal symmetry --- we may take the images under superstranslation with $t=0$, $t'=1$ and $t''=\infty$. The first operator will be precisely $O(0)$, sitting at $z=0$ (and being the highest-weight state in its R-symmetry multiplet). The third operator will be $O^\dagger(\infty)$, sitting at $z=\infty$ and being the \textit{lowest} weight state in the R-symmetry multiplet. The second operator will be sitting at $z=1$ and it will be neither the highest- nor the lowest-weight state in the R-symmetry multiplet. The symmetry algebra preserved by this configuration is generated by the four supercharges $(\sQl_A,\sQr_A)$. Following Basso, Komatsu and Vieira~\cite{Basso:2015zoa} we shall assume that this is the symmetry preserved by the ``hexagon operator''. In other words, denoting the hexagon operator by $\h$,
\begin{equation}
[\h,\sQl_A]=0\,,\qquad [\h,\sQr_A]=0\,.
\end{equation}
Indicating the form factor of $\h$ with \textit{any} state $\Psi$ as $\langle \h |\Psi\rangle$, it follows that
\begin{equation}
\label{eq:bootstrapprinciple}
\langle \h |\sQl_A|\Psi\rangle=0\,,\qquad \langle \h |\sQr_A|\Psi\rangle=0\,,\qquad
\langle \h |\mathcal{C}|\Psi\rangle=0 .
\end{equation}
The equality follows by letting $(\sQl_A,\sQr_A)$ --- or, for the third equations, a suitable anticommutator thereof --- act on the state. Vice versa, letting the (super)charges act on the ket we obtain a set of linear constaints that the hexagon form factor must obey.

\subsubsection{Vanishing of the central extension}
The bootstrap condition~\eqref{eq:bootstrapprinciple} takes a particular simple form in the case of the central charge~$\mathcal{C}$, because this acts diagonally and independently on the particles flavour.
We see from~\eqref{eq:curlyCeigenvalue} that, whenever the Ramond-Ramond coupling $h\neq 0$, $\mathcal{C}$ only annihilates physical states --- just as is the case for $\mathbf{P}$ and $\mathbf{K}$ in the spectral problem --- unless $\kappa^2=1$.
Let us recall that $\kappa$ is a free parameter in our construction, see \eqref{eq:supertranslation}; it is up to us to choose the value of $\kappa$ most suitable for the bootstrap procedure.
Following the reasoning of~\cite{Basso:2015zoa}, we must require $\kappa^2=1$, because, if not, the hexagon form factor in \eqref{eq:bootstrapprinciple} would annihilate all non-physical states, which would be too strong a requirement. In fact, we want to define an off-shell object which, like the S~matrix, may act on just a subset of the excitations that define a physical state. Henceforth we will take
\begin{equation}
\kappa=1\,.    
\end{equation}

\subsection{Bootstrapping the hexagon form factor from symmetry}
Here we will use the bootstrap principle of eq.~\eqref{eq:bootstrapprinciple} to fix as much as possible of the hexagon form factor. We will consider in particular the case where $|\Psi\rangle$ consists of a single particle, and when it consists of two. We will then propose a self-consistent ansatz for multi-particle states.

\subsubsection{One-particle states}

As discussed in section~\ref{sec:particlecontent}, we can represent the excitations of the theory, which transform under $\psu(1|1)^{\oplus4}$ c.e., in terms of tensor products of excitations in $\psu(1|1)^{\oplus2}$  c.e.\ --- for instance, for the left massive representation we have that $|Y\rangle=|\phi_\L^\B\otimes \phi_\L^\B\rangle$, $|\Psi^1\rangle=|\varphi_\L^\F\otimes \phi_\L^\B\rangle$, $|\Psi^2\rangle=|\phi_\L^\B\otimes \varphi_\L^\F\rangle$ and $|Z\rangle=|\varphi_\L^\F\otimes \varphi_\L^\F\rangle$. It is useful to rewrite the supercharges of eq.~\eqref{eq:bootstrapprinciple} in terms of the same decomposition,
\begin{equation}
\label{eq:DiagonalAlgebraGenerators}
\begin{aligned}
\sQl_1 =\sl\otimes\id+i\,\Sigma\otimes\ql, \qquad
\sQl_2 = \Sigma\otimes\sl-i\,\ql\otimes\id,\\
\sQr_1 =\qr\otimes\id+i\,\Sigma\otimes\sr,\qquad
\sQr_2 =\Sigma\otimes\qr-i\,\sr\otimes\id,
\end{aligned}
\end{equation}
where $\Sigma$ is the Fermion sign operator.
Imposing now one instance of the bootstrap equation~\eqref{eq:bootstrapprinciple} we get
\begin{equation}
    \langle\h|\sQl_1|Y(p)\rangle=0
    \quad \Rightarrow\quad
    \langle\h|\Psi^2(p)\rangle = 0\,.
\end{equation}
Similarly, it is easy to find (as expected from $\su(2)_{\bullet}$ symmetry) that $\langle\h|\Psi^1(p)\rangle = 0$. We note that, na\"ively, we have more bootstrap equations than undetermined one-particle form factors. However, they all result is one single constraint between 
\begin{equation}
\langle \h | Y(p)\rangle =i\,\frac{a_{\L}(p)}{a_{\L}(p)^*}\, \langle \h | Z(p)\rangle=i\,\langle \h | Z(p)\rangle\,,
\end{equation}
where $a_{\L}(p)$ is the representation coefficient introduced in section~\ref{sec:representations}. Note that, since the equations that we are imposing are linear, we will not be able to fix the overall normalisation of the form factor, but at best only the ratio of different elements. Working on the other representations, we find analogous results:
\begin{equation}
\begin{aligned}
&\langle \h | Y_p\rangle =i\,\langle \h | Z_p\rangle, \, \qquad
&&\langle \h | \tilde{Z}_p\rangle =-i\,\langle \h | \tilde{Y}_p\rangle, \, \\
&\langle \h | \chi^{\dot{1}}_p\rangle =i\,\langle \h | \tilde{\chi}^{\dot{1}}_p\rangle, \, \qquad
&&\langle \h | \chi^{\dot{2}}_p\rangle =-i\,\langle \h | \tilde{\chi}^{\dot{2}}_p\rangle, \,
\end{aligned}
\end{equation}
while the remaining form factors vanish,
\begin{equation}
    \langle \h | \Psi^{A}_p\rangle =0\,,\qquad
    \langle \h | \widetilde{\Psi}^{A}_p\rangle =0\,,\qquad
    \langle \h | \widetilde{T}^{A\dot{A}}_p\rangle =0\,.
\end{equation}
Without loss of generality, we normalise the form factor so that
\begin{equation}
\label{eq:oneparticlehex}
\begin{aligned}
&\langle \h | Y_p\rangle =1\,, \quad &&\langle \h | Z_p\rangle =-i\,, \qquad &&\langle \h | \tilde{Y}_p\rangle=1\,,\quad&&\langle \h |\tilde{Z}_p\rangle=-i\,, \\
&\langle \h | \chi^{\dot{1}}_p\rangle =1\,, \quad &&\langle \h | \tilde{\chi}^{\dot{1}}_p\rangle =-i\,,  \qquad &&\langle \h | {\chi}^{\dot{2}}_p \rangle=1\,,\quad&&\langle \h |\tilde{\chi}^{\dot{2}}_p \rangle=-i\,.
\end{aligned}
\end{equation}

\subsubsection{Two-particle states}
We can determine the hexagon form factor for two-particle states by explicitly evaluating the eq.~\eqref{eq:bootstrapprinciple}. At this point it is worth observing that the symmetry algebra that we are exploiting is a ``diagonal'' (in the sense of the tensor product decomposition  of section~\ref{sec:algebrafactorisation}) $\psu(1|1)^{\oplus2}$ subalgebra in $\psu(1|1)^{\oplus4}$. In this sense it is not surprising that the two-particle form-factor may be expressed in terms of the only non-trivial intertwiner of two short $\psu(1|1)^{\oplus2}$ representations, \textit{i.e.}\ the Borsato--Ohlsson-Sax--Sfondrini S~matrix~\cite{Borsato:2012ud}. This is completely analogous to what happens for the $\AdS_5\times\Sph^5$ hexagon in terms of the Beisert S~matrix.
A solution of all bootstrap equations for the two-particle form factor may be written explicitly in terms of the S-matrix elements of section~\ref{sec:Smatrix}.
Note that, as expected, we are unable to fix one overall prefactor for each choice of irreducible representations; below we shall denote such prefactors as $h(p,q)$ and postpone their discussion to section~\ref{sec:hprefactor}.

\paragraph{Form factor for two massive excitations.}
\label{sec:formfactor2particles}
We may distinguish two cases depending on whether the two exitations are left or right. When they are both left we have
\begin{equation}
\begin{aligned}
&&\langle \h | Y_p Y_q\rangle &= +A_{pq}^{\L\L},&&&\qquad
\langle \h | Z_p Z_q\rangle &= + F_{pq}^{\L\L},\\
&&\langle \h | Y_p Z_q\rangle &= -iB_{pq}^{\L\L},&&&\qquad
\langle \h | Z_p Y_q\rangle &= -i D_{pq}^{\L\L},\\
&&\langle \h | \Psi^2_p\Psi^1_q\rangle &= +i C_{pq}^{\L\L},&&&\qquad
\langle \h | \Psi^1_p\Psi^2_q\rangle &= -i C_{pq}^{\L\L}.
\end{aligned}
\end{equation}
When both particles are right we get
\begin{equation}
\begin{aligned}
&&\langle \h | \tilde{Y}_p \tilde{Y}_q\rangle &= +A_{pq}^{\R\R},\qquad
&&&\langle \h |  \tilde{Z}_p  \tilde{Z}_q\rangle &= + F_{pq}^{\R\R},\\
&&\langle \h | \tilde{Y}_p  \tilde{Z}_q\rangle &= -i B_{pq}^{\R\R},\qquad
&&&\langle \h |  \tilde{Z}_p \tilde{Y}_q\rangle &= -i D_{pq}^{\R\R},\\
&&\langle \h | \tilde{\Psi}^1_p\tilde{\Psi}^2_q\rangle &= -i C_{pq}^{\R\R},\qquad
&&&\langle \h | \tilde{\Psi}^2_p \tilde{\Psi}^1_q\rangle &= +i C_{p,q}^{\R\R}.
\end{aligned}
\end{equation}
In the case of mixed chirality, we distinguish two cases depending on the ordering of the particles. Firstly, for left--right we have
\begin{equation}
\begin{aligned}
&&\langle \h |  Y_p  \tilde{Y}_q\rangle &= + A_{pq}^{\L\R},\qquad
&&&\langle \h | \Psi^1_p \tilde{\Psi}^2_q\rangle &=- F_{pq}^{\L\R},\\
&&\langle \h |  \Psi^2_p \tilde{\Psi}^1_q\rangle &= - B_{pq}^{\L\R} ,\qquad
&&&\langle \h | Z_p  \tilde{Y}_q\rangle &= -i D_{pq}^{\L\R},\\
&&\langle \h | Y_p \tilde{Z}_q\rangle &= -i C_{pq}^{\L\R},\qquad
&&&\langle \h | Z_p \tilde{Z}_q\rangle &=  +E_{pq}^{\L\R}.
\end{aligned}
\end{equation}
Finally, for right--left we have
\begin{equation}
\begin{aligned}
&&\langle \h |  \tilde{Y}_p  Y_q\rangle &=+ A_{pq}^{\R\L},\qquad
&&&\langle \h | \tilde{\Psi}^2_p \Psi^1_q\rangle &=- F_{pq}^{\R\L},\\
&&\langle \h | \tilde{\Psi}^1_p \Psi^2_q \rangle &= - B_{pq}^{\R\L} ,\qquad
&&&\langle \h | \tilde{Z}_p  Y_q\rangle &=  -iD_{pq}^{\R\L},\\
&&\langle \h | \tilde{Y}_p Z_q\rangle &=  -iC_{pq}^{\R\L},\qquad
&&&\langle \h | \tilde{Z}_p Z_q\rangle &= + E_{pq}^{\R\L}.
\end{aligned}
\end{equation}

\paragraph{One massless and one massive particle.}
In this case we can distinguish excitations on whether they are left or right (for the massive particle) and depending on their $\su(2)_{\circ}$ charge (for the massless particle). Moreover, we can also distinguish their order. It turns out that we may write more compact formulae by explicitly making use of the one-particle form factor~$\langle\h|\chi^{\dot{A}}\rangle$. For instance, in the case of one left-massive particle and one massless particle we obtain
\begin{equation}
\begin{aligned}
&&\langle \h |  Y_p \chi^{\dot{A}}_q\rangle &= + A_{pq}^{\L\z}\,,\qquad
&&&\langle \h | Z_p \tilde{\chi}^{\dot{A}}_q\rangle &=\, +F_{pq}^{\L\z}\,,\\
&&\langle \h | Y_p \tilde{\chi}^{\dot{A}}_q \rangle &= -i B_{pq}^{\L\z} \,,\qquad
&&&\langle \h | Z_p \chi^{\dot{A}}_q\rangle &= -i D_{pq}^{\L\z}\,,\\
&&\langle \h | \Psi^2_p T^{\dot{A}1}_q\rangle &=  +iC_{pq}^{\L\z}\,,\qquad
&&&\langle \h | \Psi^1_p T^{\dot{A}2}_q\rangle &= -iC_{pq}^{\L\z}\,.
\end{aligned}
\end{equation}
Similarly, for one-right massive particle and one massless particle we have
\begin{equation}
\begin{aligned}
&&\langle \h |  \tilde{Y}_p \chi^{\dot{A}}_q\rangle &= +A_{pq}^{\R\z}\,,\qquad
&&&\langle \h | \tilde{\Psi}^2_p T^{\dot{A}1}_q\rangle&= -F_{pq}^{\R\z}\,,\\
&&\langle \h | \tilde{\Psi}^1_p T^{\dot{A}2}_q \rangle &= - B_{pq}^{\R\z} \,,\qquad
&&&\langle \h | \tilde{Z}_p \chi^{\dot{A}}_q\rangle &= -iD_{pq}^{\R\z}\,,\\
&&\langle \h | \tilde{Y}_p \tilde{\chi}^{\dot{A}}_q\rangle &=  -iC_{pq}^{\R\z}\,,\qquad
&&&\langle \h | \tilde{Z}_p \tilde{\chi}^{\dot{A}}_q\rangle &= +E_{pq}^{\R\z}\,.
\end{aligned}
\end{equation}
The possibility of writing formulae in such a compact way is a first sign of an underlying symmetry structure of the form factor which we shall investigate in the next section. To conclude here, we list the mixed-mass form factors when particles are in the reversed order, 
\begin{equation}
\begin{aligned}
&&\langle \h |  \chi^{\dot{A}}_p Y_q\rangle &= + A_{pq}^{\z\L}\,,\qquad
&&&\langle \h | \tilde{\chi}^{\dot{A}}_p Z_q\rangle &=\, +F_{pq}^{\z\L}\,,\\
&&\langle \h | \chi^{\dot{A}}_p Z_q \rangle &= -i B_{pq}^{\z\L} \,,\qquad
&&&\langle \h | \tilde{\chi}^{\dot{A}}_p Y_q\rangle &= -i D_{pq}^{\z\L}\,,\\
&&\langle \h | T^{\dot{A}2}_p \Psi^1_q\rangle &=  -iC_{pq}^{\z\L}\,,\qquad
&&&\langle \h | T^{\dot{A}1}_p \Psi^2_q\rangle &= +i C_{pq}^{\z\L}\,,
\end{aligned}
\end{equation}
and finally
\begin{equation}
\begin{aligned}
&&\langle \h |  \chi^{\dot{A}}_p \tilde{Y}_q\rangle &=+ A_{pq}^{\z\R}\,,\qquad
&&&\langle \h | T^{\dot{A}1}_p \tilde{\Psi}^2_q\rangle &=+F_{pq}^{\z\R}\,,\\
&&\langle \h | T^{\dot{A}2}_p \tilde{\Psi}^1_q \rangle& = + B_{pq}^{\z\R}\,,\qquad
&&&\langle \h | \tilde{\chi}^{\dot{A}}_p \tilde{Y}_q\rangle &= -i D_{pq}^{\z\R}\,,\\
&&\langle \h |\chi^{\dot{A}}_p \tilde{Z}_q\rangle &=  -i C_{pq}^{\z\R}\,,\qquad
&&&\langle \h | \tilde{\chi}^{\dot{A}}_p \tilde{Z}_q\rangle &= + E_{pq}^{\z\R}\,.
\end{aligned}
\end{equation}
Note that the form factor is blind to the $\su(2)_{\circ}$ index~$\dot{A}$, which is unsurprising as the algebra which we are using to constrain it commutes with $\su(2)_{\circ}$.

\paragraph{Two massless particles.}
In this case we can compactly write
\begin{equation}
\begin{aligned}
&&\langle \h |  \chi^{\dot{A}}_p \chi^{\dot{B}}_q\rangle &= + A_{pq}^{\z\z},\qquad
&&&\langle \h | \tilde{\chi}^{\dot{A}}_p \tilde{\chi}^{\dot{B}}_q\rangle &= + F_{pq}^{\z\z},\\
&&\langle \h | \chi^{\dot{A}}_p \tilde{\chi}^{\dot{B}}_q \rangle &=  -i B_{pq}^{\z\z} ,\qquad
&&&\langle \h | \tilde{\chi}^{\dot{A}}_p \chi^{\dot{B}}_q\rangle &=   -i D_{pq}^{\z\z} ,\\
&&\langle \h | T^{\dot{A}1} T^{\dot{B}2}\rangle &= +i  C_{pq}^{\z\z},\qquad
&&&\langle \h | T^{\dot{A}2} T^{\dot{B}1}\rangle& = -i C_{pq}^{\z\z} ,
\end{aligned}
\end{equation}
which again is blind to $\su(2)_\circ$.

To conclude the discussion of two-particle form factors, it is important to note that the equations~\eqref{eq:bootstrapprinciple} we have imposed are linear, so that we may obtain new solutions by multiplying each block (for instance, left--left, or left--right, \textit{etc.}) by an arbitrary function. In other words, the prefactors $\Sigma^{\L\L}_{p,q}$, $\Sigma^{\L\R}_{p,q}$, \textit{etc.}, which appear in the S~matrix elements can be changed with no effect~\eqref{eq:bootstrapprinciple}. We shall see later how they may be further constrained, see section~\ref{sec:hprefactor}.

\subsubsection{General form of the two-particle hexagon form factor}
It is possible to summarise the form of the two-particle form factor in a way that encompasses the various representations encountered thus far. Let us denote a generic $\psu(1|1)^{\oplus4}$ excitation in the tensor product form of section~\ref{sec:algebrafactorisation} as
\begin{equation}
    \Xi^{a\acute{a}}\equiv \xi^{a}\otimes \acute{\xi}^{\acute{a}}\,,
\end{equation}
where the second entry of the tensor product is distinguished by a ``prime''.
Here $\xi^a$ and $\acute{\xi}^{\acute{a}}$ could transform under any of the relevant representations which we encountered, \textit{i.e.}\ $\rho_{\pm}$, $\rho_0$ or $\rho_0'$; we absorb the information of  the representation into the indices $a$ and $\acute{a}$ to keep the notation a little lighter.
Then, using this notation, we have that
\begin{equation}
\label{eq:twoparticlehexsymbolic}
\begin{aligned}
    \langle \h | \Xi^{a\acute{a}}_p\Xi^{b\acute{b}}_q\rangle&=
    \mathbf{K}_p\mathbf{K}_q (-1)^{(F_{a}+F_{\acute{a}})F_{b}}\Big[|\xi^{b}_{q}\xi^{a}_{p}\rangle\otimes\mathbf{S}|\acute{\xi}^{\acute{a}}_{p}\acute{\xi}^{\acute{b}}_{q}\rangle\Big]\\
    &=(-1)^{(F_{a}+F_{\acute{a}})F_{b}}\, S^{\acute{a}\acute{b}}_{\acute{d}\acute{c}}(p,q)
    \mathbf{K}_p\mathbf{K}_q\Big[|\xi^{b}_{q}\xi^{a}_{p}\rangle\otimes|\acute{\xi}^{\acute{d}}_{q}\acute{\xi}^{\acute{c}}_{p}\rangle\Big],
\end{aligned}
\end{equation}
where we have introduced the ``contraction operator'' 
\begin{equation}
\label{eq:contractionop}
\begin{aligned}
    \mathbf{K}_p \equiv&\ \Big(h_Y\frac{\partial}{\partial \acute{\phi}^{\B}_{\L}(p)}\frac{\partial}{\partial \phi^{\B}_{\L}(p)}+h_Z\frac{\partial}{\partial \acute{\varphi}^{\F}_{\L}(p)}\frac{\partial}{\partial \varphi^{\F}_{\L}(p)}\Big)\\
    &+\Big(h_{\tilde{Y}}\frac{\partial}{\partial \acute{\varphi}^{\B}_{\R}(p)}\frac{\partial}{\partial \varphi^{\B}_{\R}(p)}+h_{\tilde{Z}}\frac{\partial}{\partial \acute{\phi}^{\F}_{\R}(p)}\frac{\partial}{\partial \phi^{\F}_{\R}(p)}\Big)\\
    &+\Big(h_{{\chi}^1}\frac{\partial}{\partial \acute{\phi}^{\F}_{\z}(p)}\frac{\partial}{\partial \phi^{\B}_{\z}(p)}+h_{\tilde{\chi}^1}\frac{\partial}{\partial \acute{\varphi}^{\B}_{\z}(p)}\frac{\partial}{\partial \varphi^{\F}_{\z}(p)}\Big)\\
    &+\Big(h_{{\chi}^2}\frac{\partial}{\partial \acute{\phi}^{\B}_{\z}(p)}\frac{\partial}{\partial \phi^{\F}_{\z}(p)}+h_{\tilde{\chi}^2}\frac{\partial}{\partial \acute{\varphi}^{\F}_{\z}(p)}\frac{\partial}{\partial \varphi^{\B}_{\z}(p)}\Big),
\end{aligned}
\end{equation}
where $h_Y=\langle \mathbf{h}|Y\rangle$, \textit{etc.}, are the values of the one-particle hexagon form factors of eq.~\eqref{eq:oneparticlehex}. 
Let us explain what we mean by this notation. We begin to note that $\mathbf{K}_p$ simply picks out the one-particle states with a non-trivial hexagon form factor and assigns them the value thereof, \textit{i.e.}\ $\mathbf{K}_p |\Xi^{a\acute{a}}(p)\rangle = \langle\mathbf{h}|\Xi^{a\acute{a}}\rangle$. The reason why we  go through the trouble of introducing this operator --- something not necessary in $\AdS_5\times \Sph^5$ --- is that here the one-particle hexagon form factors in the massless representations are nonvanishing for particles with Fermionic statistics. This creates a potential ambiguity for massless particles whenever we want to contract multi-particle states: note that indeed the commutator $[\mathbf{K}_{p},\mathbf{K}_q]$ does not vanish for massless particles due to the statistics. Realising the contractions in terms of the graded differential operator $\mathbf{K}_p$ will make it easier to properly account for this statistics.
Armed with this operator, let us go back to eq.~\eqref{eq:twoparticlehexsymbolic}. In the first equality we rearrange the excitations to factor out the pieces of the tensor product related to either factor of the diagonal symmetry algebra (distinguished here by the absence or presence of the prime), picking up Fermion signs as appropriate. To this end we defined
\begin{equation}
    F_{a} = \begin{cases}
    0 &\text{if}\ \xi^a\ \text{is a Boson}\\
    1 &\text{if}\ \xi^a\ \text{is a Fermion}
    \end{cases}\qquad\text{and}\quad
    F_{\acute{a}} = \begin{cases}
    0 &\text{if}\ \acute{\xi}^{\acute{a}}\ \text{is a Boson}\\
    1 &\text{if}\ \acute{\xi}^{\acute{a}}\ \text{is a Fermion}
    \end{cases}
\end{equation}
We then scatter the ``primed'' particles by using the $\psu(1|1)^{\oplus}_{\text{c.e.}}$ S~matrix in the appropriate representation (for instance, $\rho_{\L}\otimes\rho_{\L}$, $\rho_{\L}\otimes\rho_{\z}'$, \textit{etc.}). We pick up the relative S~matrix elements, which now contain a irrep-dependent prefactor $h^{\acute{a}\acute{b}}(p,q)$. Lastly, we act with the contraction operator, again keeping track of the statistics, perfectly reproducing the results which we listed above.
It is worth stressing that this prescription can also be applied to $\AdS_{5}\times\Sph^{5}$, yielding a result identical to ref.~\cite{Basso:2015zoa}.

\subsubsection{Many-particle states}
Nothing stops us from imposing eq.~\eqref{eq:bootstrapprinciple} for three- and higher-particle states. However, while for two-particle states we managed to fix the form factor completely (up to an unavoidable scalar prefactor for each choice of representations), for higher number of particles we will only be able to fix relatively few coefficients.  A better approach, following~\cite{Basso:2015zoa}, is to exploit the fact that the two-particle solution can be written in terms of a factorised S~matrix~\cite{Borsato:2012ud,Borsato:2014exa}. Then the Yang-Baxter equation allow us to write down a self-consistent ansatz with is guaranteed to satisfy all symmetry requirements. We set
\begin{equation}
\label{eq:multiparticlehexagon}
\begin{aligned}
    &\big\langle \h \big| \Xi^{a_1\acute{a}_1}_{p_1}\Xi^{a_2\acute{a}_2}_{p_2}\dots\Xi^{a_N\acute{a}_N}_{p_N}\big\rangle\equiv\\
    &\qquad\qquad\qquad\equiv
(-1)^{F_{12\cdots N}} \,
\mathbf{K}_{12\cdots N}
\Big[
\bigl| \xi^{a_N}_{p_N}\dots\xi^{a_2}_{p_2} \xi^{a_1}_{p_1} \big\rangle
\otimes
\mathbf{S}_{12\cdots N}\big| \acute{\xi}^{\acute{a}_1}_{p_1}\acute{\xi}^{\acute{a}_1}_{p_1}\dots \acute{\xi}^{\acute{a}_N}_{p_N} \big\rangle\Big]\,.
\end{aligned}
\end{equation}
where
\begin{equation}
F_{12\cdots N}\equiv\sum_{1\leq i<j\leq N } (F_{a_i}+F_{\acute{a}_i})F_{a_j}\,,\qquad
\mathbf{K}_{12\cdots N}\equiv
\mathbf{K}_{p_1}
\mathbf{K}_{p_2}\cdots
\mathbf{K}_{p_N}\,,
\end{equation}
and $\mathbf{S}_{12\cdots N}$ is the $N$-particle S~matrix, which as we remarked may be factorised owing to the Yang-Baxter equation.
Again, it is worth remarking that this formula can also be applied to the case of $\AdS_{5}\times\Sph^{5}$ and, despite the apparent difference from the proposal of ref.~\cite{Basso:2015zoa}, is perfectly equivalent to that.

\subsection{Representations of the hexagon algebra and crossing}
\label{sec:crossinghex}
In this section we will look more closely at the structure of the hexagon symmetry algebra, which is given by $\psu(1|1)^{\oplus2}$ without any central extension. This emerges as a sort of diagonal subalgebra of $\psu(1|1)^{\oplus4}$ c.e., see for instance eq.~\eqref{eq:Hexagonalgebra}.
On the other hand, we have found in section~\ref{sec:formfactor2particles} that the two-particle hexagon form factor features the Borsato--Ohlsson-Sax--Sfondrini $\psu(1|1)^{\oplus2}$ c.e.\ S~matrix~\cite{Borsato:2012ud,Borsato:2014exa}. This motivates us to investigate the tensor product decomposition of these representations more closely. 

Recall that the $\AdSST$ symmetries and representations may be factorised as described in section~\ref{sec:algebrafactorisation}. For what concerns the representations, let us use again the short hand notation introduced above. We indicate a $\psu(1|1)^{\oplus4}$ state as $\Xi^{a\acute{a}}$ and a generic $\psu(1|1)^{\oplus2}$ state as either $\xi$  or $\acute{\xi}^{\acute{a}}$ depending on how it is embedded in the $(\psu(1|1)^{\oplus2})^{\otimes 2}$ decomposition,
\begin{equation}
\label{eq:notationfactorisation}
    |\xi\rangle\equiv | \xi\otimes {1}\rangle\,,\qquad
    |\acute{\xi}\rangle\equiv | {1}\otimes\acute{\xi}\rangle\,.
\end{equation}
Let us now recall the form of the generators of the hexagon symmetry algebra $(\sQl_A,\sQr_A)$ in terms of this factorisation. Rewriting slightly eq.~\eqref{eq:DiagonalAlgebraGenerators} we get
\begin{equation}
\label{eq:rescaleddiaggenerators}
\begin{aligned}
\sQl_1 =\sl\otimes\id+i\,\Sigma\otimes\ql, \qquad
i \sQl_2 = \ql\otimes\id+i\,\Sigma\otimes\sl,\\
\sQr_1 =\qr\otimes\id+i\,\Sigma\otimes\sr,\qquad
i\sQr_2 =\sr\otimes\id+i\,\Sigma\otimes\qr,
\end{aligned}
\end{equation}
where we multiplied $\sQl_2$ and $\sQr_2$ by $i$ for later convenience. In fact, it will be slightly easier --- and completely equivalent --- to look at the representations of the algebra generated by $(\sQl_1,i\sQl_2, \sQr_1,i\sQr_2)$.

\subsubsection{Massive representations}
We want to act with the generators~\eqref{eq:rescaleddiaggenerators} on massive excitations on either side of the tensor product. We first consider states of the type $|\psi\otimes{1}\rangle$ where $\psi\in\rho_{\L}$ or $\psi\in\rho_{\R}$ is a massive $\psu(1|1)^{\oplus2}$ excitation. In the notation of eq.~\eqref{eq:notationfactorisation} we have
\begin{equation}
\label{eq:repr1sttensor}
    \begin{aligned}
      &\sQl_1 \ket{\varphi_\L^{\F}} = a^*_\L \ket{\phi_\L^{\B}} , \qquad
      &&\sQl_1 \ket{\varphi_\R^{\B}} = a^*_\R \ket{\phi_\R^{\F}} ,   \\
      i&\sQl_2 \ket{\phi_\L^{\B}} =  a_\L \ket{\varphi_\L^{\F}},\qquad
      &i&\sQl_2 \ket{\phi_\R^{\F}} =  a_\R \ket{\varphi_\R^{\B}} ,  \\
      &\sQr_1 \ket{\varphi_\L^{\F}} = b_\L \ket{\phi_\L^{\B}} , \qquad
      &&\sQr_1 \ket{\varphi_\R^{\B}} = b_\R \ket{\phi_\R^{\F}} ,\\
     i&\sQr_2 \ket{\phi_\L^{\B}} =  b^*_\L \ket{\varphi_\L^{\F}} , \qquad  &i&\sQr_2 \ket{\phi_\R^{\F}} = b^*_\R \ket{\varphi_\R^{\B}} .
    \end{aligned}
\end{equation}
Comparing with the $\psu(1|1)^{\oplus2}$ c.e.\ representations reviewed in section~\ref{sec:irrepssmallalgebra}, we see that these are precisely $\rho_{\L}$ and $\rho_{\R}$. This, of course, involves identifying $\ql=i\sQl_2$, $\qr=\sQr_1$, $\sl=\sQl_1$ and $\sr=i\sQr_2$ --- as it can readily be seen from eq.~\eqref{eq:rescaleddiaggenerators}.
Things are less trivial if we consider instead excitations of the form $|\acute{\psi}\rangle= |1\otimes \psi\rangle$. Here we find
\begin{equation}
\label{eq:repr2ndtensor}
    \begin{aligned}
&\sQl_1 \ket{\acute{\phi}_\R^{\F}} = i a_\R \ket{\acute{\varphi}_\R^{\B}} ,
\qquad       &&\sQl_1 \ket{\acute{\phi}_\L^{\B}} = i a_\L \ket{\acute{\varphi}_\L^{\F}} , \\
      i&\sQl_2 \ket{\acute{\varphi}_\R^{\B}} = ia^*_\R \ket{\acute{\phi}_\R^{\F}} , 
\qquad &i&\sQl_2 \ket{\acute{\varphi}_\L^{\F}} = ia^*_\L \ket{\acute{\phi}_\L^{\B}} , \\
      &\sQr_1 \ket{\acute{\phi}_\R^{\F}} = i b^*_\R \ket{\acute{\varphi}_\R^{\B}} ,
      \qquad&&\sQr_1 \ket{\acute{\phi}_\L^{\B}} = i b^*_\L \ket{\acute{\varphi}_\L^{\F}} ,\\
      i&\sQr_2 \ket{\acute{\varphi}_\R^{\B}} = ib_\R \ket{\acute{\phi}_\R^{\F}} ,  \qquad  &i&\sQr_2 \ket{\acute{\varphi}_\L^{\F}} = ib_\L \ket{\acute{\phi}_\L^{\B}} .
    \end{aligned}
\end{equation}
By using the definition of the crossing transformation, see section~\ref{sec:crossingparam}, we notice that the representations of eq.~\eqref{eq:repr2ndtensor} are actually the analytic continuation of those in eq.~\eqref{eq:repr1sttensor}.
Denoting crossing (respectively, anti-crossing) of a particle of momentum~$p$ as $p^{2\gamma}$ (respectively, $p^{-2\gamma}$), we have that \textit{e.g.}\ $a^*_{\L}(p^{\pm2\gamma})=\mp i a_{\R}(p)$ and $b_{\L}(p^{\pm2\gamma})=\mp i b^*_{\R}(p)$. As a consequence of these identifications we find that $|{\acute{\phi}_\R^{\F}}\rangle$ and $|{\acute{\varphi}_\R^{\B}}\rangle$ transform as the analytic continuation of $\rho_{\L}$, while $\acute{\phi}_\L^{\B}$ and $\acute{\varphi}_\L^{\F}$ transform as the analytic continuation of $\rho_{\R}$. More specifically, we may identify
\begin{equation}
\begin{aligned}
    &\ket{\acute{\phi}^\B_\L(p)} = \ket{\varphi^\B_\R(p^{-2\gamma})}\,, \qquad &&\ket{\acute{\varphi}^\F_\L(p)} = \ket{\phi^\F_\R(p^{-2\gamma})}\,,\\
    &\ket{\acute{\phi}^\F_\R(p)} = \ket{\varphi^\F_\L(p^{-2\gamma})}\,, \qquad &&\ket{\acute{\varphi}^\B_\R(p)} = \ket{\phi^\B_\L(p^{-2\gamma})}\,.
\end{aligned}
\end{equation}
It is worth observing that the $\psu(1|1)^{\oplus2}$ c.e.\ representations are not invariant under $4\gamma$-shift, but instead they pick up some minus sign~\cite{Sfondrini:2014via}. This is essentially due to the fact that the representation parameter $\eta_{*}(p)$ \eqref{eq:etaparameter} is not $4\gamma$-periodic because it is not a meromorphic function of the Zhukovski variables, see again section~\ref{sec:crossingparam}. In practice this means that after $4\gamma$ we pick up a minus sign for the Fermions,
\begin{equation}
\label{eq:4gammashift}
\begin{aligned}
    &\phi^\B_\L(p^{2\gamma}) = + \phi^\B_\L(p^{-2\gamma})\,,\qquad &&\varphi^\F_\L(p^{2\gamma}) = - \varphi^\F_\L(p^{-2\gamma}) \,,\\
    &\varphi^\B_\R(p^{2\gamma}) = + \varphi^\B_\R(p^{-2\gamma})\,,\qquad &&\phi^\F_\R(p^{2\gamma}) = - \phi^\F_\R(p^{-2\gamma})\,.
    \end{aligned}
\end{equation}
Such monodromies are well known from the study of the $\AdS_5\times\Sph^5$ S~matrix~\cite{Arutyunov:2009ga}.

We can now consider the true massive excitations of $\AdSST$, those that lie in the $\varrho_{\L}=\rho_{\L}\otimes\rho_{\L}$ and$\varrho_{\R}=\rho_{\R}\otimes\rho_{\R}$ representations. We can certainly act directly with the diagonal generators~\eqref{eq:rescaleddiaggenerators} on this representation. Alternatively, we have seen that we can identify this diagonal algebra as the Borsato--Ohlsson-Sax--Sfondrini $\psu(1|1)^{\oplus2}$ c.e.\ and the representations as the tensor products $\rho_{\L}\otimes \rho_{\R}^{-2\gamma}$ and $\rho_{\R}\otimes \rho_{\L}^{-2\gamma}$. In hindsight, this identification is quite natural because it guarantees that the central charge of $\psu(1|1)^{\oplus2}$ c.e.\  vanishes on the tensor product representations as desired. Explicitly, we write
\begin{equation}
    \begin{aligned}
        &Y_p = \phi^\B_{\L,p}\, \varphi^\B_{\R,p^{-2\gamma}}\,, \qquad &&\tilde{Z}_p = \phi^\F_{\R,p}\, \varphi^\F_{\L,p^{-2\gamma}}\,, \\
        &\Psi^1_p  = \varphi^\F_{\L,p}\, \varphi^\B_{\R,p^{-2\gamma}}\,, \qquad &&\tilde{\Psi}^1_p = \varphi^\B_{\R,p}\, \varphi^\F_{\L,p^{-2\gamma}}\,,  \\
        &\Psi^2_p  = \phi^\B_{\L,p}\, \phi^\F_{\R,p^{-2\gamma}}\,, \qquad &&\tilde{\Psi}^2_p = \phi^\F_{\R,p}\, \phi^\B_{\L,p^{-2\gamma}}\,, \\
        &Z_p = \varphi^\F_{\L,p}\, \phi^\F_{\R,p^{-2\gamma}}\,, \qquad &&\tilde{Y}_p  = \varphi^\B_{\R,p}\, \phi^\B_{\L,p^{-2\gamma}}\,.
    \end{aligned}
\end{equation}
Using these identifications, and noting the automorphism that relates $4\gamma$-shifted representations~\eqref{eq:4gammashift}, we obtain the crossing rule for the physical particles. For the left representation we have
\begin{equation}
\label{eq:crossingruleleft}
    \begin{aligned}
    &Y_p = \phi^\B_{\L,p}\, \varphi^\B_{\R,p^{-2\gamma}}\quad \xrightarrow[]{2\gamma}\quad& Y_{p^{2\gamma}}& =\phi^\B_{\L,p^{2\gamma}}\, \varphi^\B_{\R,p} =  +\varphi^\B_{\R,p}\, \phi^\B_{\L,p^{-2\gamma}} =+ \tilde{Y}_p\,, \\
    &\Psi^1_p  = \varphi^\F_{\L,p}\, \varphi^\B_{\R,p^{-2\gamma}}\quad \xrightarrow[]{2\gamma}\quad& \Psi^1_{p^{2\gamma}}& =\varphi^\F_{\L,p^{2\gamma}}\, \varphi^\B_{\R,p} =  -\varphi^\B_{\R,p}\, \varphi^\F_{\L,p^{-2\gamma}} = -\tilde{\Psi}^1_p\,, \\
    &\Psi^2_p  = \phi^\B_{\L,p}\, \phi^\F_{\R,p^{-2\gamma}}\quad \xrightarrow[]{2\gamma}\quad& \Psi^2_{p^{2\gamma}}& =\phi^\B_{\L,p^{2\gamma}}\, \phi^\F_{\R,p} =  +\phi^\F_{\R,p}\, \phi^\B_{\L,p^{-2\gamma}} = -\tilde{\Psi}^2_p\,, \\
    &Z_p = \varphi^\F_{\L,p} \phi^\F_{\R,p^{-2\gamma}}\quad \xrightarrow[]{2\gamma}\quad &Z_{p^{2\gamma}} &= \varphi^\F_{\L,p^{2\gamma}}\, \phi^\F_{\R,p} =  +\phi^\F_{\R,p}\, \varphi^\F_{\L,p^{-2\gamma}} = +\tilde{Z}_p\,.
    \end{aligned}
\end{equation}
Similarly, for the right representation we get
\begin{equation}
\label{eq:crossingruleright}
    \tilde{Z}_{p^{2\gamma}}=Z_p\,,\qquad
    \tilde{\Psi}^{1}_{p^{2\gamma}}=\Psi^1_p\,,\qquad
    \tilde{\Psi}^{2}_{p^{2\gamma}}=\Psi^2_p\,,\qquad
    \tilde{Y}_{p^{2\gamma}}=Y_p\,.
\end{equation}
Notice that these signs look different from the ones in~\eqref{eq:crossingruleleft}.

\subsubsection{Massless representations}
The argument above may be repeated for the massless representations of section~\ref{sec:representations}. Indeed we may obtain the massless representations as the $m\to0$ limit of either the right or the left representation~\cite{Borsato:2016xns}. In particular we have
\begin{equation}
    \begin{aligned}
      &\sQl_1 \ket{\varphi_\z^{\F}} = a^*_\z \ket{\phi_\z^{\B}}\, , \qquad  
      &&\sQl_1 \ket{\varphi_\z^{\B}} = a^*_\z \ket{\phi_\z^{\F}}\, ,\\
      i&\sQl_2 \ket{\phi_\z^{\B}} =  a_\z \ket{\varphi_\z^{\F}}\, , \qquad 
      &i&\sQl_2 \ket{\phi_\z^{\F}} =  a_\z \ket{\varphi_\z^{\B}}\, , \\
      &\sQr_1 \ket{\varphi_\z^{\F}} = b_\z \ket{\phi_\z^{\B}}\, , \qquad  
      &&\sQr_1 \ket{\varphi_\z^{\B}} = b_\z \ket{\phi_\z^{\F}}\, ,\\
      i&\sQr_2 \ket{\phi_\z^{\B}} =  b^*_\z \ket{\varphi_\z^{\F}}\, , \qquad 
      &i&\sQr_2 \ket{\phi_\z^{\F}} =  b^*_\z \ket{\varphi_\z^{\B}}\, ,
    \end{aligned}
\end{equation}
while for the other half of the tensor product we have
\begin{equation}
    \begin{aligned}
      &\sQl_1 \ket{\acute{\phi}_\z^{\F}} = i a_\z \ket{\acute{\varphi}_\z^{\B}}\, , 
      \qquad &&\sQl_1 \ket{\acute{\phi}_\z^{\B}} = i a_\z \ket{\acute{\varphi}_\z^{\F}}\, , \\
      i&\sQl_2 \ket{\acute{\varphi}_\z^{\B}} = ia^*_\z \ket{\acute{\phi}_\z^{\F}} \,, 
      \qquad  &i&\sQl_2 \ket{\acute{\varphi}_\z^{\F}} = ia^*_\z \ket{\acute{\phi}_\z^{\B}} \,, \\
      &\sQr_1 \ket{\acute{\phi}_\z^{\F}} = i b^*_\z \ket{\acute{\varphi}_\z^{\B}}\, ,
      \qquad  &&\sQr_1 \ket{\acute{\phi}_\z^{\B}} = i b^*_\z \ket{\acute{\varphi}_\z^{\F}}\, ,\\
      i&\sQr_2 \ket{\acute{\varphi}_\z^{\B}} = ib_\z \ket{\acute{\phi}_\z^{\F}} ,
      \qquad  &i&\sQr_2 \ket{\acute{\varphi}_\z^{\F}} = ib_\z \ket{\acute{\phi}_\z^{\B}} .
    \end{aligned}
\end{equation}
We can then identify
\begin{equation}
\label{eq:masslesschainidentif}
\begin{aligned}
    &\acute{\phi}^\B_\z(p) = -\sigma_{p}\, \varphi^\F_\z(p^{-2\gamma})\,, \qquad &&\acute{\varphi}^\B_\z(p) = -\sigma_{p}\, \phi^\F_\z(p^{-2\gamma})\,, \\
    &\acute{\phi}^\F_\z(p) = \, \varphi^\B_\z(p^{-2\gamma})\,, \qquad &&\acute{\varphi}^\F_\z(p) = \, \phi^\B_\z(p^{-2\gamma})\,.
\end{aligned}
\end{equation}
This is the analogue of what we saw above for massive representations, with one important difference: in the massive case, crossing linked left to right representations. Here, by taking the $m\to0$ limit we have that the left and right representation are isomorphic, so that we can link the massless representation to itself up to keeping track of the sign $\sigma(p)=-\text{sgn}[\sin (p/2)]$ which appears in the isomorphism. More precisely, we would have $\sigma(p^{-2\gamma})$ appearing in eq.~\eqref{eq:masslesschainidentif}, but this can be simplified to $-\sigma(p)$. 
Moreover, the massless representation parameter $\eta_{\z}(p)$ \eqref{eq:etaparameter} is $4\gamma$-periodic, see section~\ref{sec:crossingparam}. Hence, we do not pick up a minus sign for the Fermions after $4\gamma$,
\begin{equation}
\label{eq:4gammashift-massless}
\begin{aligned}
    &\phi^\B_\z(p^{2\gamma}) = + \phi^\B_\z(p^{-2\gamma})\,,\qquad &&\varphi^\F_\z(p^{2\gamma}) = + \varphi^\F_\z(p^{-2\gamma}) \,,\\
    &\varphi^\B_\z(p^{2\gamma}) = + \varphi^\B_\z(p^{-2\gamma})\,,\qquad &&\phi^\F_\z(p^{2\gamma}) = + \phi^\F_\z(p^{-2\gamma})\,.
    \end{aligned}
\end{equation}
Therefore we can write
\begin{equation}
    \begin{aligned}
        &\chi^{\dot{1}}_p = \phi^\B_{\z,p}\, \varphi^\B_{\z,p^{-2\gamma}}\,, \qquad &&\chi^{\dot{2}}_p = -i \sigma_p \phi^\F_{\z,p}\, \varphi^\F_{\z,p^{-2\gamma}}\,, \\
        &T^{\dot{1}1}_p  = \varphi^\F_{\z,p}\, \varphi^\B_{\z,p^{-2\gamma}}\,, \qquad &&T^{\dot{2}1}_p = -i \sigma_p \varphi^\B_{\z,p}\, \varphi^\F_{\z,p^{-2\gamma}}\,,  \\
        &T^{\dot{1}2}_p = -\sigma_p \phi^\B_{\z,p}\, \phi^\F_{\z,p^{-2\gamma}}\,, \qquad &&T^{\dot{2}2}_p = -i \phi^\F_{\z,p}\, \phi^\B_{\z,p^{-2\gamma}}\,, \\
        &\tilde{\chi}^{\dot{1}}_p = - \sigma_p \varphi^\F_{\z,p}\, \phi^\F_{\z,p^{-2\gamma}}\,, \qquad &&\tilde{\chi}^{\dot{2}}_p  = -i \varphi^\B_{\z,p}\, \phi^\B_{\z,p^{-2\gamma}}\,.
    \end{aligned}
\end{equation}
Using these identifications, we obtain the following crossing rule for the physical particles:
\begin{equation}
\label{eq:crossingrulemassless1}
    \begin{aligned}
    &\chi^{\dot{1}}_p = \phi^\B_{\z,p}\, \varphi^\B_{\z,p^{-2\gamma}}\quad 
    &\xrightarrow[]{2\gamma}\qquad& \chi^{\dot{1}}_{p^{2\gamma}} =\phi^\B_{\z,p^{2\gamma}}\, \varphi^\B_{\z,p} =
    &i \tilde{\chi}^{\dot{2}}_p\,, \\
    &T^{\dot{1}1}_p  = \varphi^\F_{\z,p}\, \varphi^\B_{\z,p^{-2\gamma}}\quad &\xrightarrow[]{2\gamma}\qquad& T^{\dot{1}1}_{p^{2\gamma}} =\varphi^\F_{\z,p^{2\gamma}}\, \varphi^\B_{\z,p} =
    &i \sigma_p T^{\dot{2}1}_p\,, \\
    &T^{\dot{1}2}_p  = -\sigma_p \phi^\B_{\z,p}\, \phi^\F_{\z,p^{-2\gamma}}\quad
    &\xrightarrow[]{2\gamma}\qquad& T^{\dot{1}2}_{p^{2\gamma}} =-\sigma_{p^{2\gamma}} \phi^\B_{\z,p^{2\gamma}}\, \phi^\F_{\z,p} =  
    &i \sigma_p T^{\dot{2}2}_p\,, \\
    &\tilde{\chi}^{\dot{1}}_p = -\sigma_p \varphi^\F_{\z,p} \phi^\F_{\z,p^{-2\gamma}}\quad 
    &\xrightarrow[]{2\gamma}\qquad &\tilde{\chi}^{\dot{1}}_{p^{2\gamma}} = -\sigma_{p^{2\gamma}} \varphi^\F_{\z,p^{2\gamma}}\, \phi^\F_{\z,p} = 
    &-i \chi^{\dot{2}}_p\,.
    \end{aligned}
\end{equation}
Similarly, we find
\begin{equation}
\label{eq:crossingrulemassless2}
    \chi^{\dot{2}}_{p^{2\gamma}}=i \tilde{\chi}^{\dot{1}}_p \,,\qquad
    T^{\dot{2}1}_{p^{2\gamma}}=i \sigma_p T^{\dot{1}1}_p\,,\qquad
    T^{\dot{2}2}_{p^{2\gamma}}=i \sigma_p T^{\dot{1}2}_p\,,\qquad
    \tilde{\chi}^{\dot{2}}_{p^{2\gamma}}=-i \chi^{\dot{1}}_p\,.
\end{equation}

\subsection{Constraining the scalar factors}
\label{sec:hprefactor}

\begin{figure}[t]
 \centering
 \includegraphics[width=0.6\textwidth]{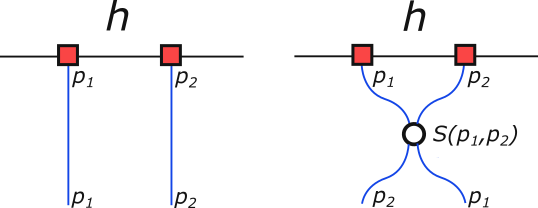}
 \caption{The Watson equation relates two expressions for the form factor, one of which involves the full S~matrix of the theory, complete with dressing factors --- in this case, $\mathbf{S}^{\AdSST}(p,q)$.}
 \label{fig:Watson}
\end{figure}
By using the diagonal $\psu(1|1)^{\oplus2}$ symmetry we have fixed the two-particle hexagon form factor, up to a scalar factor for each pair of irreducible representations. For instance, in the case of two left representations we have a prefactor $h^{\L\L}(p,q)$ which plays a role similar to that of $\Sigma^{\L\L}(p,q)$ in the S~matrix. We picked our conventions so that we may, essentially, use the S~matrix elements given in section~\ref{sec:Smatrix} up to replacing $\Sigma$ with $h$. Namely, we have the normalisation
\begin{equation}
\begin{aligned}
    &\langle \h | Y_p Y_q  \rangle = h^{\L\L}(p,q)\,,   \qquad &&\langle \h | \tilde{Y}_p \tilde{Y}_q  \rangle = h^{\R\R}(p,q) \,,\\
    &\langle \h | Y_p \tilde{Z}_q  \rangle = h^{\L\R}(p,q)\,, \qquad &&\langle \h | \tilde{Z}_p Y_q  \rangle =  h^{\R\L}(p,q) \,,
    \end{aligned}
\end{equation}
for massive particles. It is convenient to already impose left-right symmetry~\cite{Borsato:2012ud} which halves the number of independent scalar factors like in~\cite{Borsato:2013hoa}. We set
\begin{equation}
    h^{\bullet\bullet}(p,q) \equiv h^{\L\L}(p,q)= h^{\R\R}(p,q)\,,
\end{equation}
which ensures that $\langle \h | Y_p Y_q  \rangle=\langle \h | \tilde{Y}_p \tilde{Y}_q  \rangle$, and so on, and
\begin{equation}
    \tilde{h}^{\bullet\bullet}(p,q)\equiv h^{\L\R}(p,q)= e^{\frac{i}{2}(p+q)} \frac{1-x^-_{\R,p} x^-_{\L,q}}{1-x^+_{\R,p} x^+_{\L,q}} h^{\R\L}(p,q)\,,
\end{equation}
which ensures that $\langle \h | Y_p \tilde{Z}_q  \rangle=\langle \h | \tilde{Y}_p {Z}_q  \rangle$, and so on.
For massless particles, we shall assume that the prefactors are blind to the $\su(2)_\circ$ structure like it is the case for the prefactors of the S~matrix~\cite{Borsato:2014exa,Borsato:2014hja}. (Strictly speaking, this is something that would need to be verified against perturbative results.) In this case, we have a single massless dressing factor, which appears as
\begin{equation}
    \langle\h|\chi^{\dot{A}}_p\chi^{\dot{B}}_q\rangle=h^{\circ\circ}(p,q)\,.
\end{equation}
Finally, we have a pair of dressing factors related to processes that involve one massive and one massless particle, namely
\begin{equation}
    \langle\h|Y_p\chi^{\dot{A}}_q\rangle = \langle\h|\tilde{Y}_p\chi^{\dot{A}}_q\rangle =  h^{\bullet\circ}(p,q)\,,
    \qquad
    \langle\h|\chi^{\dot{A}}_pY_q\rangle = \langle\h|\chi^{\dot{A}}_p\tilde{Y}_q\rangle =  h^{\circ\bullet}(p,q)\,,
\end{equation}
where we exploited both $\su(2)_{\circ}$ and left-right symmetry. We will see below how additional physical constraints allow us to make a proposal for these five prefactors.

\subsubsection{The Watson equation}
The first physical constraint that we consider is the Watson equation which relates the original $\psu(1|1)^{\oplus4}$ S~matrix and the hexagon form factor, see figure~\ref{fig:Watson}. In formulae it says that we may swap a pair of particles in the form factor by means of the S~matrix,
\begin{equation}
    \langle \h |\Xi^{a_1\acute{a}_1}_{p_1}\cdots
    \Xi^{a_j\acute{a}_j}_{p_j}\Xi^{a_{j+1}\acute{a}_{j+1}}_{p_{j+1}}
    \cdots\rangle = 
    \langle \h |\mathbf{S}^{\text{AdS}_3\times\Sph^3\times\Tor^4}_{j,j+1}|\Xi^{a_1\prime{a}_1}_{p_1}\cdots
    \Xi^{a_j\acute{a}_j}_{p_j}\Xi^{a_{j+1}\acute{a}_{j+1}}_{p_{j+1}}
    \cdots\rangle\,,
\end{equation}
where $\mathbf{S}^{\AdSST}$ is the full $\psu(1|1)^{\oplus4}$ S~matrix of~\cite{Borsato:2014exa} (see also appendix~\ref{app:Smatrix}) complete with its dressing factors.
Clearly, owing to our factorised ansatz, it is sufficient to impose the constraint for the two-particle form factor. The constraint is a matrix equation, whose only non-trivial part is the overall normalisation --- though it is well worth to check the whole matrix equation to ensure that the form factor and the  $\psu(1|1)^{\oplus4}$ S~matrix are compatible, \textit{i.e.}\ are written in the same basis. We find the following conditions
\begin{equation}
\label{eq:watson}
\begin{gathered}
    \frac{h^{\bullet\bullet}(p,q)}{h^{\bullet\bullet}(q,p)} = \big[\Sigma^{\bullet\bullet}(p,q)\big]^2\,,
    \qquad
    \frac{\tilde{h}^{\bullet\bullet}(p,q)}{\tilde{h}^{\bullet\bullet}(q,p)} = \big[\tilde{\Sigma}^{\bullet\bullet}(p,q)\big]^2\,,\\
    \frac{h^{\circ\circ}(p,q)}{h^{\circ\circ}(q,p)} = -\big[\Sigma^{\circ\circ}(p,q)\big]^2\,,\\
    \frac{h^{\bullet\circ}_*(p,q)}{h^{\circ\bullet}_*(q,p)} = \big[\Sigma^{\bullet\circ}_*(p,q)\big]^2\,,
    \qquad
    \frac{{h}^{\circ\bullet}_*(p,q)}{{h}^{\bullet\circ}_*(q,p)} = \big[{\Sigma}^{\circ\bullet}_*(p,q)\big]^2\,,
\end{gathered}
\end{equation}
where the subscript  $*$ stands for either L or R. Here the minus sign for masless modes can be understood by recalling that, when looking at highest-weight states, we are scattering Fermions.
All in all, these condition are akin to the antisymmetry conditions that braiding unitarity imposes on~$\Sigma$.

\subsubsection{Decoupling condition and crossing}
\begin{figure}[t]
 \centering
 \includegraphics[width=0.7\textwidth]{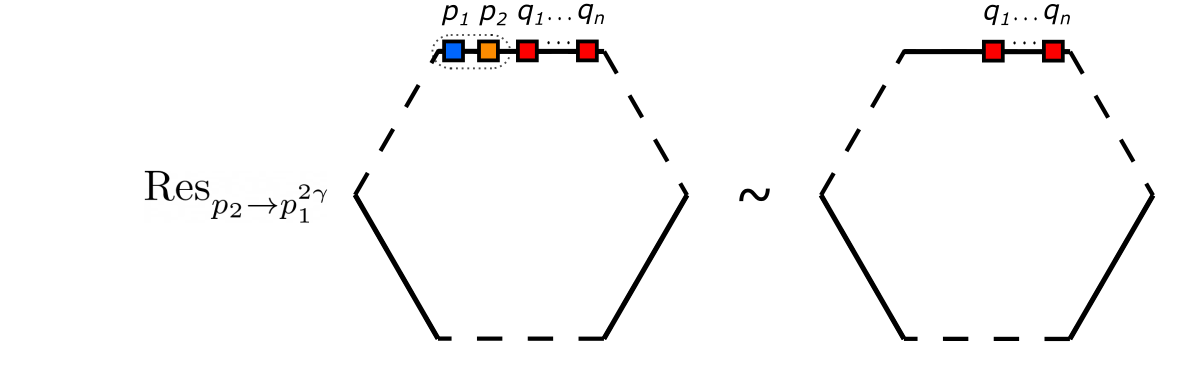}
 \caption{The decoupling condition. When the particles in the form factor feature a particle-antiparticle pair (here, those with momenta $p_1$ and $p_2$, the latter in the crossed channel), those decouple: the form factor diverges as a simple pole whose residue is the form factor of the remaining particles ($q_1,\dots,q_n$ in the figure).}
 \label{fig:Decoupling}
\end{figure}
One further condition that we may impose on the form factor is that, whenever two particles form a singlet, they decouple. 
Note that in order to have a singlet, the particles' momenta $p_1$ and $p_2$ cannot both be physical. The singlet must in particular be annihilated by the momentum and energy operators, which means that $p_1+p_2=0$ and $E(p_1)+E(p_2)=0$, \textit{i.e.}\ one of the momenta is crossed, $p_1=p_2^{\pm2\gamma}$.
Then, when the momenta satisfy the bound-state condition, the form factor has a pole whose residue is
\begin{equation}
    \text{Res}_{p_2\to p_1^{2\gamma}}\langle \h | \Xi^{a_1\acute{a}_1}_{p_1}\Xi^{a_2\acute{a}_2}_{p_2}\cdots \Xi^{a_N\acute{a}_N}_{p_N}\rangle = C^{a_1\acute{a}_1,a_2\acute{a}_2} \langle \h | \Xi^{a_3\acute{a}_3}_{p_3}\cdots \Xi^{a_N\acute{a}_N}_{p_N}\rangle\,,
\end{equation}
where $C^{a_1\acute{a}_1,a_2\acute{a}_2} $ projects onto the singlet representation. 
Crucially, $C^{a_1\acute{a}_1,a_2\acute{a}_2} $ must be independent of $p_3,\dots p_N$. Using the factorised form of the hexagon form factor~\eqref{eq:multiparticlehexagon} one can see that this boils down to the requirement that for the $\psu(1|1)^{\oplus2}$ S~matrix scattering with the singlet is inconsequential,
\begin{equation}
    \mathbf{S}_{12}\mathbf{S}_{23} \Big(C_{\acute{a}_1,\acute{a}_2}|\acute{\xi}^{\acute{a}_1}(p_1) \acute{\xi}^{\acute{a}_2}(p_1^{2\gamma}) \acute{\xi}^{\acute{a}_3}(p_3)\rangle \Big) = C_{\acute{a}_1,\acute{a}_2}\,|\acute{\xi}^{\acute{a}_3}(p_3)\acute{\xi}^{\acute{a}_1}(p_1) \acute{\xi}^{\acute{a}_2}(p_1^{2\gamma}) \rangle\,,
\end{equation}
where the scattering phase on the right-hand side is precisely equal to one. This is the crossing equation for the $\psu(1|1)^{\oplus2}$ S~matrix as derived in ref.~\cite{Borsato:2012ud}; the relationship between the $\psu(1|1)^{\oplus2}$ projector~$C_{\acute{a}_1,\acute{a}_2}$ and the $\psu(1|1)^{\oplus4}$ projector $C_{a_1\acute{a}_1,a_2\acute{a}_2}$ was discussed in ref.~\cite{Borsato:2013qpa}.
In other words, the hexagon form factor will satisfy the decoupling condition as long as the $\psu(1|1)^{\oplus2}$ S~matrix, normalised in terms of $h(p,q)$s, satisfies the crossing equation. This gives the following constraints of the hexagon prefactors in the massive sector,
\begin{equation}
\label{eq:crossingmassive}
\begin{aligned}
    &h^{\bullet\bullet}(p,q) \tilde{h}^{\bullet\bullet}(p^{2\gamma},q)= h^{\bullet\bullet}(p,q) \tilde{h}^{\bullet\bullet}(p,q^{-2\gamma}) = \left[e^{-\frac{i}{2}p} \frac{x^+_{*,p}-x^-_{*,q}}{x^-_{*,p}-x^-_{*,q}} \right]^{-1} , \\
    &h^{\bullet\bullet}(p^{2\gamma},q) \tilde{h}^{\bullet\bullet}(p,q) = h^{\bullet\bullet}(p,q^{-2\gamma}) \tilde{h}^{\bullet\bullet}(p,q)= \left[ e^{-\frac{i}{2}p} \frac{1- x^+_{*,p} x^+_{*,q}}{1- x^-_{*,p} x^+_{*,q}}\right]^{-1},
    \end{aligned}
\end{equation}
which like in the case of the $\psu(1|1)^{\oplus2}$ S~matrix has non-trivial double crossing equations. For instance, by crossing the first line by $2\gamma$ and dividing it by the second line we have
\begin{equation}
\label{eq:fourgammamonodromy}
    \frac{\tilde{h}^{\bullet\bullet}(p^{4\gamma},q)}{\tilde{h}^{\bullet\bullet}(p,q)}=\frac{1-x^+_{*,p}x^+_{*,q}}{1-x^-_{*,p}x^+_{*,q}}\frac{1-x^-_{*,p}x^-_{*,q}}{1-x^+_{*,p}x^-_{*,q}}\,,
\end{equation}
\textit{i.e.}, a non-trivial monodromy. Similarly,  $h^{\bullet\bullet}(p^{4\gamma},q)\neq h^{\bullet\bullet}(p,q)$.
In the massless sector we have
\begin{equation}
\label{eq:crossingmassless}
\begin{aligned}
    h^{\z\z}(p,q) h^{\z\z}(p^{2\gamma},q) = \left[ e^{\frac{i}{2}q} \frac{x^+_{\z,p}-x^-_{\z,q}}{x^+_{\z,p}-x^+_{\z,q}} \right]^{-1} \,,\\
    h^{\z\z}(p,q) h^{\z\z}(p,q^{-2\gamma}) = \left[ e^{-\frac{i}{2}p} \frac{x^+_{\z,p}-x^-_{\z,q}}{x^-_{\z,p}-x^-_{\z,q}} \right]^{-1}\,,
\end{aligned}
\end{equation}
and in the mixed-mass sector 
\begin{equation}
    \begin{aligned}
    h^{\bullet \circ}_{\R \z}(p^{2\gamma},q) h^{\bullet \circ}_{\L \z}(p,q) &= e^{-i\frac{q}{2}} \frac{x^+_{\L,p} - x^+_{\z,q}}{x^+_{\L,p} - x^-_{\z,q}}\,,\\
    h^{\circ \bullet}_{\z \L}(p^{2\gamma},q) h^{\circ \bullet}_{\z \L}(p,q) &= e^{-i\frac{q}{2}} \frac{x^+_{\z,p} - x^+_{\L,q}}{x^+_{\z,p} - x^-_{\L,q}}\,,\\
    h^{\bullet \circ}_{\L \z}(p^{2\gamma},q) h^{\bullet \circ}_{\R \z}(p,q) &= e^{-i\frac{q}{2}} \frac{1-x^-_{\R,p} x^+_{\z,q}}{1-x^-_{\R,p}x^-_{\z,q}}\,,\\
    h^{\circ \bullet}_{\z \R}(p^{2\gamma},q) h^{\circ \bullet}_{\z \R}(p,q) &= e^{+i\frac{q}{2}} \frac{1-x^+_{\z,p} x^-_{\R,q}}{1-x^+_{\z,p}x^+_{\R,q}}\,.
    \end{aligned}
\end{equation}

\subsubsection{Cyclic invariance}
\begin{figure}[t]
 \centering
 \includegraphics[width=0.85\textwidth]{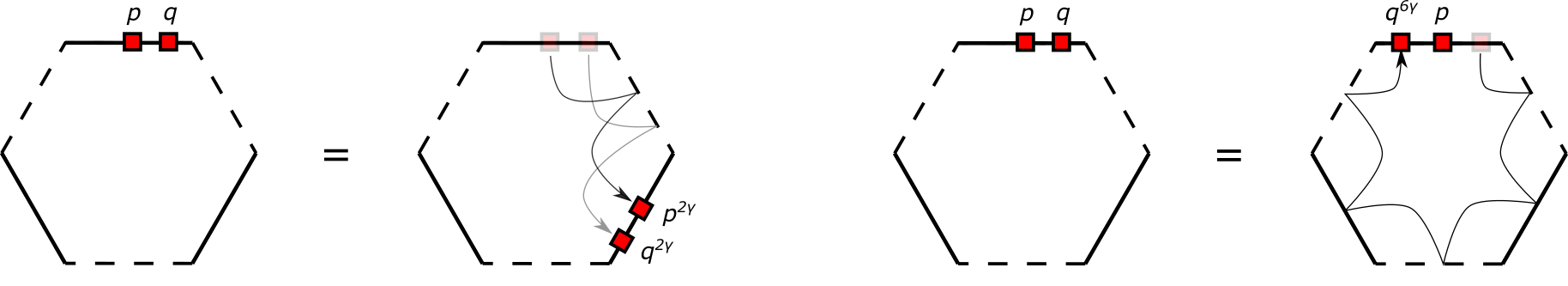}
 \caption{Consistency conditions for the hexagon construction. Left: cyclically relabeling the edges of the hexagon amount to ``move'' all excitations by $2\gamma$ in the sense of section~\ref{sec:crossinghex}. Right: we can also move one or more individual particles by $6\gamma$ in such a way to obtain a different ordering of momenta, which in turn can be simplified using the Watson equation (see above).}
 \label{fig:cyclicity}
\end{figure}
One last set of conditions can be found by using the fact that none of the three operators appearing in the three-point function should play any special role. In the hexagon formalism, it is postulated that ``moving'' an excitation from an edge to another amounts to a $2\gamma$ crossing transformation. Hence we should have that the scalar factor (as well as the matrix part of the S~matrix) remain invariant under $2\gamma$-shifting all momenta, in particular
\begin{equation}
    h^{\bullet\bullet}(p,q)= h^{\bullet\bullet}(p^{2\gamma},q^{2\gamma})\,,
\end{equation}
see figure~\ref{fig:cyclicity} left, and similarly for other sectors. This, as well as the other similar conditions, are actually a consequence of the crossing equations above, as it can be seen by using the first equality in~\eqref{eq:crossingmassive}. Similar conditions for other processes can be similarly proved, also using the monodromies of the matrix part of the $\psu(1|1)^{\oplus2}$ S~matrix~\cite{Borsato:2012ud}, \textit{cf.}~\cite{Arutyunov:2009ga}.

A more interesting condition arises if we pick a two-particle process and we cycle only the second particle all around the hexagon, see figure~\ref{fig:cyclicity} right. In the case of the massive left representation, this gives for the highest-weight states
\begin{equation}
    \langle \h | Y(p)Y(q)\rangle =
    \langle \h | \tilde{Y}(q^{6\gamma})Y(p)\rangle\,,
\end{equation}
where we used eqs.~\eqref{eq:crossingruleleft} and~\eqref{eq:crossingruleright}. Evaluating this explicitly, we get 
\begin{equation}
    h^{\bullet\bullet}(p,q)=e^{-iq/2} \frac{x^{-}_{\L}(p)-x^{+}_{\L}(q)}{x^{-}_{\L}(p)-x^{-}_{\L}(q)}\,\tilde{h}^{\bullet\bullet}(q^{6\gamma},p)\,,
\end{equation}
this formula may be further simplified by using the monodromy condition~\eqref{eq:fourgammamonodromy} and the crossing equation itself~\eqref{eq:crossingmassive}, yielding finally
\begin{equation}
    h^{\bullet\bullet}(p,q)h^{\bullet\bullet}(q,p)= \frac{x^{+}_{\L}(p)-x^{+}_{\L}(q)}{x^{-}_{\L}(p)-x^{+}_{\L}(q)}\frac{x^{-}_{\L}(p)-x^{-}_{\L}(q)}{x^{+}_{\L}(p)-x^{-}_{\L}(q)}\,.
\end{equation}
Combining this equation with the Watson equation~\eqref{eq:watson} one can fix $h^{\bullet\bullet}(p,q)$ up to an overall sign.

Similarly, $\tilde{h}^{\bullet\bullet}(p,q)$ should remain invariant under $2\gamma$-shiftings
\begin{equation}
    \tilde{h}^{\bullet\bullet}(p,q)= \tilde{h}^{\bullet\bullet}(p^{2\gamma},q^{2\gamma})\,,
\end{equation}
The highest-weight state form-factor leads to
\begin{equation}
    \langle \h | Y(p) \tilde{Z}(q)\rangle =
    \langle \h | Z(q^{6\gamma})Y(p)\rangle\,,
\end{equation}
which can be written explicitly as
\begin{equation}
    \tilde{h}^{\bullet\bullet}(p,q)=e^{ip/2} \frac{1-x^-_{\L}(p) x^-_{\R}(q)}{1-x^{+}_{\L}(p) x^{-}_{\R}(q)}\,h^{\bullet\bullet}(q^{6\gamma},p)\,,
\end{equation}
Using the crossing equations, this leads to
\begin{equation}
    \tilde{h}^{\bullet\bullet}(p,q) \tilde{h}^{\bullet\bullet}(q,p)= \frac{1-x^+_{\L}(p) x^-_{\R}(q)}{1-x^{-}_{\L}(p) x^{-}_{\R}(q)} \frac{1-x^-_{\L}(p) x^+_{\R}(q)}{1-x^{+}_{\L}(p) x^{+}_{\R}(q)}\,.
\end{equation}
This condition takes care of all cyclicity requirements on the hexagon for massive particles.

In the massless case we have
\begin{equation}
    \langle \h | \chi(p) \chi(q)\rangle =
    i \langle \h | \tilde{\chi}(q^{6\gamma})\chi(p)\rangle\,,
\end{equation}
which reads explicitly
\begin{equation}
    h^{\circ\circ}(p,q)= e^{ip/2} \frac{x^-_{\z}(p) - x^+_{\z}(q)}{x^{+}_{\z}(p) -x^{+}_{\z}(q)}\,h^{\circ\circ}(q^{6\gamma},p)\,.
\end{equation}
Using the crossing equation for $h^{\circ\circ}(p,q)$ leads to
\begin{equation}
    h^{\circ\circ}(p,q) h^{\circ\circ}(q,p)= \frac{x^+_{\z}(p) -x^+_{\z}(q)}{x^{+}_{\z}(p)- x^{-}_{\z}(q)} \frac{x^-_{\z}(p) -x^-_{\z}(q)}{x^{-}_{\z}(p) -x^{+}_{\z}(q)}\,.
\end{equation}
It is interesting to note that, in this case, upon cycling the massless Bosons, we sometimes pick up an overall minus sign. This is not surprising because, for an odd number of particles, the Hexagon has Fermion statistic. As a result, when cycling a massless Boson around a Fermionic number of objects with an overall Fermion statistic, we must account for an additional minus sign.

For the remaining dressing factors, one finds similar equations, namely
\begin{equation}
    h^{\bullet\circ}_{\L \z}(p,q) h^{\circ\bullet}_{\z \L}(q,p) = \frac{x^{+}_{\L}(p)-x^{+}_{\z}(q)}{x^{-}_{\L}(p)-x^{+}_{\z}(q)} \frac{x^{-}_{\L}(p)-x^{-}_{\z}(q)}{x^{+}_{\L}(p)-x^{-}_{\z}(q)}\,,
\end{equation}
and
\begin{equation}
    h^{\bullet\circ}_{\R \z}(p,q) h^{\circ\bullet}_{\z \R}(q,p) = \frac{1-x^{-}_{\R}(p) x^{+}_{\z}(q)}{1-x^{+}_{\R}(p) x^{+}_{\z}(q)} \frac{1-x^{+}_{\R}(p) x^{-}_{\z}(q)}{1-x^{-}_{\R}(p) x^{-}_{\z}(q)}\,.
\end{equation}
Once again, when cycling a massless Boson around a Fermion-statistic object we pick up a minus sign.

\subsubsection{Solution for the scalar factors}
We can now put together the conditions we encountered to write down a solution for the square of the various pre-factors $h^{**}(p,q)$. Taking the square root, we write
\begin{equation}
\begin{aligned}
    h^{\bullet\bullet}(p,q)&= \frac{e^{i(p+q)/2}}{\sigma^{\bullet\bullet}(p,q)}\sqrt{\frac{[x^+_{*}(p)-x^+_{*}(q)][x^-_{*}(p)-x^-_{*}(q)]}{[x^+_{*}(p)-x^-_{*}(q)]^2}}
    \sqrt{\frac{1-\frac{1}{x^{-}_{*}(p)x^{+}_{*}(p)}}
    {1-\frac{1}{x^{+}_{*}(p)x^{-}_{*}(p)}}}\,,\\
    \tilde{h}^{\bullet\bullet}(p,q)&= \frac{e^{-i \frac{q}{2}}}{\tilde{\sigma}^{\bullet \bullet}(p,q)} \frac{1-\frac{1}{x^-_{*}(p)x^+_{*}(q)}}{1-\frac{1}{x^+_{*}(p)x^+_{*}(q)}}\,,
\end{aligned}
\end{equation}
where we have chosen the branches so that in the BMN limit (schematically, $k=0$, $p\sim p/h$, and $h\to\infty$) the prefactor reduces to plus one. The non-trivial monodromy of the prefactor is due to~$\sigma^{\bullet\bullet}(p,q)$, which is known for pure-Ramond-Ramond backgrounds and was given in ref.~\cite{Borsato:2013hoa}.

Similarly, in the massless sector we find
\begin{equation}
    h^{\circ\circ}(p,q)= \frac{e^{i(p-q)/4}}{\sigma^{\circ\circ}(p,q)}\sqrt{\frac{[x^+_{\z}(p)-x^+_{\z}(q)][x^-_{\z}(p)-x^-_{\z}(q)]}{[x^+_{\z}(p)-x^-_{\z}(q)]^2}}\,.
\end{equation}

In the mixed-mass sector we find the prefactors
\begin{equation}
    \begin{aligned}
        h^{\bullet \circ}_{\L \z}(p,q) &= e^{+i \frac{p}{4}} \sqrt{ \frac{x^{-}_{\L}(p)-x^+_{\z}(q)}{x^{-}_{\L}(p)-x^-_{\z}(q)} \zeta(p,q)} \frac{1}{\sigma^{\bullet \circ}_{\L \z}(p,q)}\,,\\
        h^{\circ \bullet}_{\z \L}(p,q) &= e^{-i \frac{q}{4}} \sqrt{\frac{x^{+}_{\z}(p)-x^+_{\L}(q)}{x^{+}_{\z}(p)-x^-_{\L}(q)} \zeta(p,q)} \frac{1}{\sigma^{\circ \bullet}_{\z \L}(p,q)}\,,\\
        h^{\bullet \circ}_{\R \z}(p,q) &= e^{-i(\frac{p}{4}+\frac{q}{2})} \sqrt{  \frac{[1-x^-_{\R}(p)x^+_{\z}(q)][1-x^+_{\R}(p)x^+_{\z}(q)]}{[1-x^-_{\R}(p) x^-_{\z}(q)]^2} \tilde{\zeta}(p,q) } \frac{1}{\sigma^{\bullet \circ}_{\R \z}(p,q)}\,,\\
        h^{\circ \bullet}_{\z \R}(p,q) &= e^{+i(\frac{p}{2}+\frac{q}{4})} \sqrt{ \frac{[1-x^-_{\z}(p) x^+_{\R}(q)][1-x^-_{\z}(p) x^-_{\R}(q)]}{[1-x^+_{\z}(p) x^+_{\R}(q)]^2} \tilde{\zeta}(p,q)} \frac{1}{\sigma^{\circ \bullet}_{\z \R}(p,q)}\,,
    \end{aligned}
\end{equation}
where we introduced the functions
\begin{equation}
\begin{aligned}
    \zeta(p,q) &= \sqrt{\frac{x^-_{*,p}-x^-_{*,q}}{x^+_{*,p}-x^-_{*,q}} \frac{x^+_{*,p}-x^+_{*,q}}{x^-_{*,p}-x^+_{*,q}}} \,,\\ \tilde{\zeta}(p,q)&= \sqrt{\frac{1-x^+_{*,p} x^+_{*,q}}{1-x^+_{*,p} x^-_{*,q}} \frac{1-x^-_{*,p} x^-_{*,q}}{1-x^-_{*,p} x^+_{*,q}}}.
    \end{aligned}
\end{equation}

\section{Some protected three-point functions}
\label{sec:check}
As a test of our proposal it is good to compute some explicit result and compare it with the literature. For generic (non-protected) operators, we would only be able to carry out such a comparison in the case of pure-NS-NS backgrounds. Things are much simpler if we consider three-point functions of protected (half-BPS) operators, which are themselves protected by supersymmetry~\cite{Baggio:2012rr}.

\subsection{Definition of the correlation functions}
Let us start by reviewing the structure of the half-BPS states, which is substantially richer here than for $\AdS_5\times\Sph^5$ (because overall here we have less supersymmetry).

\subsubsection{Structure of protected operators}
While in $\AdS_5\times\Sph^5$ we have exactly one BPS operator for each value of the ``orbital'' R-charge $J$  (with energy $J$ owing to the BPS bound) this is not the case here. First of all, we have two $\su(2)$ (left and right) orbital quantum numbers, which we indicate by $(J,\tilde{J})$. These are the eigenvalues of the highest-weight state in the BPS representation under $(\Jl_3,\Jr_3)$, respectively. Recall that the $\psu(1,1|2)^{\oplus2}$ BPS bound gives $-\Ll_0=\Jl_3$ and $-\Lr_0=\Jr_3$. Then, for every positive integer value of $j$ we have the following diamond of BPS multiplets, indicated here by the charge of their highest-weight states:
\begin{equation}
\begin{gathered}
\big(\tfrac{j-1}{2},\tfrac{j-1}{2}\big)\\
\big(\tfrac{j}{2},\tfrac{j-1}{2}\big)^{\dot{A}}\qquad \big(\tfrac{j-1}{2},\tfrac{j}{2}\big)^{\dot{A}}\\
\big(\tfrac{j+1}{2},\tfrac{j-1}{2}\big)\qquad \big(\tfrac{j}{2},\tfrac{j}{2}\big)^{\dot{A}\dot{B}}\qquad\big(\tfrac{j-1}{2},\tfrac{j+1}{2}\big)\\
\big(\tfrac{j+1}{2},\tfrac{j}{2}\big)^{\dot{A}}\qquad \big(\tfrac{j}{2},\tfrac{j+1}{2}\big)^{\dot{A}}\\
\big(\tfrac{j+1}{2},\tfrac{j+1}{2}\big)
\end{gathered}
\end{equation}
for a total of 16 multiplets. The dotted indices indicate that some of these states transform in the $\mathbf{2}$ or $\mathbf{2}\otimes\mathbf{2}$ representation $\su(2)_{\circ}$.
This structure can be related to the Hodge diamond of $\Tor^4$ or to a Clifford module generated by four Fermion zero-modes.
In particular, looking at the dual CFT, these multiplets may be identified with those arising from the symmetric-product orbifold CFT of $\Tor^4$, $\text{Sym}^N \Tor^4$. Using the notation of ref.~\cite{Pakman:2007hn} (which will be convenient for what follows), the diamond looks like this:
\begin{equation}
\label{eq:hodge-CFT}
\begin{gathered}
    \mathbb{V}^{--}_{j}\\
    \mathbb{V}^{\dot{A}-}_{j}\qquad
    \mathbb{V}^{-\dot{A}}_{j}\\
    \mathbb{V}^{+-}_{j}\qquad
    \mathbb{V}^{\dot{A}\dot{B}}_{j}\qquad
    \mathbb{V}^{-+}_{j}\\
    \mathbb{V}^{+\dot{A}}_{j}\qquad
    \mathbb{V}^{\dot{A}+}_{j}\\
    \mathbb{V}^{++}_{j}
\end{gathered}
\end{equation}
where the subscript index $j$ in $\mathbb{V}^{**}_j$ denotes the length of the permutation cycle of the operator.

\renewcommand{\arraystretch}{1.5}
\begin{table}[t]
\begin{center}
\begin{tabular}{c|c c}
Magnon& $\qquad \Jl^3\qquad$&$\qquad \Jr^3\qquad$\\
\hline
$\lim_{p\to 0^+}|\chi^1(p)\rangle$& $-\frac{1}{2}$&0\\
$\lim_{p\to 0^-}|\chi^2(p)\rangle$& 0&$+\frac{1}{2}$\\
$\lim_{p\to 0^+}|\tilde{\chi}^1(p)\rangle$& $+\frac{1}{2}$&0\\
$\lim_{p\to 0^-}|\tilde{\chi}^2(p)\rangle$& 0&$-\frac{1}{2}$\\
\end{tabular}
\caption{The $\su(2)_{\L}\oplus\su(2)_{\R}$ charge of massless particles.
The Fermions $|\chi^{\dot{A}}(p)\rangle$ have $\su(2)$ spin under  $(\Jl^3-\Jr^3)$ equal to $-1/2$, while $|\tilde{\chi}^{\dot{A}}(p)\rangle$ have $+1/2$. Given that they all have $M=0$, the $\su(1,1)$ spin $(-\Ll_0+\Lr_0)$ follows, cf.~\eqref{eq:centralcharges}. This is also consistent with the fact that $|\tilde{\chi}^{\dot{A}}(p)\rangle=\Ql^1\Ql^2|\chi^{\dot{A}}(p)\rangle$. However, the particles are chiral depending on the sign of $\sin(p/2)$, \textit{cf.}~\eqref{eq:dispersion}. Hence, in different momentum regions they will be annihilated by either $-\Ll_0$ and $\Jl^3$ or by $-\Lr_0$ and $\Jr^3$. Keeping that into account, we propose the following identification of the massless modes.
}\label{tab:zeromodes}
\end{center}
\end{table}
\renewcommand{\arraystretch}{1}

In the language of integrability that we have so far used,  one state is the BMN vacuum $|0\rangle$, featuring no particles at all, while the remaining can be constructed by inserting on top of the vacuum the massless Fermions $\chi^{\dot{A}}(p)$ and $\tilde{\chi}^{\dot{A}}(p)$ at zero momentum \cite{Baggio:2017kza}.
The zero-modes which we can use have charges under $(\Jl_3,\Jr_3)$ as in table~\ref{tab:zeromodes}~\cite{Dei:2018mfl} and, owing to Pauli's principle, yield precisely 16 states. Note that, unlike the zero-modes of \textit{massive} states, the zero-modes of $\chi^{\dot{A}}(p)$ and $\tilde{\chi}^{\dot{A}}(p)$ do not yield $\psu(1,1|2)^{\oplus2}$ descendants, but genuinely new $\psu(1,1|2)^{\oplus2}$ multiplets.
Based on table~\ref{tab:zeromodes}, the highest-weight states can be identified as it follows:
\begin{equation}
\label{eq:hodge-integrable}
    \begin{gathered}
        |\chi^1\tilde{\chi}^2\rangle\\
        \big(|\chi^1\tilde{\chi}^1\tilde{\chi}^2\rangle,|\tilde{\chi}^2\rangle\big) \qquad
        \big(|{\chi}^1\rangle,|\chi^1{\chi}^2\tilde{\chi}^2\rangle\big)\\
        |\tilde{\chi}^1\tilde{\chi}^2\rangle\qquad |0\rangle\oplus\big(|\chi^1\tilde{\chi}^1\rangle ,|\chi^1\tilde{\chi}^1\chi^2\tilde{\chi}^2\rangle,|\chi^2\tilde{\chi}^2\rangle\big)\qquad |{\chi}^1{\chi}^2\rangle\\
        \big(|\tilde{\chi}^1\rangle, |{\chi}^2\tilde{\chi}^1\tilde{\chi}^2\rangle\big)
        \qquad \big(|{\chi}^1{\chi}^2\tilde{\chi}^1\rangle,|{\chi}^2\rangle\big)\\
        |\chi^2\tilde{\chi}^1\rangle
    \end{gathered}
\end{equation}
where in the middle of the Hodge diamond we have distinguished the $\mathbf{0}$ and $\mathbf{3}$ representation of $\su(2)_{\circ}$. It should be emphasised that the number of magnons (the ``length'' of the operators) is not a quantum number here, and it is not preserved by the $\su(2)_{\circ}$ action.
All various magnons are at zero momentum as in table~\ref{tab:zeromodes}.
Despite the nice structure, we should be careful with identifying multiplets from~\eqref{eq:hodge-CFT} to~\eqref{eq:hodge-integrable}.
From the above we see that most of the half-BPS multiplets can mix among themselves when going from the integrability description. In fact, there are several multiplets with the exact same charge. For instance, the states $\mathbb{V}^{--}_{j+1}$, $\varepsilon_{\dot{A}\dot{B}}\mathbb{V}^{\dot{A}\dot{B}}_{j}$ and $\mathbb{V}^{++}_{j-1}$ have the same charges and therefore, we cannot distinguish the relative multiples just by their quantum numbers. All of them could in principle mix with $|0\rangle, |\chi^2\tilde{\chi}^1\rangle$ and $|\chi^1\tilde{\chi}^2\rangle$.
Fortunately, the multiplets $\mathbb{V}^{+-}_j$ and $\mathbb{V}^{-+}_j$ do not mix --- neither among themselves nor with any other half-BPS multiplet --- so that it is quite convenient to focus on them.

\subsubsection{The correlation functions}

We focus on the three-point functions that may be constructed out of operators in the multiplets of $\mathbb{V}^{+-}_j$ and $\mathbb{V}^{-+}_j$ for appropriate values of $j$. Broadly, speaking, they fall in two categories: three-point functions involving all operators from the same type of multiplets, and those involving three-point functions with two operators from one type of multiplet and the third from the other --- the other combinations follow from exchanging the left and right algebra.
These correlation functions are well-known in the literature~\cite{Gaberdiel:2007vu,Dabholkar:2007ey,Pakman:2007hn}. We take a look at the result as written in ref.~\cite{Pakman:2007hn}, which has the advantage of being presented quite compactly and including the Clebsch-Gordan coefficients. 
Recall from section~\ref{sec:hexagonalgebra} that in our construction we want one of the operators to be the highest-weight state, one to be the lowest-weight state, and one to have zero magnetic $\su(2)$ quantum number --- \textit{i.e.}, to be the $\su(2)$ descendant ``in the middle'' of the multiplet.

The result of~\cite{Pakman:2007hn} reads, in particular
\begin{equation}
\label{eq:pakmanseverresult}
\begin{aligned}
    \langle\mathbb{V}^{-+}_{j_1}\mathbb{V}^{-+}_{j_2}\mathbb{V}^{-+}_{j_3}\rangle =-\frac{1}{4\sqrt{N}} D_{J_1J_2J_3}D_{\tilde{J}_1\tilde{J}_2\tilde{J}_3}\frac{(j_1+j_2+j_3-1)(j_1+j_2+j_3+1)}{\sqrt{j_1j_2j_3}}\,,\\
    \langle\mathbb{V}^{-+}_{j_1}\mathbb{V}^{-+}_{j_2}\mathbb{V}^{+-}_{j_3}\rangle =-\frac{1}{4\sqrt{N}} D_{J_1J_2J_3}D_{\tilde{J}_1\tilde{J}_2\tilde{J}_3}\frac{(j_1+j_2-j_3-1)(j_1+j_2-j_3+1)}{\sqrt{j_1j_2j_3}}\,,
\end{aligned}
\end{equation}
where in the first line $J_k=j_k+1$ and $\bar{J}_k=j_k-1$, and in the second line the same holds except for operator~3, for which instead $J_3=j_3-1$ and $\bar{J}_3=j_3+1$.
The factors $D_{J_1J_2J_3}$ and $D_{\tilde{J}_1\tilde{J}_2\tilde{J}_3}$ depend also on the magnetic $\su(2)$ charges, \textit{i.e.} on the $\Jl^3$ and $\Jr^3$ charges of the operators, respectively. Recall that operator~$1$ is a highest-weight state, operator~$3$ is a lowest weight state, and operator~$2$ has vanishing orbital quantum numbers. All in all, for our configuration of states are simply given by
\begin{equation}
    D_{J_1J_2J_3} = (-1)^{J_2+2J_3}\frac{J_2!}{\sqrt{(2J_2)!}}\,.
\end{equation}
The prefactor $1/\sqrt{N}$ is an overall normalisation common to all three-point functions --- $N$ is the number of copies in the symmetric product orbifold CFT $\text{Sym}^N(\Tor^4)$.
Note that in practice, for the three-point function to be non-vanishing, we want to specialise~\eqref{eq:pakmanseverresult} to the case $J_3=J_1$, $\tilde{J}_3=\tilde{J}_1$.

\subsection{Hexagon computation}
\label{sec:hexagoncomputation}
We will describe here how to use the formalism which we developed in order to reproduce the result~\eqref{eq:pakmanseverresult}. It should be stressed that the integrability machinery is suitable to compute non-protected correlation functions --- this is just intended as a relatively simple check of our proposal.


\begin{figure}[t]
 \centering
 \includegraphics[width=1\textwidth]{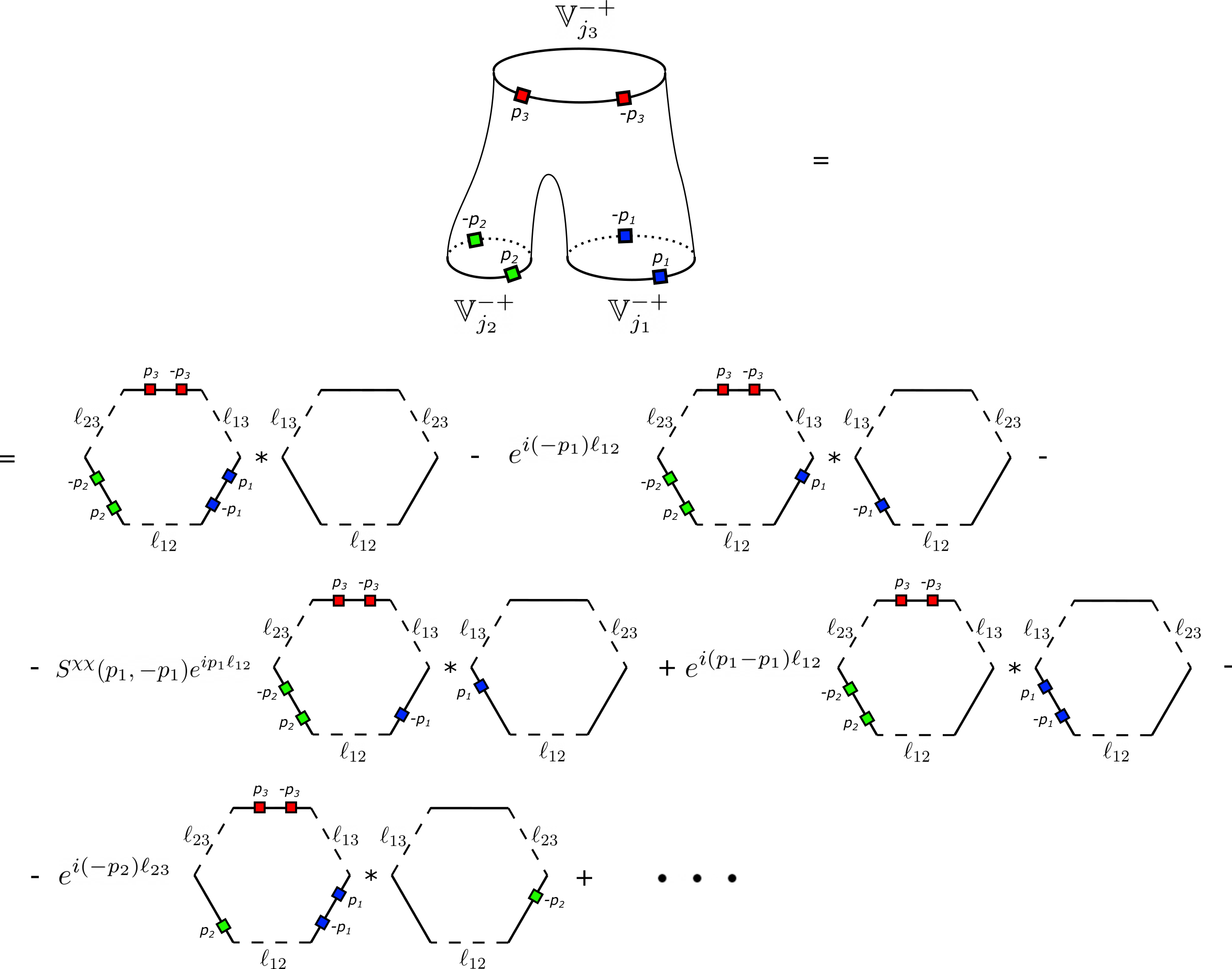}
 \caption{We represent schematically some of the terms contributing to the hexagon computation of the three-point function. We cut the three-point functions in two hexagons, one corresponding to the ``front'' of it and one to the ``back'' (the cut runs parallel to the surface of the page). Then, we have to sum over all possible ways of distributing each pair of particles over the two patches, for a total of $(2^{2})^{3}=64$ possibilities; in the figure we only write the first $2^2=4$ terms relative to moving around~$\{p_1,-p_1\}$ (in blue), and one term relative to moving~$\{p_2,-p_2\}$ (in green). The various terms have to be weighted as in eq.~\eqref{eq:hexweights}.}
 \label{fig:3ptHex}
\end{figure}
The operators of interest are those related to $\mathbb{V}^{-+}_{j}$ and $\mathbb{V}^{+-}_{j}$, namely
\begin{equation}
    \mathbb{V}^{-+}_{j} \sim \lim_{p\to0^+}\lim_{q\to0^-} |\tilde{\chi}^1(p)\tilde{\chi}^2(q)\rangle\,,
    \qquad
    \mathbb{V}^{-+}_{j} \sim \lim_{p\to0^+}\lim_{q\to0^-} |{\chi}^1(p){\chi}^2(q)\rangle\,,
\end{equation}
constructed over a vacuum of total R-charge~$j$. The expression above stresses that the zero-momentum magnons described above should be treated with some care --- we will see that indeed singularities may arise from the $p,q\to0$ limit. This is not surprising, given that among other things the dispersion relation is singular at that point, see eq.~\eqref{eq:dispersion}. It turns out that things may be simplified a little, namely we can take the limit on the two momenta symmetrically,
\begin{equation}
    \mathbb{V}^{+-}_{j} \sim \lim_{p\to0^+} |\tilde{\chi}^1(p)\tilde{\chi}^2(-p)\rangle\,,
    \qquad
    \mathbb{V}^{-+}_{j} \sim \lim_{p\to0^+} |{\chi}^1(p){\chi}^2(-p)\rangle\,,
\end{equation}

We are interested in inserting three such operators on the three distinguished edges of the hexagon, which we have labeled with $0\gamma$, $2\gamma$ and $4\gamma$. Hence we have to consider the following excitations
\begin{equation}
\label{eq:symmetriccase-particles}
    \langle\mathbb{V}^{-+}_{j_1}\mathbb{V}^{-+}_{j_2}\mathbb{V}^{-+}_{j_3}\rangle \sim \Big(\{\chi^1(p_1),\chi^2(-p_1)\},\, \{\chi^1(p_2),\chi^2(-p_2)\},\, \{\chi^1(p_3),\chi^2(-p_3)\}\Big)\,.
\end{equation}
Here and from now on, we leave the $p_j$s generic. We will see later how to take the limit. Similarly, we have
\begin{equation}
\label{eq:nonsymmetriccase-particles}
    \langle\mathbb{V}^{-+}_{j_1}\mathbb{V}^{-+}_{j_2}\mathbb{V}^{+-}_{j_3}\rangle \sim
    \Big(\{\chi^1(p_1),\chi^2(-p_1)\},\, \{\chi^1(p_2),\chi^2(-p_2)\},\, \{\tilde{\chi}^1(p_3),\tilde{\chi}^2(-p_3)\}\Big)\,.
\end{equation}

The hexagon prescription~\cite{Basso:2015zoa} requires us to partition the three sets of excitations identified above in all possible ways over the two hexagonal patches of worldsheet, see figure~\ref{fig:3ptHex}. Let us consider the case of~\eqref{eq:symmetriccase-particles}. Then we have three sets $X_1=\{{\chi}^1(p_1),{\chi}^2(-p_1)\}$, $X_2=\{\tilde{\chi}^2(p_2),\tilde{\chi}^1(-p_2)\}$ and $X_3=\{{\chi}^1(p_3),{\chi}^2(-p_3)\}$. Accordingly, we sum over all partitions of the form $X_j=\alpha_j\cup\bar{\alpha}_j$ obtaining
\begin{equation}
\label{eq:hexweights}
    \Big(\prod_{j=1}^3\sum_{X_j=\alpha_j\cup\bar{\alpha}_j} (-1)^{\bar{\alpha_j}} w_{\alpha_j,\bar{\alpha_j}} \Big) \langle \mathbf{h}|\alpha_{1}^{4\gamma}\alpha_{2}^{2\gamma}\alpha_{3}^{0\gamma}\rangle\,
    \langle \mathbf{h}|\bar{\alpha}_{2}^{4\gamma}\bar{\alpha}_{1}^{2\gamma}\bar{\alpha}_{3}^{0\gamma}\rangle\,,
\end{equation}
where the different ordering of the partitions is due to the different orientation of the two hexagonal patches (which is necessary to glue them back to give a three-point function). Accordingly, we have also indicated how the particles have to be analytically continued on the various edges. It is worth emphasising that, following the rules of section~\ref{sec:crossinghex}, a $2\gamma$-shift results in a flavour change, \textit{e.g.}~$\chi^1(p^{2\gamma}) =i\tilde{\chi}^{2}(p)$.
Finally, the sum is weighted by the factor $w_{\alpha,\bar{\alpha}}$, which takes the form 
\begin{equation}
w_{\alpha_1,\bar{\alpha}_1}=
 \begin{cases}
    1 & \alpha=\{{\chi}^1(p_1){\chi}^2(-p_1)\},\qquad \bar{\alpha}=\emptyset\\
    e^{i (-p_1)\ell_{12}} & \alpha=\{{\chi}^1(p_1)\},\qquad \bar{\alpha}=\{{\chi}^2(-p_1)\}\\
    S^{\chi\chi}(p_1,-p_1)e^{i (p_1)\ell_{12}} & \alpha=\{{\chi}^2(-p_1)\},\qquad \bar{\alpha}=\{{\chi}^1(p_1)\}\\
    e^{i (p_1-p_1)\ell_{12}}=1 & \alpha=\emptyset,\qquad \bar{\alpha}=\{{\chi}^1(p_1){\chi}^2(-p_1)\}
\end{cases}
\end{equation}
The expression further simplifies in the small-$p$ limit because $S^{\chi\chi}(p,-p)\to1$. Here we have introduced the ``bridge length''~\cite{Basso:2015zoa} $\ell_{23}$; we have
\begin{equation}
    \ell_{12}=\frac{j_1+j_2-j_3}{2}\,,\qquad
    \ell_{23}=\frac{j_2+j_3-j_1}{2}\,,\qquad
    \ell_{13}=\frac{j_1+j_3-j_2}{2}\,.
\end{equation}
Similar formulae hold for the weight factors for the other partitions, up to cycling the indices $1,2$ and~$3$. Furthermore, it is also true that $S^{\tilde{\chi}\tilde{\chi}}(p,-p)\to1$. Finally, it should be noted that there is some confusion in the literature concerning the signs which should be assigned to a given partition, especially when the permuted particles are Fermionic~\cite{Basso:2015zoa,Caetano:2016keh}. In our case we will impose that the signs satisfy all relevant self-consistency and symmetry conditions, at which point we will be able to obtain the result and match the existing literature. 

\subsubsection{Limit procedure}
As we have mentioned, the limit $p_j\to0$ will require some care. We can expect two types of singular behaviour: one arises because of possible singularities at $p=0$, while the other is due to a pair of momenta approaching each other, $p_j=p_k$. Recall from the discussion of crossing (section~\ref{sec:crossinghex}) that a particle-antiparticle  pair results in a pole; this is what will happen when, \textit{e.g.}, $p_2\to p_1$ in our setup.
There is one further complication that we should bear in mind: the hexagon formalism should not depend on the details of how we construct the external states --- for instance, it should not depend on the ordering of the particles within each state. This is indeed the case, but only as long as the particles in each state satisfy the Bethe equations. In other words, in order to have a consistent formalism we need to require $p_1, p_2$ and $p_3$ to obey the Bethe equations. These are very simple in our setup, because we are interested in a limit where particles behave as free, \textit{i.e.}~$S^{\chi\chi}(p,-p)=S^{\tilde{\chi}\tilde{\chi}}(p,-p)=1$. Still, they do impose three non-trivial conditions,
\begin{equation}
\label{eq:freebethe}
    e^{i p_k j_k}=1\qquad \Rightarrow\qquad
    p_k = \frac{2\pi \nu_k}{j_k}\,,\quad \nu_k\in\mathbb{Z}\,,\qquad k=1,2,3\,.
\end{equation}
This discrete structure calls for a little more care in taking the coincident-momenta limit. To this end we introduce a small $\varepsilon>0$ and three real numbers $\epsilon_k$ and redefine
\begin{equation}
    j_k\to \frac{j_k}{1+\varepsilon\,\epsilon_k}\,,\qquad
    p_k\to p_k (1+\varepsilon\,\epsilon_k)\,,
\end{equation}
which leaves the Bethe equations~\eqref{eq:freebethe} unchanged. In this language, we can get the coincident-momenta limit by setting
\begin{equation}
    p_1=p_2=p_3\,,\qquad\varepsilon\to0\,.
\end{equation}
This will also provide us with a check of our construction: the limit should be independent from $\epsilon_1,\epsilon_2$ and $\epsilon_3$.

In practice, in our computation it will be useful to consider one additional limit. The structure constants for the three-point correlation functions of protected operators are themselves protected~\cite{Baggio:2015vxa}. As a result, we may choose any value of $h,k$ that we want. From eq.~\eqref{eq:xpm} we note that, for massless particles, kinematics only depends on the ratio $h/k$ (up to an overall factor of $k$ which washes out of all S-matrix elements). Hence it is convenient to take the limit $h/k\to0$ with $k$ arbitrary. The upshot is that, in this way, we may rewrite all the ingredients necessary for the computation in terms of the new variables
\begin{equation}
\label{eq:yvariables}
y^\pm(p) \equiv e^{\pm i \frac{p}{2}} \, \frac{\sin\bigl(\frac{p}{2}\bigr)}{\frac{p}{2}}\,,
\end{equation}
which play the role of~$x^{\pm}_{\z}$. In terms of these, we can easily rewrite the various S-matrix elements necessary for the hexagon computation, including the relevant scalar factor. For instance, we have 
\begin{equation}
\begin{aligned}
h^{\circ\circ}(p_2^{2 \gamma}, p_3^{0 \gamma})\, A(p_2^{2 \gamma}, p_3^{0 \gamma}) & \quad\to\quad \tilde{h}_{23} e^{\frac{i}{4}(p_2 + p_3)} \, (y_3^- - y_2^-) \, ,  \\
h^{\circ\circ}(p_2^{2 \gamma}, p_3^{0 \gamma})\, B(p_2^{2 \gamma}, p_3^{0 \gamma}) & \quad\to\quad \tilde{h}_{23} e^{\frac{i}{4}(p_3 - p_2)} \, (y_3^- - y_2^+) \, , \\
h^{\circ\circ}(p_2^{2 \gamma}, p_3^{0 \gamma}) \,C(p_2^{2 \gamma}, p_3^{0 \gamma}) & \quad\to\quad \tilde{h}_{23} \tilde{\gamma}_2 \, \tilde{\gamma}_3 \, , \\
h^{\circ\circ}(p_2^{2 \gamma}, p_3^{0 \gamma})\, D(p_2^{2 \gamma}, p_3^{0 \gamma}) & \quad\to\quad \tilde{h}_{23} e^{\frac{i}{4}(p_2 - p_3)} \, (y_3^+ - y_2^-) \, , \\
h^{\circ\circ}(p_2^{2 \gamma}, p_3^{0 \gamma})\, E(p_2^{2 \gamma}, p_3^{0 \gamma}) &\quad\to\quad \tilde{h}_{23} \tilde{\gamma}_2 \, \tilde{\gamma}_3  \, , \\
h^{\circ\circ}(p_2^{2 \gamma}, p_3^{0 \gamma})\, F(p_2^{2 \gamma}, p_3^{0 \gamma}) & \quad\to\quad \tilde{h}_{23}\, e^{-\frac{i}{4}(p_2 + p_3)} \, (y_2^+ - y_3^+)  \, ,
\end{aligned}
\end{equation}
when both momenta have the same sign. Here 
\begin{equation}
\tilde \gamma_j \, := \, \sqrt{i(y^-_j-y^+_j)} \, , \qquad \tilde h_{23} \, = \, \frac{\text{sgn}(p_2-p_3)}{\sqrt{(y^-_2-y^-_3)(y^+_2-y^+_3)}}\,.
\end{equation}
The expressions become even simpler when momenta have opposite signs: in that case the reflection part of the S~matrix vanishes ($C=E=0$) and one is left with a free S~matrix, up to frame factors --- exactly how it was argued in refs.~\cite{Baggio:2018rpv,Dei:2018mfl}.

\subsubsection{Computation of the form factor}

We now turn to the computation of the hexagon form factors for the two correlators of interest~(\ref{eq:symmetriccase-particles}--\ref{eq:nonsymmetriccase-particles}). For the correlator involving three identical states, we can expect a singularity when any pair of momenta become singular. Hence the most singular part of~\eqref{eq:symmetriccase-particles} should go like
\begin{equation}
\label{eq:divergence-symmetric}
    \frac{1}{\varepsilon^6}\frac{1}{(\epsilon_1-\epsilon_2)^2(\epsilon_2-\epsilon_3)^2(\epsilon_1-\epsilon_3)^2}\,.
\end{equation}
Conversely, for the correlator of eq.~\eqref{eq:nonsymmetriccase-particles} we expect a pole from the decoupling condition only for operators one and two --- the third operators being different --- so that it will go like
\begin{equation}
\label{eq:divergence-nonsymmetric}
    \frac{1}{\varepsilon^2}\frac{1}{(\epsilon_1-\epsilon_2)^2}\,.
\end{equation}
A first obvious issue to address is how to resolve this mismatch, given that both correlators should eventually give a finite result, possibly up to an overall factor.
Let us start from the completely symmetric case of eq.~\eqref{eq:symmetriccase-particles}. Among all various ways of partitioning the particles, the one yielding the highest $O(\varepsilon^{-6})$ singularity occurs when the three particles with positive momenta $\{p_1,p_2,p_3\}$ sit on one hexagon, $\{-p_1,-p_2,-p_3\}$ sit on the other, or when they all sit on the same hexagon. In the former case, when \textit{e.g.}\ $\{p_1,p_2,p_3\}$ are on the ``front'' hexagon we pick up a numerator proportional to the following polynomial~$\mathcal{P}$ in~$y_k^\pm$:
\begin{equation}
\begin{aligned}
\mathcal{P} =& +  y^-_1 y^-_2 y^+_1 - y^-_2 y^-_3 y^+_1 - y^-_1 y^-_3 y^+_2 + 
 y^-_2 y^-_3 y^+_2 - y^-_2 y^+_1 y^+_2 + y^-_3 y^+_1 y^+_2  \\
 &-  y^-_1 y^-_2 y^+_3 + y^-_1 y^-_3 y^+_3 - y^-_1 y^+_1 y^+_3 + 
 y^-_2 y^+_1 y^+_3 + y^-_1 y^+_2 y^+_3 - y^-_3 y^+_2 y^+_3\,.
\end{aligned}
\end{equation}
Conversely, when $\{p_1,p_2,p_3\}$ are on the ``back'' hexagon we pick the complex conjugate~$\mathcal{P}^*$, which is obtained from $\mathcal{P}$ by swapping $y^{\pm}_k\leftrightarrow y^{\mp}_k$. Repeating the computation for $\{-p_1,-p_2,-p_3\}$ we come to the conclusion that the full result is proportional to $\mathcal{P}\mathcal{P}^*$; similarly, when all particles are on the same hexagon we get $\mathcal{P}^2$ or $(\mathcal{P}^*)^2$.
It is useful to introduce the quantity
\begin{equation}
    \Delta^\mp_{ij} \equiv y^\mp_i - y^\mp_j\,,
\end{equation}
in terms of which we can encode the $y^\pm_k$ dependence in all but one variable, say $y^\pm_1$. Then we have
\begin{equation}
\begin{aligned}
\mathcal{P}  =\,& (\Delta^-_{12} \Delta^+_{12} -  \Delta^-_{12} \Delta^+_{13} + \Delta^-_{13} \Delta^+_{13}) \, y^-_1  - (\Delta^-_{12} \Delta^+_{12} - \Delta^-_{12} \Delta^+_{13} + \Delta^-_{13} \Delta^+_{13}) \, y^+_1\\
 &-\Delta^-_{12} \Delta^-_{13} \Delta^+_{12} + \Delta^-_{13} \Delta^+_{12} \Delta^+_{13}\,,
\end{aligned}
\end{equation}
and similarly for~$\mathcal{P}^*$. We see that in the coincidence limit when $p_2\to p_1$ and $p_3\to p_1$, the numerator goes like $O(\varepsilon^4)$. In conclusion, the term which na\"ively would be the most divergent~\eqref{eq:divergence-symmetric} eventually goes like~$O(\varepsilon^{-2})$, exactly like eq.~\eqref{eq:divergence-nonsymmetric}. By way of example, if e.g. the particle with momentum $+p_1$ is moved from the partition $\{p_1,p_2,p_3\}, \, \{-p_1,-p_2,-p_3\}$ into the other hexagon such as to obtain $\{p_2,p_3\}, \, \{p_1,-p_1,-p_2,-p_3\}$ the back hexagon runs up only a simple pole, while the front one still is maximally singular. On the other hand, we lose $\mathcal{P}$ from the back hexagon, while $\mathcal{P}^*$ will still arise on the front one. Again, the result has only a second order pole.

Carrying out the computation to the end, we find that in the coincident-momenta limit the result for the symmetric correlator ~\eqref{eq:symmetriccase-particles} is proportional to
\begin{equation}
\begin{aligned}
&\big(j_1+j_2+j_3\big)^2  \frac{(y^--y^+)^4}{(y^- y^+)^2 \, (1-y^-)(1-y^+)} \\
&\qquad -  \frac{(8 \, d^2 - d^4 - 10 \, d^2 s + d^4 s - 8 \, s^2 + 7 \, d^2 s^2 + 12 \, s^3 - 
 2 \, d^2 s^3 - 6 \, s^4 + s^5)^2}{64 \, (y^- y^+)^2 \, (1-y^-)^3 (1-y^+)^3}\,,
\end{aligned} \label{resOne}
\end{equation}
with $s \, = \, y^-+y^+, \, d \, = \, y^--y^+$. Notice that this expression is real. In the limit $p\to0$ this finally gives
\begin{equation}
4  \big(j_1+j_2+j_3-1\big)\big(j_1+j_2+j_3+1\big)  p^2 + \dots\,,
\end{equation}
as expected from eq.~\eqref{eq:pakmanseverresult}.
In a similar way we can compute the hexagon form factor for the correlator~\eqref{eq:nonsymmetriccase-particles}, where the third operator is different from the other two. In this case, the result is proportional to
\begin{equation}
\begin{aligned}
&\big(2 \, \ell_{12}\big)^2  \frac{(y^--y^+)^4}{(y^- y^+)^2 \, (1-y^-)(1-y^+)} \\
&\qquad -  \frac{(8 \, d^2 - 4 \, d^4 - 16 \, d^2 s + d^4 s - 8 \, s^2 + 10 \, d^2 s^2 + 12 \, s^3 - 
 2 \, d^2 s^3 - 6 \, s^4 + s^5)^2}{64 \, (y^- y^+)^2 \, (1-y^-)^3 (1-y^+)^3}\,.
\end{aligned} \label{resTwo}
 \end{equation}
The resulting $p\to0$ limit is
\begin{equation}
 -4 \, (1 - 2 \, \ell_{12})(1 - 2 \, \ell_{12}) \, p^2 + \ldots\,,
\end{equation}
which matches with~\eqref{eq:pakmanseverresult}.
In particular, if we disregard the overall normalisations, we find that the ratio of the two families of correlation functions match for arbitrary $j_1$, $j_2$ and~$j_3$.

We have not mentioned a selection rule concerning the results \eqref{resOne} and \eqref{resTwo} and their limits: in the whole discussion it was assumed that the $J_i$ are such that the $l_{ij}$ are integer as suggested by perturbative field theory. Further, we have to distinguish the cases $\sum \nu_k \, \in \, 2 \mathbb{Z}$ for which formulae \eqref{resOne}, \eqref{resTwo} are valid, and $\sum \nu_k \, \in \, 2 \mathbb{Z} + 1$ for which the correlators actually vanish.

To conclude this discussion, we note that our result only relied on the ``asymptotic'' part of the hexagon, without accounting for wrapping effects.  This can be done in this formalism order by order~\cite{Basso:2015zoa, Eden:2015ija,Basso:2015eqa,Basso:2017muf} by considering L\"uscher-type corrections. It is natural to ask why our result nonetheless matches those in the literature. This is because we are dealing with half-BPS states, or equivalently precisely with states that are composed of zero-momentum excitations only. The argument was first noted in refs.~\cite{Borsato:2016kbm,Baggio:2017kza} in the context of the computation of the spectrum for the very same operators. Essentially, the transfer matrix appearing in the computation of (arbitrarily high) wrapping effects only involves zero-momentum excitations. As such it get precisely the same and opposite corrections for Fermionic and Bosonic wrapping effect, leading to a complete cancellation of wrapping.

\section{Conclusions and outlook}
\label{sec:conclusions}
In this article we have seen that the hexagon approach for the computation of three-point functions by integrability set out in ref.~\cite{Basso:2015zoa} can also be applied to $\AdSST$. This is the first example of an integrable superstring background with this feature other than the original $\AdS_5\times\Sph^5$. The main aim of this paper was to perform the bootstrap procedure for the hexagon form factor, check its internal consistency and perform a basic check of the resulting machinery. In this regard, we have been successful. There are now many interesting directions that should be studied.

Our framework can be used to study background with a mixture of NSNS and RR background fluxes. In a sense, the case of pure-RR background fluxes seems simplest because in that case we know all the dressing factors~\cite{Borsato:2013hoa,Borsato:2016xns} and much of the intuition from $\AdS_5\times\Sph^5$ may be exploited; this is also the case that naturally corresponds to the D1-D5 brane systems, which is of interest in holography. Conversely, the pure-NSNS case would also be very interesting to study, as in that case we should be able to make contact with the computation of correlation functions by worldsheet CFT techniques~\cite{Maldacena:2001km}. The main obstacle in this case is that we do not know the scalar factors; however, given the relative simplicity of the system at the pure-NSNS point --- which is quite apparent when studying the spectral problem~\cite{Baggio:2018gct,Dei:2018mfl} --- it is possible that we could make an educated guess for them.

The last few years saw a spectacular development in the use of worldsheet-CFT approaches to understand pure-NSNS backgrounds, which is particularly powerful for the level $k=1$ theory~\cite{Giribet:2018ada,Gaberdiel:2018rqv, Eberhardt:2018ouy, Eberhardt:2020bgq,Gaberdiel:2020ycd}. Most of these new development deal with the ``long-string'' part of the WZW spectrum, \textit{i.e.}\ with the part emerging from continuous representations --- at $k=1$, this actually constitutes the whole spectrum. Conversely, our analysis here applies to short strings, which emerge from \textit{discrete} representations. This is not surprising because our analysis is generically valid for RR-flux or mixed-flux backgrounds; for these backgrounds, there are no long strings. It would be extremely interesting to take the NSNS limit and try to recover the long-string spectrum from the short-string one. Recently, this was argued at the level of the spectrum in ref.~\cite{Sfondrini:2020ovj}. It would be very interesting to do this for correlation functions. It is curious how the long-string spectrum, which can be studied in amazing detail with CFT techniques at $k=1$~\cite{Eberhardt:2020bgq,Gaberdiel:2020ycd} (owing to the existence of a free-field representation) is so subtle to incorporate in the integrability description: it is certainly something worthy of further investigation.

The most-general case of mixed-flux backgrounds will possibly be the most challenging, as once again the scalar factors are unknown and probably highly nontrivial, see also ref.~\cite{Babichenko:2014yaa}.

Another interesting point is how to incorporate finite-size (``wrapping'') effects, which is the bane of most integrability approaches. In $\AdS_5\times\Sph^5$ this can be done order-by-order~\cite{Basso:2015zoa, Eden:2015ija,Basso:2015eqa,Basso:2017muf}. Here it is likely that things are more complicated, at least in general, due to the presence of massless modes~\cite{Abbott:2015pps}. However, we expect that in the pure-NSNS case we should be able to deal quite easily with all wrapping effects, due to the simple structure highlighted in ref.~\cite{Dei:2018mfl}. In fact, studying wrapping effect in this context may well be a training ground for incorporating them in more general backgrounds.

It is worth emphasising that the hexagon formalism may be used also to construct higher-point correlation functions~\cite{Eden:2016xvg,Fleury:2016ykk} as well as non-planar correlators~\cite{Eden:2017ozn,Bargheer:2017nne,Bargheer:2019kxb}. This gives another setup in which wrapping may be manageable, namely the correlation functions of BPS operators. The simplest case is that of a four-point function, which would show a very non-trivial dependence on the conformal and R-symmetry cross-ratios. In $\AdS_5\times\Sph^5$ it is possible to study this in quite some detail at small 't~Hooft coupling, see refs.~\cite{Fleury:2016ykk,Fleury:2017eph}. Can we perform a similar study here? If so, this would undoubtedly shed new lights on the structure of interactions at generic points of at AdS$_3$/CFT$_2$ moduli space~\cite{Maldacena:2015iua}.

Finally, it is natural to wonder which other backgrounds are amenable to this bootstrap approach. Two natural candidates from the point of view of integrability are $\AdS_4\times \text{CP}^3$~\cite{Klose:2010ki} and $\AdS_3\times\Sph^3\times\Sph^3\times\Sph^1$~\cite{Borsato:2015mma}. The main obstacle which we encounter here is that neither of these backgrounds has a factorised symmetry algebra --- unlike the case of $\AdS_5\times\Sph^5$ where one could identify a diagonal $\su(2|2)$, and of $\AdSST$ where we found a diagonal $\su(1|1)^{\oplus2}$. All the same, these backgrounds are all integrable as far as the spectral problem is concerned, and their integrable structure is remarkably similar. It would almost seem unnatural if their correlation functions cannot be bootstrapped. The same goes for the various integrable deformations of all these setups that one may consider. Among those, it would be particularly interesting to consider ``quantum'' deformations~\cite{Beisert:2008tw}, whose geometric description~\cite{Delduc:2013qra,Arutyunov:2013ega} was recently shown to include a consistent string background~\cite{Hoare:2018ngg, Seibold:2020ywq}.

We hope to return in the near future to some of these intriguing questions.

\section*{Acknowledgements}
We thank Riccardo Borsato, Andrea Dei, Marius de Leeuw, Sergey Frolov and Fiona K.~Seibold for discussions. AS thanks Riccardo Borsato, Olof Ohlsson Sax, Bodgan Stefa\'nski jr.~and Alessandro Torrielli for past collaboration and discussions on many points related to this article.
The authors are grateful to ETH Zurich for hospitality at various stages of this project.
BE and AS acknowledge support from the Swiss National Science Foundation under Spark grant n.~190657. DlP acknowledges support by the Stiftung der Deutschen Wirtschaft.

\appendix
\section{Explicit form of the full S~Matrix}
\label{app:Smatrix}
In order to check that the bootstrapped two-particle hexagon form factor obeys the Watson equation, we must use the explicit form of the S-matrix in the different sectors. This is know in the literature~\cite{Borsato:2013qpa,Borsato:2013qpa,Lloyd:2014bsa} but the explicit expressions are somewhat scattered between different papers that have slightly different notations. Hence we collect it here. The full $\mathrm{psu}(1|1)^4$ S~matrix can be obtained by taking the graded tensor product~\cite{Borsato:2013qpa} of two copies of the $\mathrm{psu}(1|1)^2$ S~matrix of~\cite{Borsato:2012ud},
\begin{equation}
    \mathbf{S} = S \hat{\otimes} \acute{S},
\end{equation}
which can be defined in terms of the matrix elements by
\begin{equation}
    (\mathcal{M} \hat{\otimes} \acute{\mathcal{M}})^{I \acute{I},J\acute{J}}_{K \acute{K},L\acute{L}} = (-1)^{F_{\acute{K}}F_L+F_J F_{\acute{I}}} \mathcal{M}^{IJ}_{KL} \acute{\mathcal{M}}^{\acute{I}\acute{J}}_{\acute{K}\acute{L}}.
\end{equation}
We recall our convention for $\epsilon^{ab}$ here
\begin{equation}
    \epsilon^{12}=-\epsilon^{21}=-\epsilon_{12}=\epsilon_{21}= 1.
\end{equation}

\subsection{The massive sector}
\subsubsection{Left-left scattering}
\begin{equation}
\begin{aligned}
    \mathbf{S}|Y_p Y_q\rangle &= +A_{pq}^{\L\L} A_{pq}^{\L\L} |Y_q Y_p \rangle,\\
    \mathbf{S}|Z_p Z_q\rangle &= +F_{pq}^{\L\L} F_{pq}^{\L\L} |Z_q Z_p \rangle,\\
    \mathbf{S}|Y_p Z_q\rangle &= +C_{pq}^{\L\L} C_{pq}^{\L\L} |Y_q Z_p \rangle + B_{pq}^{\L\L} B_{pq}^{\L\L} |Z_q Y_p \rangle - B_{pq}^{\L\L} C_{pq}^{\L\L} \epsilon_{ab} |\Psi^a_q \Psi^b_p \rangle,\\
    \mathbf{S}|Z_p Y_q\rangle &= +D_{pq}^{\L\L} D_{pq}^{\L\L} |Y_q Z_p \rangle + E_{pq}^{\L\L} E_{pq}^{\L\L} |Z_q Y_p \rangle - D_{pq}^{\L\L} E_{pq}^{\L\L} \epsilon_{ab} |\Psi^a_q \Psi^b_p \rangle,\\
    \mathbf{S}|Y_p \Psi^a_q\rangle &= +A_{pq}^{\L\L} C_{pq}^{\L\L} |Y_p \Psi^a_q\rangle + A_{pq}^{\L\L} B_{pq}^{\L\L} |\Psi^a_q Y_p\rangle,\\ 
    \mathbf{S}|Z_p \Psi^a_q\rangle &= +E_{pq}^{\L\L} F_{pq}^{\L\L} |Z_p \Psi^a_q\rangle - D_{pq}^{\L\L} F_{pq}^{\L\L} |\Psi^a_q Z_p\rangle,\\ 
    \mathbf{S}|\Psi^a_p Y_q\rangle &= +A_{pq}^{\L\L} D_{pq}^{\L\L} |Y_q \Psi^a_p\rangle + A_{pq}^{\L\L} E_{pq}^{\L\L} |\Psi^a_q Y_p\rangle,\\ 
    \mathbf{S}|\Psi^a_p Z_q\rangle &= -B_{pq}^{\L\L} F_{pq}^{\L\L} |Z_q \Psi^a_p\rangle + C_{pq}^{\L\L} F_{pq}^{\L\L} |\Psi^a_q Z_p\rangle,\\ 
    \mathbf{S}|\Psi^a_p \Psi^b_q\rangle &= \delta^{ab} A_{pq}^{\L\L} F_{pq}^{\L\L} |\Psi^b_q \Psi^a_p\rangle + \epsilon^{ab} (C_{pq}^{\L\L} D_{pq}^{\L\L} |Y_q Z_p \rangle +B_{pq}^{\L\L} E_{pq}^{\L\L} |Z_q Y_p \rangle ) \\
    & \quad + (1-\delta^{ab}) (C_{pq}^{\L\L} E_{pq}^{\L\L} |\Psi^a_q \Psi^b_p \rangle - B_{pq}^{\L\L} D_{pq}^{\L\L} |\Psi^b_q \Psi^a_p \rangle ).
\end{aligned}
\end{equation}
The last process can be further simplified by using the identity 
\begin{equation}
    C_{pq}^{\L\L} E_{pq}^{\L\L} - B_{pq}^{\L\L} D_{pq}^{\L\L} = A_{pq}^{\L\L} F_{pq}^{\L\L}.
\end{equation}

\subsubsection{Right-right scattering}
\begin{equation}
\begin{aligned}
    \mathbf{S}|\tilde{Y}_p \tilde{Y}_q\rangle &= +A_{pq}^{\R\R} A_{pq}^{\R\R} |\tilde{Y}_q \tilde{Y}_p \rangle,\\
    \mathbf{S}|\tilde{Z}_p \tilde{Z}_q\rangle &= +F_{pq}^{\R\R} F_{pq}^{\R\R} |\tilde{Z}_q \tilde{Z}_p \rangle,\\
    \mathbf{S}|\tilde{Y}_p \tilde{Z}_q\rangle &= +C_{pq}^{\R\R} C_{pq}^{\R\R} |\tilde{Y}_q \tilde{Z}_p \rangle + B_{pq}^{\R\R} B_{pq}^{\R\R} |\tilde{Z}_q \tilde{Y}_p \rangle - B_{pq}^{\R\R} C_{pq}^{\R\R} \epsilon_{ab} |\tilde{\Psi}^a_q \tilde{\Psi}^b_p \rangle,\\
    \mathbf{S}|\tilde{Z}_p \tilde{Y}_q\rangle &= +D_{pq}^{\R\R} D_{pq}^{\R\R} |\tilde{Y}_q \tilde{Z}_p \rangle + E_{pq}^{\R\R} E_{pq}^{\R\R} |\tilde{Z}_q \tilde{Y}_p \rangle - D_{pq}^{\R\R} E_{pq}^{\R\R} \epsilon_{ab} |\tilde{\Psi}^a_q \tilde{\Psi}^b_p \rangle,\\
    \mathbf{S}|\tilde{Y}_p \tilde{\Psi}^a_q\rangle &= +A_{pq}^{\R\R} C_{pq}^{\R\R} |\tilde{Y}_p \tilde{\Psi}^a_q\rangle + A_{pq}^{\R\R} B_{pq}^{\R\R} |\tilde{\Psi}^a_q \tilde{Y}_p\rangle,\\ 
    \mathbf{S}|\tilde{Z}_p \tilde{\Psi}^a_q\rangle &= +E_{pq}^{\R\R} F_{pq}^{\R\R} |\tilde{Z}_p \tilde{\Psi}^a_q\rangle - D_{pq}^{\R\R} F_{pq}^{\R\R} |\tilde{\Psi}^a_q \tilde{Z}_p\rangle,\\ 
    \mathbf{S}|\tilde{\Psi}^a_p \tilde{Y}_q\rangle &= +A_{pq}^{\R\R} D_{pq}^{\R\R} |\tilde{Y}_q \tilde{\Psi}^a_p\rangle + A_{pq}^{\R\R} E_{pq}^{\R\R} |\tilde{\Psi}^a_q \tilde{Y}_p\rangle,\\ 
    \mathbf{S}|\tilde{\Psi}^a_p \tilde{Z}_q\rangle &= -B_{pq}^{\R\R} F_{pq}^{\R\R} |\tilde{Z}_q \tilde{\Psi}^a_p\rangle + C_{pq}^{\R\R} F_{pq}^{\R\R} |\tilde{\Psi}^a_q \tilde{Z}_p\rangle,\\ 
    \mathbf{S}|\tilde{\Psi}^a_p \tilde{\Psi}^b_q\rangle &= \delta^{ab} A_{pq}^{\R\R} F_{pq}^{\R\R} |\tilde{\Psi}^b_q \tilde{\Psi}^a_p\rangle + \epsilon^{ab} (C_{pq}^{\R\R} D_{pq}^{\R\R} |\tilde{Y}_q \tilde{Z}_p \rangle +B_{pq}^{\R\R} E_{pq}^{\R\R} |\tilde{Z}_q \tilde{Y}_p \rangle ) \\
    & \quad + (1-\delta^{ab}) (C_{pq}^{\R\R} E_{pq}^{\R\R} |\tilde{\Psi}^a_q \tilde{\Psi}^b_p \rangle - B_{pq}^{\R\R} D_{pq}^{\R\R} |\tilde{\Psi}^b_q \tilde{\Psi}^a_p \rangle ).
\end{aligned}
\end{equation}
The last process can be further simplified by using the identity 
\begin{equation}
    C_{pq}^{\R\R} E_{pq}^{\R\R} - B_{pq}^{\R\R} D_{pq}^{\R\R} = A_{pq}^{\R\R} F_{pq}^{\R\R}.
\end{equation}

\subsubsection{Left-right scattering}
\begin{equation}
\begin{aligned}
    \mathbf{S}|Y_p \tilde{Y}_q\rangle &= +A_{pq}^{\L\R} A_{pq}^{\L\R} |\tilde{Y}_q Y_p\rangle - B_{pq}^{\L\R} B_{pq}^{\L\R} |\tilde{Z}_q Z_p\rangle - A_{pq}^{\L\R} B_{pq}^{\L\R} \epsilon_{ab}|\tilde{\Psi}^a_{q} \Psi^b_p\rangle,\\
    \mathbf{S}|Z_p \tilde{Z}_q\rangle &= -F_{pq}^{\L\R} F_{pq}^{\L\R} |\tilde{Y}_q Y_p\rangle + E_{pq}^{\L\R} E_{pq}^{\L\R} |\tilde{Z}_q Z_p\rangle + E_{pq}^{\L\R} F_{pq}^{\L\R} \epsilon_{ab} |\tilde{\Psi}^a_{q} \Psi^b_p\rangle,\\
    \mathbf{S}|Y_p \tilde{Z}_q\rangle &= +C_{pq}^{\L\R} C_{pq}^{\L\R} |\tilde{Z}_q Y_p\rangle,\\
    \mathbf{S}|Z_p \tilde{Y}_q\rangle &= +D_{pq}^{\L\R} D_{pq}^{\L\R} |\tilde{Y}_q Z_p\rangle,\\
    \mathbf{S}|Y_p \tilde{\Psi}^a_{q}\rangle &= +A_{pq}^{\L\R} C_{pq}^{\L\R} |\tilde{\Psi}^a_{q} Y_p\rangle - B_{pq}^{\L\R} C_{pq}^{\L\R} |\tilde{Z}_q \Psi^a_p\rangle,\\
    \mathbf{S}|Z_p \tilde{\Psi}^a_{q}\rangle &= -D_{pq}^{\L\R} E_{pq}^{\L\R} |\tilde{\Psi}^a_{q} Z_p\rangle + D_{pq}^{\L\R} F_{pq}^{\L\R} |\tilde{Y}_q \Psi^a_p\rangle,\\
    \mathbf{S}|\Psi^a_p \tilde{Y}_q\rangle &= +A_{pq}^{\L\R} D_{pq}^{\L\R} |\tilde{Y}_q \Psi^a_p\rangle - B_{pq}^{\L\R} D_{pq}^{\L\R} |\tilde{\Psi}^a_{q} Z_p\rangle,\\
    \mathbf{S}|\Psi^a_p \tilde{Z}_q\rangle &= -C_{pq}^{\L\R} E_{pq}^{\L\R} |\tilde{Z}_q \Psi^a_p\rangle + C_{pq}^{\L\R} F_{pq}^{\L\R} |\tilde{\Psi}^a_{q} Y_p\rangle,\\
    \mathbf{S}|\Psi^a_p \tilde{\Psi}^b_{q}\rangle &= -C_{pq}^{\L\R} D_{pq}^{\L\R} \delta^{ab}|\tilde{\Psi}^b_{q} \Psi^a_p\rangle - \epsilon^{ab} (A_{pq}^{\L\R} F_{pq}^{\L\R} |\tilde{Y}_q Y_p\rangle - B_{pq}^{\L\R} E_{pq}^{\L\R} |\tilde{Z}_q Z_p\rangle) \\
    & \quad +(1-\delta^{ab}) (A_{pq}^{\L\R} E_{pq}^{\L\R} |\tilde{\Psi}^b_{q} \Psi^a_p\rangle - B_{pq}^{\L\R} F_{pq}^{\L\R} |\tilde{\Psi}^a_{q} \Psi^b_p\rangle ).
\end{aligned}
\end{equation}
The last process can be further simplified by using the identity 
\begin{equation}
    A_{pq}^{\L\R} E_{pq}^{\L\R} - B_{pq}^{\L\R} F_{pq}^{\L\R} = -C_{pq}^{\L\R} D_{pq}^{\L\R}.
\end{equation}

\subsubsection{Right-left scattering}
\begin{equation}
\begin{aligned}
    \mathbf{S}|\tilde{Y}_p Y_q\rangle &= +A_{pq}^{\R\L} A_{pq}^{\R\L} |Y_q \tilde{Y}_p\rangle - B_{pq}^{\R\L} B_{pq}^{\R\L} |Z_q \tilde{Z}_p\rangle + A_{pq}^{\R\L} B_{pq}^{\R\L} \epsilon_{ab} |\Psi^a_{q} \tilde{\Psi}^b_p\rangle,\\
    \mathbf{S}|\tilde{Z}_p Z_q\rangle &= -F_{pq}^{\R\L} F_{pq}^{\R\L} |Y_q \tilde{Y}_p\rangle + E_{pq}^{\R\L} E_{pq}^{\R\L} |Z_q \tilde{Z}_p\rangle - E_{pq}^{\R\L} F_{pq}^{\R\L} \epsilon_{ab} |\Psi^a_{q} \tilde{\Psi}^b_p\rangle,\\
    \mathbf{S}|\tilde{Y}_p Z_q\rangle &= +C_{pq}^{\R\L} C_{pq}^{\R\L} |Z_q \tilde{Y}_p\rangle,\\
    \mathbf{S}|\tilde{Z}_p Y_q\rangle &= +D_{pq}^{\R\L} D_{pq}^{\R\L} |Y_q \tilde{Z}_p\rangle,\\
    \mathbf{S}|\tilde{Y}_p \Psi^a_{q}\rangle &= +A_{pq}^{\R\L} C_{pq}^{\R\L} |\Psi^a_{q} \tilde{Y}_p\rangle + B_{pq}^{\R\L} C_{pq}^{\R\L} |Z_q \tilde{\Psi}^a_p\rangle,\\
    \mathbf{S}|\tilde{Z}_p \Psi^a_{q}\rangle &= -D_{pq}^{\R\L} E_{pq}^{\R\L} |\Psi^a_{q} \tilde{Z}_p\rangle - D_{pq}^{\R\L} F_{pq}^{\R\L} |Y_q \tilde{\Psi}^a_p\rangle,\\
    \mathbf{S}|\tilde{\Psi}^a_p Y_q\rangle &= +A_{pq}^{\R\L} D_{pq}^{\R\L} |Y_q \tilde{\Psi}^a_p\rangle + B_{pq}^{\R\L} D_{pq}^{\R\L} |\Psi^a_{q} \tilde{Z}_p\rangle,\\
    \mathbf{S}|\tilde{\Psi}^a_p Z_q\rangle &= -C_{pq}^{\R\L} E_{pq}^{\R\L} |Z_q \tilde{\Psi}^a_p\rangle - C_{pq}^{\R\L} F_{pq}^{\R\L} |\Psi^a_{q} \tilde{Y}_p\rangle,\\
    \mathbf{S}|\tilde{\Psi}^a_p \Psi^b_{q}\rangle &= -C_{pq}^{\R\L} D_{pq}^{\R\L} \delta^{ab}|\Psi^b_{q} \tilde{\Psi}^a_p\rangle - \epsilon^{ab} (A_{pq}^{\R\L} F_{pq}^{\R\L} |Y_q \tilde{Y}_p\rangle - B_{pq}^{\R\L} E_{pq}^{\R\L} |Z_q \tilde{Z}_p\rangle) \\
    & \quad +(1-\delta^{ab}) (A_{pq}^{\R\L} E_{pq}^{\R\L} |\Psi^b_{q} \tilde{\Psi}^a_p\rangle - B_{pq}^{\R\L} F_{pq}^{\R\L} |\Psi^a_{q} \tilde{\Psi}^b_p\rangle ).
\end{aligned}
\end{equation}
The last process can be further simplified by using the identity 
\begin{equation}
    A_{pq}^{\R\L} E_{pq}^{\R\L} - B_{pq}^{\R\L} F_{pq}^{\R\L} = -C_{pq}^{\R\L} D_{pq}^{\R\L}.
\end{equation}

\subsection{The mixed-mass sector}
\subsubsection{Left-massless scattering}
\begin{equation}
\begin{aligned}
    \mathbf{S}|Z_p T^{\dot{A}a}_q\rangle &= - D_{pq}^{\L\z} F_{pq}^{\L\z} |T^{\dot{A}a}_q Z_p \rangle - E_{pq}^{\L\z} F_{pq}^{\L\z} |\tilde{\chi}^{\dot{A}}_q \Psi^a_p \rangle,\\
    \mathbf{S}|Y_p T^{\dot{A}a}_q\rangle &= +A_{pq}^{\L\z} B_{pq}^{\L\z} |T^{\dot{A}a}_q Y_p \rangle - A_{pq}^{\L\z} C_{pq}^{\L\z} |\chi^{\dot{A}}_q \Psi^a_p \rangle,\\
    \mathbf{S}|\Psi^a_p \tilde{\chi}^{\dot{A}}_q\rangle &= + B_{pq}^{\L\z} F_{pq}^{\L\z} |\tilde{\chi}^{\dot{A}}_q \Psi^a_p \rangle + C_{pq}^{\L\z} F_{pq}^{\L\z} |T^{\dot{A}a}_q Z_p \rangle,\\
    \mathbf{S}|\Psi^a_p \chi^{\dot{A}}_q\rangle &= -A_{pq}^{\L\z} D_{pq}^{\L\z} |\chi^{\dot{A}}_q \Psi^a_p \rangle + A_{pq}^{\L\z} E_{pq}^{\L\z} |T^{\dot{A}a}_q Y_p \rangle,\\
    \mathbf{S}|Z_p \tilde{\chi}^{\dot{A}}_q\rangle &= +F_{pq}^{\L\z} F_{pq}^{\L\z} |\tilde{\chi}^{\dot{A}}_q Z_p \rangle,\\
    \mathbf{S}|Y_p \chi^{\dot{A}}_q\rangle &= +A_{pq}^{\L\z} A_{pq}^{\L\z} |\chi^{\dot{A}}_q Y_p \rangle,\\
    \mathbf{S}|Z_p \chi^{\dot{A}}_q\rangle &= +D_{pq}^{\L\z} D_{pq}^{\L\z} |\chi^{\dot{A}}_q Z_p \rangle +E_{pq}^{\L\z} E_{pq}^{\L\z} |\tilde{\chi}^{\dot{A}}_q Y_p \rangle +D_{pq}^{\L\z} E_{pq}^{\L\z} \epsilon_{ab} |T^{\dot{A}a}_q \Psi^b_p \rangle,\\
    \mathbf{S}|Y_p \tilde{\chi}^{\dot{A}}_q\rangle &= +B_{pq}^{\L\z} B_{pq}^{\L\z} |\tilde{\chi}^{\dot{A}}_q Y_p \rangle +C_{pq}^{\L\z} C_{pq}^{\L\z} |\chi^{\dot{A}}_q Z_p \rangle +B_{pq}^{\L\z} C_{pq}^{\L\z} \epsilon_{ab} |T^{\dot{A}a}_q \Psi^b_p \rangle,\\
    \mathbf{S}|\Psi^a_p T^{\dot{A}b}_q\rangle &= - \delta^{ab} A_{pq}^{\L\z} F_{pq}^{\L\z} |T^{\dot{A}a}_q \Psi^b_p \rangle + \epsilon^{ab}(C_{pq}^{\L\z} D_{pq}^{\L\z} |\chi^{\dot{A}}_q Z_p \rangle + B_{pq}^{\L\z} E_{pq}^{\L\z} |\tilde{\chi}^{\dot{A}}_q Y_p \rangle ) \\
    &\quad +(1-\delta^{ab})(B_{pq}^{\L\z} D_{pq}^{\L\z} |T^{\dot{A}a}_q \Psi^b_p \rangle - C_{pq}^{\L\z} E_{pq}^{\L\z} |T^{\dot{A}b}_q \Psi^a_p \rangle),
\end{aligned}
\end{equation}
The last process can be further simplified by using the identity 
\begin{equation}
    C_{pq}^{\L\z} E_{pq}^{\L\z} - B_{pq}^{\L\z} D_{pq}^{\L\z} = A_{pq}^{\L\z} F_{pq}^{\L\z}.
\end{equation}

\subsubsection{Massless-left scattering}
\begin{equation}
\begin{aligned}
    \mathbf{S}|T^{\dot{A}a}_p Z_q\rangle &= -B_{pq}^{\z\L} F_{pq}^{\z\L} |Z_q T^{\dot{A}a}_p \rangle + C_{pq}^{\z\L} F_{pq}^{\z\L} |\Psi^a_q \tilde{\chi}^{\dot{A}}_p \rangle,\\
    \mathbf{S}|T^{\dot{A}a}_p Y_q\rangle &= +A_{pq}^{\z\L} D_{pq}^{\z\L} |Y_q T^{\dot{A}a}_p \rangle + A_{pq}^{\z\L} E_{pq}^{\z\L} |\Psi^a_q \chi^{\dot{A}}_p \rangle,\\
    \mathbf{S}|\tilde{\chi}^{\dot{A}}_p \Psi^a_q\rangle &= +D_{pq}^{\z\L} F_{pq}^{\z\L} |\Psi^a_q \tilde{\chi}^{\dot{A}}_p \rangle - E_{pq}^{\z\L} F_{pq}^{\z\L} |Z_q T^{\dot{A}a}_p \rangle,\\
    \mathbf{S}|\chi^{\dot{A}}_p \Psi^a_q\rangle &= -A_{pq}^{\z\L} B_{pq}^{\z\L} |\Psi^a_q \chi^{\dot{A}}_p \rangle - A_{pq}^{\z\L} C_{pq}^{\z\L} |Y_q T^{\dot{A}a}_p \rangle,\\
    \mathbf{S}|\tilde{\chi}^{\dot{A}}_p Z_q\rangle &= +F_{pq}^{\z\L} F_{pq}^{\z\L} |Z_q \tilde{\chi}^{\dot{A}}_p \rangle,\\
    \mathbf{S}|\chi^{\dot{A}}_p Y_q\rangle &= +A_{pq}^{\z\L} A_{pq}^{\z\L} |Y_q \chi^{\dot{A}}_p \rangle,\\
    \mathbf{S}|\chi^{\dot{A}}_p Z_q\rangle &= +B_{pq}^{\z\L} B_{pq}^{\z\L} |Z_q \chi^{\dot{A}}_p \rangle +C_{pq}^{\z\L} C_{pq}^{\z\L} |Y_q \tilde{\chi}^{\dot{A}}_p \rangle - B_{pq}^{\z\L} C_{pq}^{\z\L} \epsilon_{ab} |\Psi^a_q T^{\dot{A}b}_p \rangle,\\
    \mathbf{S}|\tilde{\chi}^{\dot{A}}_p Y_q\rangle &= +D_{pq}^{\z\L} D_{pq}^{\z\L} |Y_q \tilde{\chi}^{\dot{A}}_p \rangle +E_{pq}^{\z\L} E_{pq}^{\z\L} |Z_q \chi^{\dot{A}}_p \rangle -D_{pq}^{\z\L} E_{pq}^{\z\L} \epsilon_{ab} |\Psi^a_q T^{\dot{A}b}_p \rangle,\\
    \mathbf{S}|T^{\dot{A}a}_p \Psi^b_q\rangle &= - \delta^{ab} A_{pq}^{\z\L} F_{pq}^{\z\L} |\Psi^b_q T^{\dot{A}a}_p \rangle - \epsilon^{ab}(B_{pq}^{\z\L} E_{pq}^{\z\L}  |Z_q \chi^{\dot{A}}_p \rangle +C_{pq}^{\z\L} D_{pq}^{\z\L} |Y_q \tilde{\chi}^{\dot{A}}_p \rangle ) \\
    &\quad +(1-\delta^{ab})(D_{pq}^{\z\L} B_{pq}^{\z\L} |\Psi^b_q T^{\dot{A}a}_p \rangle -E_{pq}^{\z\L} C_{pq}^{\z\L} |\Psi^a_q T^{\dot{A}b}_p \rangle),
\end{aligned}
\end{equation}
The last process can be further simplified by using the identity 
\begin{equation}
    C_{pq}^{\z\L} E_{pq}^{\z\L} - B_{pq}^{\z\L} D_{pq}^{\z\L} = A_{pq}^{\z\L} F_{pq}^{\z\L}.
\end{equation}

\subsubsection{Right-massless scattering}
\begin{equation}
\begin{aligned}
    \mathbf{S}|\tilde{Z}_p T^{\dot{A}a}_q\rangle &= -D_{pq}^{\R\z} E_{pq}^{\R\z} |T^{\dot{A}a}_q \tilde{Z}_p \rangle + D_{pq}^{\R\z} F_{pq}^{\R\z} |\chi^{\dot{A}}_q \tilde{\Psi}^a_p \rangle,\\
    \mathbf{S}|\tilde{Y}_p T^{\dot{A}a}_q\rangle &= +A_{pq}^{\R\z} C_{pq}^{\R\z} |T^{\dot{A}a}_q \tilde{Y}_p \rangle - B_{pq}^{\R\z} C_{pq}^{\R\z} |\tilde{\chi}^{\dot{A}}_q \tilde{\Psi}^a_p \rangle,\\
    \mathbf{S}|\tilde{\Psi}^a_p \chi^{\dot{A}}_q\rangle &= -  A_{pq}^{\R\z} D_{pq}^{\R\z} |\chi^{\dot{A}}_q \tilde{\Psi}^a_p \rangle + B_{pq}^{\R\z} D_{pq}^{\R\z} |T^{\dot{A}a}_q \tilde{Z}_p \rangle,\\
    \mathbf{S}|\tilde{\Psi}^a_p \tilde{\chi}^{\dot{A}}_q\rangle &= +C_{pq}^{\R\z} E_{pq}^{\R\z} |\tilde{\chi}^{\dot{A}}_q \tilde{\Psi}^a_p \rangle - C_{pq}^{\R\z} F_{pq}^{\R\z} |T^{\dot{A}a}_q \tilde{Y}_p \rangle,\\
    \mathbf{S}|\tilde{Z}_p \chi^{\dot{A}}_q\rangle &= +D_{pq}^{\R\z} D_{pq}^{\R\z} |\chi^{\dot{A}}_q \tilde{Z}_p \rangle,\\
    \mathbf{S}|\tilde{Y}_p \tilde{\chi}^{\dot{A}}_q\rangle &= +C_{pq}^{\R\z} C_{pq}^{\R\z} |\tilde{\chi}^{\dot{A}}_q \tilde{Y}_p \rangle,\\
    \mathbf{S}|\tilde{Z}_p \tilde{\chi}^{\dot{A}}_q\rangle &= +E_{pq}^{\R\z} E_{pq}^{\R\z} |\tilde{\chi}^{\dot{A}}_q \tilde{Z}_p \rangle -F_{pq}^{\R\z} F_{pq}^{\R\z} |\chi^{\dot{A}}_q \tilde{Y}_p \rangle +F_{pq}^{\R\z} E_{pq}^{\R\z} \epsilon_{ab} |T^{\dot{A}a}_q \tilde{\Psi}^b_p \rangle,\\
    \mathbf{S}|\tilde{Y}_p \chi^{\dot{A}}_q\rangle &= +A_{pq}^{\R\z} A_{pq}^{\R\z} |\chi^{\dot{A}}_q \tilde{Y}_p \rangle -B_{pq}^{\R\z} B_{pq}^{\R\z} |\tilde{\chi}^{\dot{A}}_q \tilde{Z}_p \rangle -B_{pq}^{\R\z} A_{pq}^{\R\z} \epsilon_{ab} |T^{\dot{A}a}_q \tilde{\Psi}^b_p \rangle,\\
    \mathbf{S}|\tilde{\Psi}^a_p T^{\dot{A}b}_q\rangle &= +C_{pq}^{\R\z} D_{pq}^{\R\z} \delta^{ab} |T^{\dot{A}a}_q \tilde{\Psi}^b_p \rangle + \epsilon^{ab}(A_{pq}^{\R\z} F_{pq}^{\R\z} |\chi^{\dot{A}}_q \tilde{Y}_p \rangle - B_{pq}^{\R\z} E_{pq}^{\R\z}  |\tilde{\chi}^{\dot{A}}_q \tilde{Z}_p \rangle) \\
    &\quad + (1-\delta^{ab}) (B_{pq}^{\R\z} F_{pq}^{\R\z} |T^{\dot{A}a}_q \tilde{\Psi}^b_p \rangle -A_{pq}^{\R\z} E_{pq}^{\R\z} |T^{\dot{A}b}_q \tilde{\Psi}^a_p \rangle).
\end{aligned}
\end{equation}
The last process can be further simplified by using the identity 
\begin{equation}
    B_{pq}^{\R\z} F_{pq}^{\R\z} - A_{pq}^{\R\z} E_{pq}^{\R\z} = C_{pq}^{\R\z} D_{pq}^{\R\z}.
\end{equation}

\subsubsection{Massless-right scattering}
\begin{equation}
\begin{aligned}
    \mathbf{S}|T^{\dot{A}a}_p \tilde{Z}_q\rangle &= -C_{pq}^{\z\R} E_{pq}^{\z\R} |\tilde{Z}_q T^{\dot{A}a}_p \rangle + C_{pq}^{\z\R} F_{pq}^{\z\R} |\tilde{\Psi}^a_q \chi^{\dot{A}}_p \rangle,\\
    \mathbf{S}|T^{\dot{A}a}_p \tilde{Y}_q\rangle &= +A_{pq}^{\z\R} D_{pq}^{\z\R} |\tilde{Y}_q T^{\dot{A}a}_p \rangle - B_{pq}^{\z\R} D_{pq}^{\z\R} |\tilde{\Psi}^a_q \tilde{\chi}^{\dot{A}}_p \rangle,\\
    \mathbf{S}|\chi^{\dot{A}}_p \tilde{\Psi}^a_q\rangle &= - A_{pq}^{\z\R} C_{pq}^{\z\R} |\tilde{\Psi}^a_q \chi^{\dot{A}}_p \rangle + B_{pq}^{\z\R} C_{pq}^{\z\R} |\tilde{Z}_q T^{\dot{A}a}_p \rangle,\\
    \mathbf{S}|\tilde{\chi}^{\dot{A}}_p \tilde{\Psi}^a_q\rangle &= +D_{pq}^{\z\R} E_{pq}^{\z\R} |\tilde{\Psi}^a_q \tilde{\chi}^{\dot{A}}_p \rangle - D_{pq}^{\z\R} F_{pq}^{\z\R} |\tilde{Y}_q T^{\dot{A}a}_p \rangle,\\
    \mathbf{S}|\chi^{\dot{A}}_p \tilde{Z}_q\rangle &= +C_{pq}^{\z\R} C_{pq}^{\z\R} |\tilde{Z}_q \chi^{\dot{A}}_p \rangle,\\
    \mathbf{S}|\tilde{\chi}^{\dot{A}}_p \tilde{Y}_q\rangle &= +D_{pq}^{\z\R} D_{pq}^{\z\R} |\tilde{Y}_q \tilde{\chi}^{\dot{A}}_p \rangle,\\
    \mathbf{S}|\tilde{\chi}^{\dot{A}}_p \tilde{Z}_q\rangle &= +E_{pq}^{\z\R} E_{pq}^{\z\R} |\tilde{Z}_q \tilde{\chi}^{\dot{A}}_p \rangle -F_{pq}^{\z\R} F_{pq}^{\z\R} |\tilde{Y}_q \chi^{\dot{A}}_p \rangle +F_{pq}^{\z\R} E_{pq}^{\z\R} \epsilon_{ab} |\tilde{\Psi}^a_q T^{\dot{A}b}_p \rangle,\\
    \mathbf{S}|\chi^{\dot{A}}_p \tilde{Y}_q\rangle &= +A_{pq}^{\z\R} A_{pq}^{\z\R} |\tilde{Y}_q \chi^{\dot{A}}_p \rangle -B_{pq}^{\z\R} B_{pq}^{\z\R} |\tilde{Z}_q \tilde{\chi}^{\dot{A}}_p \rangle - A_{pq}^{\z\R} B_{pq}^{\z\R} \epsilon_{ab} |\tilde{\Psi}^a_q T^{\dot{A}b}_p \rangle,\\
    \mathbf{S}|T^{\dot{A}a}_p \tilde{\Psi}^b_q\rangle &= +C_{pq}^{\z\R} D_{pq}^{\z\R} \delta^{ab} |\tilde{\Psi}^a_q T^{\dot{A}b}_p \rangle + \epsilon^{ab}(A_{pq}^{\z\R} F_{pq}^{\z\R} |\tilde{Y}_q \chi^{\dot{A}}_p \rangle - B_{pq}^{\z\R} E_{pq}^{\z\R}  |\tilde{Z}_q \tilde{\chi}^{\dot{A}}_p \rangle )  \\
    &\quad + (1-\delta^{ab}) (B_{pq}^{\z\R} F_{pq}^{\z\R} |\tilde{\Psi}^a_q T^{\dot{A}b}_p \rangle -A_{pq}^{\z\R} E_{pq}^{\z\R} |\tilde{\Psi}^b_q T^{\dot{A}a}_p \rangle).
\end{aligned}
\end{equation}
The last process can be further simplified by using the identity 
\begin{equation}
    B_{pq}^{\z\R} F_{pq}^{\z\R} - A_{pq}^{\z\R} E_{pq}^{\z\R} = C_{pq}^{\z\R} D_{pq}^{\z\R}.
\end{equation}

\subsection{The massless sector}
\begin{equation}
\begin{aligned}
    \mathbf{S}|T^{\dot{A}a}_p T^{\dot{B}b}_q\rangle &= -A_{pq}^{\z\z} F_{pq}^{\z\z} \delta^{ab} |T^{\dot{B}a}_q T^{\dot{A}b}_p \rangle + \epsilon^{ab}(C_{pq}^{\z\z} D_{pq}^{\z\z} |\chi^{\dot{B}}_q \tilde{\chi}^{\dot{A}}_p \rangle + B_{pq}^{\z\z} E_{pq}^{\z\z} |\tilde{\chi}^{\dot{B}}_q \chi^{\dot{A}}_p \rangle) \\
    &\quad -(1-\delta^{ab}) (C_{pq}^{\z\z} E_{pq}^{\z\z} |T^{\dot{B}a}_q T^{\dot{A}b}_p \rangle - B_{pq}^{\z\z} D_{pq}^{\z\z} |T^{\dot{B}b}_q T^{\dot{A}a}_p \rangle ),\\
    \mathbf{S}|T^{\dot{A}a}_p \tilde{\chi}^{\dot{B}}_q\rangle &= -B_{pq}^{\z\z} F_{pq}^{\z\z} |\tilde{\chi}^{\dot{B}}_q T^{\dot{A}a}_p \rangle - C_{pq}^{\z\z} F_{pq}^{\z\z} |T^{\dot{B}a}_q \tilde{\chi}^{\dot{A}}_p \rangle,\\
    \mathbf{S}|\tilde{\chi}^{\dot{A}}_p T^{\dot{B}a}_q\rangle &= - D_{pq}^{\z\z} F_{pq}^{\z\z} |T^{\dot{B}a}_q \tilde{\chi}^{\dot{A}}_p \rangle - E_{pq}^{\z\z} F_{pq}^{\z\z} |\tilde{\chi}^{\dot{B}}_p T^{\dot{A}a}_p  \rangle,\\
    \mathbf{S}|T^{\dot{A}a}_p \chi^{\dot{B}}_q\rangle &= +A_{pq}^{\z\z} D_{pq}^{\z\z} |\chi^{\dot{B}}_q T^{\dot{A}a}_p \rangle - A_{pq}^{\z\z} E_{pq}^{\z\z} |T^{\dot{B}a}_q \chi^{\dot{A}}_p \rangle,\\
    \mathbf{S}|\chi^{\dot{A}}_p T^{\dot{B}a}_q\rangle &= +A_{pq}^{\z\z} B_{pq}^{\z\z} |T^{\dot{B}a}_q \chi^{\dot{A}}_p \rangle - A_{pq}^{\z\z} C_{pq}^{\z\z} |\chi^{\dot{B}}_p T^{\dot{A}a}_p  \rangle,\\
    \mathbf{S}|\tilde{\chi}^{\dot{A}}_p \tilde{\chi}^{\dot{B}}_q\rangle &= -F_{pq}^{\z\z} F_{pq}^{\z\z} |\tilde{\chi}^{\dot{B}}_q \tilde{\chi}^{\dot{A}}_p \rangle,\\
    \mathbf{S}|\chi^{\dot{A}}_p \chi^{\dot{B}}_q\rangle &= -A_{pq}^{\z\z} A_{pq}^{\z\z} |\chi^{\dot{B}}_q \chi^{\dot{A}}_p \rangle,\\
    \mathbf{S}|\tilde{\chi}^{\dot{A}}_p \chi^{\dot{B}}_q\rangle &= -D_{pq}^{\z\z} D_{pq}^{\z\z} |\chi^{\dot{B}}_q \tilde{\chi}^{\dot{A}}_p \rangle - E_{pq}^{\z\z} E_{pq}^{\z\z} |\tilde{\chi}^{\dot{B}}_q \chi^{\dot{A}}_p \rangle - E_{pq}^{\z\z} D_{pq}^{\z\z} \epsilon_{ab} |T^{\dot{B}a}_q T^{\dot{A}b}_p \rangle,\\
    \mathbf{S}|\chi^{\dot{A}}_p \tilde{\chi}^{\dot{B}}_q\rangle &= -B_{pq}^{\z\z} B_{pq}^{\z\z} |\tilde{\chi}^{\dot{B}}_q \chi^{\dot{A}}_p \rangle - C_{pq}^{\z\z} C_{pq}^{\z\z} |\chi^{\dot{B}}_q \tilde{\chi}^{\dot{A}}_p \rangle - B_{pq}^{\z\z} C_{pq}^{\z\z} \epsilon_{ab} |T^{\dot{B}a}_q T^{\dot{A}b}_p \rangle.
\end{aligned}
\end{equation}
The first process can be further simplified by using the identity 
\begin{equation}
    C_{pq}^{\z\z} E_{pq}^{\z\z} - B_{pq}^{\z\z} D_{pq}^{\z\z} = A_{pq}^{\z\z} F_{pq}^{\z\z}.
\end{equation}

\bibliography{refs}

\makeatletter \@ifundefined{Sphere}{\newcommand{\Sphere}{\text{S}}}{}
  \@ifundefined{AdS}{\newcommand{\AdS}{\text{AdS}}}{}
  \@ifundefined{CFT}{\newcommand{\CFT}{\text{CFT}}}{}
  \@ifundefined{CP}{\newcommand{\CP}{\text{CP}}}{}
  \@ifundefined{Torus}{\newcommand{\Torus}{\text{T}}}{}
  \@ifundefined{superN}{\newcommand{\superN}{\mathcal{N}}}{}
  \@ifundefined{grpOSp}{\newcommand{\grpOSp}{\mathrm{OSp}}}{}
  \@ifundefined{grpPSU}{\newcommand{\grpPSU}{\mathrm{PSU}}}{}
  \@ifundefined{grpSU}{\newcommand{\grpSU}{\mathrm{SU}}}{}
  \@ifundefined{grpU}{\newcommand{\grpU}{\mathrm{U}}}{}
  \@ifundefined{grpD}{\newcommand{\grpD}{\mathrm{D}}}{}
  \@ifundefined{grpSL}{\newcommand{\grpSL}{\mathrm{SL}}}{}
  \@ifundefined{grpSp}{\newcommand{\grpSp}{\mathrm{Sp}}}{}
  \@ifundefined{grpUSp}{\newcommand{\grpUSp}{\mathrm{USp}}}{}
  \@ifundefined{grpSO}{\newcommand{\grpSO}{\mathrm{SO}}}{}
  \@ifundefined{grpO}{\newcommand{\grpO}{\mathrm{O}}}{}
  \@ifundefined{algOSp}{\newcommand{\algOSp}{\mathfrak{osp}}}{}
  \@ifundefined{algPSU}{\newcommand{\algPSU}{\mathfrak{psu}}}{}
  \@ifundefined{algSU}{\newcommand{\algSU}{\mathfrak{su}}}{}
  \@ifundefined{algSp}{\newcommand{\algSp}{\mathfrak{sp}}}{}
  \@ifundefined{algSL}{\newcommand{\algSL}{\mathfrak{sl}}}{}
  \@ifundefined{algGL}{\newcommand{\algGL}{\mathfrak{gl}}}{}
  \@ifundefined{algU}{\newcommand{\algU}{\mathfrak{u}}}{}
  \@ifundefined{algSO}{\newcommand{\algSO}{\mathfrak{so}}}{}
  \@ifundefined{algO}{\newcommand{\algO}{\mathfrak{o}}}{}
  \@ifundefined{Integers}{\newcommand{\Integers}{\mathbb{Z}}}{}
  \@ifundefined{Reals}{\newcommand{\Reals}{\mathbb{R}}}{} \makeatother

\providecommand{\href}[2]{#2}\begingroup\raggedright\begin{thebibliography}{10}

\bibitem{Maldacena:1997re}
J.~M. Maldacena, \emph{The large {N} limit of superconformal field theories and
  supergravity}, {\emph{Adv. Theor. Math. Phys.} {\bfseries 2} (1998) 231}
  [\href{https://arxiv.org/abs/hep-th/9711200}{{\ttfamily hep-th/9711200}}].

\bibitem{Witten:1998qj}
E.~Witten, \emph{Anti-de {S}itter space and holography}, {\emph{Adv. Theor.
  Math. Phys.} {\bfseries 2} (1998) 253}
  [\href{https://arxiv.org/abs/hep-th/9802150}{{\ttfamily hep-th/9802150}}].

\bibitem{Gubser:1998bc}
S.~S. Gubser, I.~R. Klebanov and A.~M. Polyakov, \emph{Gauge theory correlators
  from non-critical string theory},
  \href{https://doi.org/10.1016/S0370-2693(98)00377-3}{\emph{Phys. Lett.}
  {\bfseries B428} (1998) 105}
  [\href{https://arxiv.org/abs/hep-th/9802109}{{\ttfamily hep-th/9802109}}].

\bibitem{tHooft:1973alw}
G.~'t~Hooft, \emph{{A Planar Diagram Theory for Strong Interactions}},
  \href{https://doi.org/10.1016/0550-3213(74)90154-0}{\emph{Nucl. Phys. B}
  {\bfseries 72} (1974) 461}.

\bibitem{Minahan:2002ve}
J.~A. Minahan and K.~Zarembo, \emph{The {B}ethe-ansatz for {$\superN = 4$}
  super {Y}ang-{M}ills}, {\emph{JHEP} {\bfseries 0303} (2003) 013}
  [\href{https://arxiv.org/abs/hep-th/0212208}{{\ttfamily hep-th/0212208}}].

\bibitem{Bethe:1931hc}
H.~Bethe, \emph{On the theory of metals. 1. {E}igenvalues and eigenfunctions
  for the linear atomic chain}, {\emph{Z. Phys.} {\bfseries 71} (1931) 205}.

\bibitem{Faddeev:1996iy}
L.~D. Faddeev, \emph{How algebraic {B}ethe ansatz works for integrable model},
  \href{https://arxiv.org/abs/hep-th/9605187}{{\ttfamily hep-th/9605187}}.

\bibitem{Ambjorn:2005wa}
J.~Ambj{\o}rn, R.~A. Janik and C.~Kristjansen, \emph{Wrapping interactions and
  a new source of corrections to the spin-chain/string duality},
  \href{https://doi.org/10.1016/j.nuclphysb.2005.12.007}{\emph{Nucl. Phys.}
  {\bfseries B736} (2006) 288}
  [\href{https://arxiv.org/abs/hep-th/0510171}{{\ttfamily hep-th/0510171}}].

\bibitem{Arutyunov:2007tc}
G.~Arutyunov and S.~Frolov, \emph{On string {S}-matrix, bound states and
  {TBA}}, \href{https://doi.org/10.1088/1126-6708/2007/12/024}{\emph{JHEP}
  {\bfseries 0712} (2007) 024}
  [\href{https://arxiv.org/abs/0710.1568}{{\ttfamily 0710.1568}}].

\bibitem{Gromov:2013pga}
N.~Gromov, V.~Kazakov, S.~Leurent and D.~Volin, \emph{Quantum spectral curve
  for {AdS$_{5}$/CFT$_{4}$}},
  \href{https://doi.org/10.1103/PhysRevLett.112.011602}{\emph{Phys. Rev. Lett.}
  {\bfseries 112} (2014) 011602}
  [\href{https://arxiv.org/abs/1305.1939}{{\ttfamily 1305.1939}}].

\bibitem{Arutyunov:2009ga}
G.~Arutyunov and S.~Frolov, \emph{Foundations of the {$\AdS_{5} \times
  \Sphere^5$} superstring. part {I}},
  \href{https://doi.org/10.1088/1751-8113/42/25/254003}{\emph{J. Phys. A}
  {\bfseries A42} (2009) 254003}
  [\href{https://arxiv.org/abs/0901.4937}{{\ttfamily 0901.4937}}].

\bibitem{Beisert:2010jr}
N.~Beisert, C.~Ahn, L.~F. Alday, Z.~Bajnok, J.~M. Drummond, L.~Freyhult et~al.,
  \emph{{Review of AdS/CFT Integrability: An Overview}},
  \href{https://doi.org/10.1007/s11005-011-0529-2}{\emph{Lett. Math. Phys.}
  {\bfseries 99} (2012) 3} [\href{https://arxiv.org/abs/1012.3982}{{\ttfamily
  1012.3982}}].

\bibitem{Basso:2015zoa}
B.~Basso, S.~Komatsu and P.~Vieira, \emph{Structure constants and integrable
  bootstrap in planar {$\superN = 4$} {SYM} theory},
  \href{https://arxiv.org/abs/1505.06745}{{\ttfamily 1505.06745}}.

\bibitem{Beisert:2005tm}
N.~Beisert, \emph{The {$\algSU(2|2)$} dynamic {$S$}-matrix}, {\emph{Adv. Theor.
  Math. Phys.} {\bfseries 12} (2008) 945}
  [\href{https://arxiv.org/abs/hep-th/0511082}{{\ttfamily hep-th/0511082}}].

\bibitem{Luscher:1985dn}
M.~L{\"u}scher, \emph{Volume dependence of the energy spectrum in massive
  quantum field theories. 1. {S}table particle states},
  \href{https://doi.org/10.1007/BF01211589}{\emph{Commun. Math. Phys.}
  {\bfseries 104} (1986) 177}.

\bibitem{Luscher:1986pf}
M.~L{\"u}scher, \emph{Volume dependence of the energy spectrum in massive
  quantum field theories. 2. {S}cattering states},
  \href{https://doi.org/10.1007/BF01211097}{\emph{Commun. Math. Phys.}
  {\bfseries 105} (1986) 153}.

\bibitem{Eden:2015ija}
B.~Eden and A.~Sfondrini, \emph{{Three-point functions in ${\cal N}=4$ SYM: the
  hexagon proposal at three loops}},
  \href{https://doi.org/10.1007/JHEP02(2016)165}{\emph{JHEP} {\bfseries 02}
  (2016) 165} [\href{https://arxiv.org/abs/1510.01242}{{\ttfamily
  1510.01242}}].

\bibitem{Basso:2015eqa}
B.~Basso, V.~Goncalves, S.~Komatsu and P.~Vieira, \emph{{Gluing Hexagons at
  Three Loops}},
  \href{https://doi.org/10.1016/j.nuclphysb.2016.04.020}{\emph{Nucl. Phys.}
  {\bfseries B907} (2016) 695}
  [\href{https://arxiv.org/abs/1510.01683}{{\ttfamily 1510.01683}}].

\bibitem{Basso:2017muf}
B.~Basso, V.~Goncalves and S.~Komatsu, \emph{{Structure constants at wrapping
  order}},  \href{https://arxiv.org/abs/1702.02154}{{\ttfamily 1702.02154}}.

\bibitem{Eden:2016xvg}
B.~Eden and A.~Sfondrini, \emph{{Tessellating cushions: four-point functions in
  $\mathcal{N} $ = 4 SYM}},
  \href{https://doi.org/10.1007/JHEP10(2017)098}{\emph{JHEP} {\bfseries 10}
  (2017) 098} [\href{https://arxiv.org/abs/1611.05436}{{\ttfamily
  1611.05436}}].

\bibitem{Fleury:2016ykk}
T.~Fleury and S.~Komatsu, \emph{{Hexagonalization of Correlation Functions}},
  \href{https://doi.org/10.1007/JHEP01(2017)130}{\emph{JHEP} {\bfseries 01}
  (2017) 130} [\href{https://arxiv.org/abs/1611.05577}{{\ttfamily
  1611.05577}}].

\bibitem{Eden:2017ozn}
B.~Eden, Y.~Jiang, D.~le~Plat and A.~Sfondrini, \emph{{Colour-dressed hexagon
  tessellations for correlation functions and non-planar corrections}},
  \href{https://doi.org/10.1007/JHEP02(2018)170}{\emph{JHEP} {\bfseries 02}
  (2018) 170} [\href{https://arxiv.org/abs/1710.10212}{{\ttfamily
  1710.10212}}].

\bibitem{Bargheer:2017nne}
T.~Bargheer, J.~Caetano, T.~Fleury, S.~Komatsu and P.~Vieira, \emph{{Handling
  Handles I: Nonplanar Integrability}},
  \href{https://arxiv.org/abs/1711.05326}{{\ttfamily 1711.05326}}.

\bibitem{Bargheer:2019kxb}
T.~Bargheer, F.~Coronado and P.~Vieira, \emph{{Octagons I: Combinatorics and
  Non-Planar Resummations}},
  \href{https://doi.org/10.1007/JHEP08(2019)162}{\emph{JHEP} {\bfseries 08}
  (2019) 162} [\href{https://arxiv.org/abs/1904.00965}{{\ttfamily
  1904.00965}}].

\bibitem{OhlssonSax:2018hgc}
O.~Ohlsson~Sax and B.~Stefa{\'n}ski, \emph{{Closed strings and moduli in
  AdS$_{3}$/CFT$_{2}$}},
  \href{https://doi.org/10.1007/JHEP05(2018)101}{\emph{JHEP} {\bfseries 05}
  (2018) 101} [\href{https://arxiv.org/abs/1804.02023}{{\ttfamily
  1804.02023}}].

\bibitem{Maldacena:2000hw}
J.~M. Maldacena and H.~Ooguri, \emph{Strings in {$\AdS_{3}$} and
  {$\grpSL(2,R)$} {WZW} model. {I}},
  \href{https://doi.org/10.1063/1.1377273}{\emph{J. Math. Phys.} {\bfseries 42}
  (2001) 2929} [\href{https://arxiv.org/abs/hep-th/0001053}{{\ttfamily
  hep-th/0001053}}].

\bibitem{Maldacena:2000kv}
J.~M. Maldacena, H.~Ooguri and J.~Son, \emph{Strings in {$\AdS_{3}$} and the
  {$\grpSL(2,R)$} {WZW} model. {II}: {E}uclidean black hole},
  \href{https://doi.org/10.1063/1.1377039}{\emph{J. Math. Phys.} {\bfseries 42}
  (2001) 2961} [\href{https://arxiv.org/abs/hep-th/0005183}{{\ttfamily
  hep-th/0005183}}].

\bibitem{Maldacena:2001km}
J.~M. Maldacena and H.~Ooguri, \emph{Strings in {$\AdS_{3}$} and the
  {$\grpSL(2,R)$} {WZW} model.~{III}: Correlation functions},
  \href{https://doi.org/10.1103/PhysRevD.65.106006}{\emph{Phys. Rev.}
  {\bfseries D65} (2002) 106006}
  [\href{https://arxiv.org/abs/hep-th/0111180}{{\ttfamily hep-th/0111180}}].

\bibitem{Berkovits:1999im}
N.~Berkovits, C.~Vafa and E.~Witten, \emph{Conformal field theory of {AdS}
  background with {R}amond-{R}amond flux}, {\emph{JHEP} {\bfseries 9903} (1999)
  018} [\href{https://arxiv.org/abs/hep-th/9902098}{{\ttfamily
  hep-th/9902098}}].

\bibitem{Giribet:2018ada}
G.~Giribet, C.~Hull, M.~Kleban, M.~Porrati and E.~Rabinovici,
  \emph{{Superstrings on AdS$_{3}$ at $k = 1$}},
  \href{https://doi.org/10.1007/JHEP08(2018)204}{\emph{JHEP} {\bfseries 08}
  (2018) 204} [\href{https://arxiv.org/abs/1803.04420}{{\ttfamily
  1803.04420}}].

\bibitem{Gaberdiel:2018rqv}
M.~R. Gaberdiel and R.~Gopakumar, \emph{{Tensionless string spectra on
  AdS$_{3}$}}, \href{https://doi.org/10.1007/JHEP05(2018)085}{\emph{JHEP}
  {\bfseries 05} (2018) 085}
  [\href{https://arxiv.org/abs/1803.04423}{{\ttfamily 1803.04423}}].

\bibitem{Eberhardt:2018ouy}
L.~Eberhardt, M.~R. Gaberdiel and R.~Gopakumar, \emph{{The Worldsheet Dual of
  the Symmetric Product CFT}},
  \href{https://doi.org/10.1007/JHEP04(2019)103}{\emph{JHEP} {\bfseries 04}
  (2019) 103} [\href{https://arxiv.org/abs/1812.01007}{{\ttfamily
  1812.01007}}].

\bibitem{Eberhardt:2020bgq}
L.~Eberhardt, \emph{{Partition functions of the tensionless string}},
  \href{https://doi.org/10.1007/JHEP03(2021)176}{\emph{JHEP} {\bfseries 03}
  (2021) 176} [\href{https://arxiv.org/abs/2008.07533}{{\ttfamily
  2008.07533}}].

\bibitem{Gaberdiel:2020ycd}
M.~R. Gaberdiel, R.~Gopakumar, B.~Knighton and P.~Maity, \emph{{From symmetric
  product CFTs to AdS$_{3}$}},
  \href{https://doi.org/10.1007/JHEP05(2021)073}{\emph{JHEP} {\bfseries 05}
  (2021) 073} [\href{https://arxiv.org/abs/2011.10038}{{\ttfamily
  2011.10038}}].

\bibitem{Babichenko:2009dk}
A.~Babichenko, B.~Stefa{\'n}ski, jr. and K.~Zarembo, \emph{Integrability and
  the {$\AdS_{3}/\CFT_{2}$} correspondence},
  \href{https://doi.org/10.1007/JHEP03(2010)058}{\emph{JHEP} {\bfseries 1003}
  (2010) 058} [\href{https://arxiv.org/abs/0912.1723}{{\ttfamily 0912.1723}}].

\bibitem{Cagnazzo:2012se}
A.~Cagnazzo and K.~Zarembo, \emph{{B}-field in {$\AdS_{3}/\CFT_{2}$}
  correspondence and integrability},
  \href{https://doi.org/10.1007/JHEP11(2012)133,
  10.1007/JHEP04(2013)003}{\emph{JHEP} {\bfseries 1211} (2012) 133}
  [\href{https://arxiv.org/abs/1209.4049}{{\ttfamily 1209.4049}}].

\bibitem{Sfondrini:2014via}
A.~Sfondrini, \emph{Towards integrability for {$\AdS_{3}/\CFT_{2}$}},
  \href{https://doi.org/10.1088/1751-8113/48/2/023001}{\emph{J. Phys.}
  {\bfseries A48} (2015) 023001}
  [\href{https://arxiv.org/abs/1406.2971}{{\ttfamily 1406.2971}}].

\bibitem{Hoare:2013pma}
B.~Hoare and A.~A. Tseytlin, \emph{On string theory on {$\AdS_{3} \times
  \Sphere^3 \times \Torus^4$} with mixed 3-form flux: tree-level {S}-matrix},
  \href{https://doi.org/10.1016/j.nuclphysb.2013.05.005}{\emph{Nucl. Phys.}
  {\bfseries B873} (2013) 682}
  [\href{https://arxiv.org/abs/1303.1037}{{\ttfamily 1303.1037}}].

\bibitem{Hoare:2013lja}
B.~Hoare, A.~Stepanchuk and A.~Tseytlin, \emph{Giant magnon solution and
  dispersion relation in string theory in {$\AdS_{3} \times \Sphere^3 \times
  \Torus^4$} with mixed flux},
  \href{https://doi.org/10.1016/j.nuclphysb.2013.12.011}{\emph{Nucl. Phys.}
  {\bfseries B879} (2014) 318}
  [\href{https://arxiv.org/abs/1311.1794}{{\ttfamily 1311.1794}}].

\bibitem{Lloyd:2014bsa}
T.~Lloyd, O.~Ohlsson~Sax, A.~Sfondrini and B.~Stefa{\'n}ski, jr., \emph{The
  complete worldsheet {S} matrix of superstrings on {$\AdS_{3} \times \Sphere^3
  \times \Torus^4$} with mixed three-form flux},
  \href{https://doi.org/10.1016/j.nuclphysb.2014.12.019}{\emph{Nucl. Phys.}
  {\bfseries B891} (2015) 570}
  [\href{https://arxiv.org/abs/1410.0866}{{\ttfamily 1410.0866}}].

\bibitem{Abbott:2015pps}
M.~C. Abbott and I.~Aniceto, \emph{{Massless L\"uscher terms and the
  limitations of the $\AdS_{3}$ asymptotic Bethe ansatz}},
  \href{https://doi.org/10.1103/PhysRevD.93.106006}{\emph{Phys. Rev.}
  {\bfseries D93} (2016) 106006}
  [\href{https://arxiv.org/abs/1512.08761}{{\ttfamily 1512.08761}}].

\bibitem{Baggio:2018gct}
M.~Baggio and A.~Sfondrini, \emph{{Strings on NS-NS Backgrounds as Integrable
  Deformations}}, \href{https://doi.org/10.1103/PhysRevD.98.021902}{\emph{Phys.
  Rev.} {\bfseries D98} (2018) 021902}
  [\href{https://arxiv.org/abs/1804.01998}{{\ttfamily 1804.01998}}].

\bibitem{Dei:2018mfl}
A.~Dei and A.~Sfondrini, \emph{{Integrable spin chain for stringy
  Wess-Zumino-Witten models}},
  \href{https://doi.org/10.1007/JHEP07(2018)109}{\emph{JHEP} {\bfseries 07}
  (2018) 109} [\href{https://arxiv.org/abs/1806.00422}{{\ttfamily
  1806.00422}}].

\bibitem{Sfondrini:2020ovj}
A.~Sfondrini, \emph{{Long Strings and Symmetric Product Orbifold from the
  AdS$_3$ Bethe Equations}},
  \href{https://arxiv.org/abs/2010.02782}{{\ttfamily 2010.02782}}.

\bibitem{Eberhardt:2019ywk}
L.~Eberhardt, M.~R. Gaberdiel and R.~Gopakumar, \emph{{Deriving the
  AdS$_{3}$/CFT$_{2}$ correspondence}},
  \href{https://doi.org/10.1007/JHEP02(2020)136}{\emph{JHEP} {\bfseries 02}
  (2020) 136} [\href{https://arxiv.org/abs/1911.00378}{{\ttfamily
  1911.00378}}].

\bibitem{Larsen:1999uk}
F.~Larsen and E.~J. Martinec, \emph{{$\grpU(1)$} charges and moduli in the
  {D1}-{D5} system}, {\emph{JHEP} {\bfseries 9906} (1999) 019}
  [\href{https://arxiv.org/abs/hep-th/9905064}{{\ttfamily hep-th/9905064}}].

\bibitem{Borsato:2013qpa}
R.~Borsato, O.~Ohlsson~Sax, A.~Sfondrini, B.~Stefa{\'n}ski, jr. and
  A.~Torrielli, \emph{The all-loop integrable spin-chain for strings on
  {$\AdS_{3} \times \Sphere^3 \times \Torus^4$}: the massive sector},
  \href{https://doi.org/10.1007/JHEP08(2013)043}{\emph{JHEP} {\bfseries 1308}
  (2013) 043} [\href{https://arxiv.org/abs/1303.5995}{{\ttfamily 1303.5995}}].

\bibitem{Borsato:2014exa}
R.~Borsato, O.~Ohlsson~Sax, A.~Sfondrini and B.~Stefa{\'n}ski, jr.,
  \emph{Towards the all-loop worldsheet {S} matrix for {$\AdS_{3} \times
  \Sphere^3 \times \Torus^4$}},
  \href{https://doi.org/10.1103/PhysRevLett.113.131601}{\emph{Phys. Rev. Lett.}
  {\bfseries 113} (2014) 131601}
  [\href{https://arxiv.org/abs/1403.4543}{{\ttfamily 1403.4543}}].

\bibitem{Borsato:2014hja}
R.~Borsato, O.~Ohlsson~Sax, A.~Sfondrini and B.~Stefa{\'n}ski, jr, \emph{The
  complete {$\AdS_{3} \times \Sphere^3 \times \Torus^4$} worldsheet
  {S}-matrix}, \href{https://doi.org/10.1007/JHEP10(2014)066}{\emph{JHEP}
  {\bfseries 1410} (2014) 66}
  [\href{https://arxiv.org/abs/1406.0453}{{\ttfamily 1406.0453}}].

\bibitem{Borsato:2013hoa}
R.~Borsato, O.~Ohlsson~Sax, A.~Sfondrini, B.~Stefa{\'n}ski, jr. and
  A.~Torrielli, \emph{Dressing phases of {$\AdS_{3}/\CFT_{2}$}},
  \href{https://doi.org/10.1103/PhysRevD.88.066004}{\emph{Phys. Rev.}
  {\bfseries D88} (2013) 066004}
  [\href{https://arxiv.org/abs/1306.2512}{{\ttfamily 1306.2512}}].

\bibitem{Borsato:2016xns}
R.~Borsato, O.~Ohlsson~Sax, A.~Sfondrini, B.~Stefa{\'n}ski, jr. and
  A.~Torrielli, \emph{{On the dressing factors, Bethe equations and Yangian
  symmetry of strings on {$\AdS_{3} \times \Sphere^3 \times \Torus^4$}}},
  \href{https://doi.org/10.1088/1751-8121/50/2/024004}{\emph{J. Phys.}
  {\bfseries A50} (2017) 024004}
  [\href{https://arxiv.org/abs/1607.00914}{{\ttfamily 1607.00914}}].

\bibitem{Fontanella:2019baq}
A.~Fontanella and A.~Torrielli, \emph{{Geometry of Massless Scattering in
  Integrable Superstring}},
  \href{https://doi.org/10.1007/JHEP06(2019)116}{\emph{JHEP} {\bfseries 06}
  (2019) 116} [\href{https://arxiv.org/abs/1903.10759}{{\ttfamily
  1903.10759}}].

\bibitem{Babichenko:2014yaa}
A.~Babichenko, A.~Dekel and O.~Ohlsson~Sax, \emph{Finite-gap equations for
  strings on {$\AdS_{3} \times \Sphere^3 \times \Torus^4$} with mixed 3-form
  flux}, \href{https://doi.org/10.1007/JHEP11(2014)122}{\emph{JHEP} {\bfseries
  1411} (2014) 122} [\href{https://arxiv.org/abs/1405.6087}{{\ttfamily
  1405.6087}}].

\bibitem{Arutyunov:2006ak}
G.~Arutyunov, S.~Frolov, J.~Plefka and M.~Zamaklar, \emph{The off-shell
  symmetry algebra of the light-cone {$\AdS_{5} \times \Sphere^5$}
  superstring}, \href{https://doi.org/10.1088/1751-8113/40/13/018}{\emph{J.
  Phys.} {\bfseries A40} (2007) 3583}
  [\href{https://arxiv.org/abs/hep-th/0609157}{{\ttfamily hep-th/0609157}}].

\bibitem{Borsato:2012ud}
R.~Borsato, O.~Ohlsson~Sax and A.~Sfondrini, \emph{A dynamic {$\algSU(1|1)^2$}
  {S}-matrix for {$\AdS_{3}/\CFT_{2}$}},
  \href{https://doi.org/10.1007/JHEP04(2013)113}{\emph{JHEP} {\bfseries 1304}
  (2013) 113} [\href{https://arxiv.org/abs/1211.5119}{{\ttfamily 1211.5119}}].

\bibitem{Borsato:2012ss}
R.~Borsato, O.~Ohlsson~Sax and A.~Sfondrini, \emph{All-loop {B}ethe ansatz
  equations for {$\AdS_{3}/\CFT_{2}$}},
  \href{https://doi.org/10.1007/JHEP04(2013)116}{\emph{JHEP} {\bfseries 1304}
  (2013) 116} [\href{https://arxiv.org/abs/1212.0505}{{\ttfamily 1212.0505}}].

\bibitem{Drukker:2008pi}
N.~Drukker and J.~Plefka, \emph{{The Structure of n-point functions of chiral
  primary operators in N=4 super Yang-Mills at one-loop}},
  \href{https://doi.org/10.1088/1126-6708/2009/04/001}{\emph{JHEP} {\bfseries
  04} (2009) 001} [\href{https://arxiv.org/abs/0812.3341}{{\ttfamily
  0812.3341}}].

\bibitem{Baggio:2012rr}
M.~Baggio, J.~de~Boer and K.~Papadodimas, \emph{A non-renormalization theorem
  for chiral primary 3-point functions},
  \href{https://doi.org/10.1007/JHEP07(2012)137}{\emph{JHEP} {\bfseries 07}
  (2012) 137} [\href{https://arxiv.org/abs/1203.1036}{{\ttfamily 1203.1036}}].

\bibitem{Pakman:2007hn}
A.~Pakman and A.~Sever, \emph{Exact {$\superN = 4$} correlators of
  {$\AdS_{3}/\CFT_{2}$}},
  \href{https://doi.org/10.1016/j.physletb.2007.06.041}{\emph{Phys. Lett.}
  {\bfseries B652} (2007) 60}
  [\href{https://arxiv.org/abs/0704.3040}{{\ttfamily 0704.3040}}].

\bibitem{Baggio:2017kza}
M.~Baggio, O.~Ohlsson~Sax, A.~Sfondrini, B.~Stefa{\'n}ski, jr. and
  A.~Torrielli, \emph{{Protected string spectrum in AdS$_{3}$/CFT$_{2}$ from
  worldsheet integrability}},
  \href{https://doi.org/10.1007/JHEP04(2017)091}{\emph{JHEP} {\bfseries 04}
  (2017) 091} [\href{https://arxiv.org/abs/1701.03501}{{\ttfamily
  1701.03501}}].

\bibitem{Gaberdiel:2007vu}
M.~R. Gaberdiel and I.~Kirsch, \emph{Worldsheet correlators in
  {$\AdS_{3}/\CFT_{2}$}},
  \href{https://doi.org/10.1088/1126-6708/2007/04/050}{\emph{JHEP} {\bfseries
  0704} (2007) 050} [\href{https://arxiv.org/abs/hep-th/0703001}{{\ttfamily
  hep-th/0703001}}].

\bibitem{Dabholkar:2007ey}
A.~Dabholkar and A.~Pakman, \emph{Exact chiral ring of {$\AdS_{3}/\CFT_{2}$}},
  \href{https://doi.org/10.4310/ATMP.2009.v13.n2.a2}{\emph{Adv. Theor. Math.
  Phys.} {\bfseries 13} (2009) 409}
  [\href{https://arxiv.org/abs/hep-th/0703022}{{\ttfamily hep-th/0703022}}].

\bibitem{Caetano:2016keh}
J.~Caetano and T.~Fleury, \emph{{Fermionic Correlators from Integrability}},
  \href{https://doi.org/10.1007/JHEP09(2016)010}{\emph{JHEP} {\bfseries 09}
  (2016) 010} [\href{https://arxiv.org/abs/1607.02542}{{\ttfamily
  1607.02542}}].

\bibitem{Baggio:2015vxa}
M.~Baggio, V.~Niarchos and K.~Papadodimas, \emph{{On exact correlation
  functions in SU(N) $ \mathcal{N}=2 $ superconformal QCD}},
  \href{https://doi.org/10.1007/JHEP11(2015)198}{\emph{JHEP} {\bfseries 11}
  (2015) 198} [\href{https://arxiv.org/abs/1508.03077}{{\ttfamily
  1508.03077}}].

\bibitem{Baggio:2018rpv}
M.~Baggio, A.~Sfondrini, G.~Tartaglino-Mazzucchelli and H.~Walsh, \emph{{On
  $T\bar{T}$ deformations and supersymmetry}},
  \href{https://arxiv.org/abs/1811.00533}{{\ttfamily 1811.00533}}.

\bibitem{Borsato:2016kbm}
R.~Borsato, O.~Ohlsson~Sax, A.~Sfondrini and B.~Stefa{\'n}ski, \emph{{On the
  spectrum of {$\AdS_{3} \times \Sphere^3 \times \Torus^4$} strings with
  Ramond-Ramond flux}},
  \href{https://doi.org/10.1088/1751-8113/49/41/41LT03}{\emph{J. Phys.}
  {\bfseries A49} (2016) 41LT03}
  [\href{https://arxiv.org/abs/1605.00518}{{\ttfamily 1605.00518}}].

\bibitem{Fleury:2017eph}
T.~Fleury and S.~Komatsu, \emph{{Hexagonalization of Correlation Functions II:
  Two-Particle Contributions}},
  \href{https://doi.org/10.1007/JHEP02(2018)177}{\emph{JHEP} {\bfseries 02}
  (2018) 177} [\href{https://arxiv.org/abs/1711.05327}{{\ttfamily
  1711.05327}}].

\bibitem{Maldacena:2015iua}
J.~Maldacena, D.~Simmons-Duffin and A.~Zhiboedov, \emph{{Looking for a bulk
  point}}, \href{https://doi.org/10.1007/JHEP01(2017)013}{\emph{JHEP}
  {\bfseries 01} (2017) 013}
  [\href{https://arxiv.org/abs/1509.03612}{{\ttfamily 1509.03612}}].

\bibitem{Klose:2010ki}
T.~Klose, \emph{Review of {AdS/CFT} integrability, {C}hapter {IV.3}: {$\superN
  = 6$} {C}hern-{S}imons and strings on {$\AdS_{4} \times \CP^3$}},
  \href{https://doi.org/10.1007/s11005-011-0520-y}{\emph{Lett. Math. Phys.}
  {\bfseries 99} (2010) 401} [\href{https://arxiv.org/abs/1012.3999}{{\ttfamily
  1012.3999}}].

\bibitem{Borsato:2015mma}
R.~Borsato, O.~Ohlsson~Sax, A.~Sfondrini and B.~Stefa{\'n}ski, jr., \emph{The
  {$\AdS_{3} \times \Sphere^3 \times \Sphere^3 \times \Sphere^1$} worldsheet
  {S} matrix}, \href{https://doi.org/10.1088/1751-8113/48/41/415401}{\emph{J.
  Phys.} {\bfseries A48} (2015) 415401}
  [\href{https://arxiv.org/abs/1506.00218}{{\ttfamily 1506.00218}}].

\bibitem{Beisert:2008tw}
N.~Beisert and P.~Koroteev, \emph{Quantum deformations of the one-dimensional
  {H}ubbard model},
  \href{https://doi.org/10.1088/1751-8113/41/25/255204}{\emph{J. Phys.}
  {\bfseries A41} (2008) 255204}
  [\href{https://arxiv.org/abs/0802.0777}{{\ttfamily 0802.0777}}].

\bibitem{Delduc:2013qra}
F.~Delduc, M.~Magro and B.~Vicedo, \emph{An integrable deformation of the
  {$\AdS_{5} \times \Sphere^5$} superstring action},
  \href{https://doi.org/10.1103/PhysRevLett.112.051601}{\emph{Phys. Rev. Lett.}
  {\bfseries 112} (2014) 051601}
  [\href{https://arxiv.org/abs/1309.5850}{{\ttfamily 1309.5850}}].

\bibitem{Arutyunov:2013ega}
G.~Arutyunov, R.~Borsato and S.~Frolov, \emph{{S}-matrix for strings on
  {$\eta$}-deformed {$\AdS_{5} \times \Sphere^5$}},
  \href{https://doi.org/10.1007/JHEP04(2014)002}{\emph{JHEP} {\bfseries 1404}
  (2014) 002} [\href{https://arxiv.org/abs/1312.3542}{{\ttfamily 1312.3542}}].

\bibitem{Hoare:2018ngg}
B.~Hoare and F.~K. Seibold, \emph{{Supergravity backgrounds of the
  $\eta$-deformed AdS$_2 \times S^2 \times T^6 $ and AdS$_5 \times S^5$
  superstrings}}, \href{https://doi.org/10.1007/JHEP01(2019)125}{\emph{JHEP}
  {\bfseries 01} (2019) 125}
  [\href{https://arxiv.org/abs/1811.07841}{{\ttfamily 1811.07841}}].

\bibitem{Seibold:2020ywq}
F.~K. Seibold, S.~J. van Tongeren and Y.~Zimmermann, \emph{{The twisted story
  of worldsheet scattering in $\eta$-deformed $AdS_5 \times S^5$}},
  \href{https://doi.org/10.1007/JHEP12(2020)043}{\emph{JHEP} {\bfseries 12}
  (2020) 043} [\href{https://arxiv.org/abs/2007.09136}{{\ttfamily
  2007.09136}}].

\end{thebibliography}\endgroup
\end{document}